\documentclass[twocolumn]{aastex631}
\usepackage{natbib}
\usepackage{color}
\def    \apjl  		{\rm {ApJL}}
\def    \apj  		{\rm {ApJ}}
\def    \mnras  	{\rm {MNRAS}}
\def    \araa  		{\rm {ARA\& A}}
\def    \apjl  		{\rm {ApJL}}

\def	\cm		{\,{\rm {cm}}}
\def	\K		{\,{\rm K}}
\def	\g		{\,{\rm {g}}}

\def \bea {\begin{eqnarray}}
\def \ena {\end{eqnarray}}

\def	\B	{{\rm B}}

\def	\C	{{\rm C}}

\def	\cm	{\,{\rm cm}}
\def	\d	{{\rm d}}

\def	\D	{{\rm D}}

\def	\g	{\,{\rm g}}

\def	\G	{{\rm G}}

\def	\H	{{\rm H}}

\def	\micron	{\mu{\rm m}}

\def	\N	{{\rm N}}

\def	\s	{\,{\rm s}}

\def	\V	{{\rm V}}

\newcommand{\ratio}{{\langle B_{t}^{2} \rangle}/{\langle B_{0}^{2} \rangle}}

\usepackage{amsmath}
\renewcommand{\vec}[1]{\boldsymbol{#1}}

\begin{document}
\shorttitle{Magnetic fields and gas kinematics in 30 Doradus}
\shortauthors{Tram et al.}
\title{SOFIA observations of 30 Doradus: II - Magnetic fields and large scale gas kinematics}

\author{Le Ngoc Tram}
\affiliation{Max-Planck-Institut f\"ur Radioastronomie, Auf dem H\"ugel 69, 53121, Bonn, Germany, \href{mailto:nle@mpifr-bonn.mpg.de}{nle@mpifr-bonn.mpg.de}}

\author{Lars Bonne}
\affiliation{Stratospheric Observatory for Infrared Astronomy, Universities Space Research Association, NASA Ames Research Center, MS 232-11, Moffett Field, 94035 CA, USA}

\author{Yue Hu}
\affiliation{Department of Physics, University of Wisconsin-Madison, Madison, WI 53706, USA}
\affiliation{Department of Astronomy, University of Wisconsin-Madison, Madison, WI 53706, USA}

\author{Enrique Lopez-Rodriguez}
\affiliation{Kavli Institute for Particle Astrophysics and Cosmology (KIPAC), Stanford University, Stanford, CA 94305, USA}

\author{Jordan A. Guerra}
\affiliation{Department of Physics, Villanova University, 800 E. Lancaster Ave., Villanova, PA 19085, USA}

\author{Pierre Lesaffre}
\affiliation{Laboratoire de Physique de l’École normal supérieur, ENS, Université PSL, CNRS, Sorbonne Université, Université de Paris, France}
\affiliation{Observatoire de Paris, PSL University, Sorbonne Université, LERMA, F-75014, Paris, France}

\author{Antoine Gusdorf}
\affiliation{Laboratoire de Physique de l’École normal supérieur, ENS, Université PSL, CNRS, Sorbonne Université, Université de Paris, France}
\affiliation{Observatoire de Paris, PSL University, Sorbonne Université, LERMA, F-75014, Paris, France}

\author{Thiem Hoang}
\affiliation{Korea Astronomy and Space Science Institute, Daejeon 34055, South Korea}
\affiliation{Korea University of Science and Technology, 217 Gajeong-ro, Yuseong-gu, Daejeon, 34113, South Korea}

\author{Min-Young Lee}
\affiliation{Korea Astronomy and Space Science Institute, Daejeon 34055, South Korea}

\author{Alex Lazarian}
\affiliation{Department of Astronomy, University of Wisconsin-Madison, Madison, WI 53706, USA}
\affiliation{Centro de Investigación en Astronomía, Universidad Bernardo O’Higgins, Santiago, General Gana 1760, 8370993, Chile}

\author{B-G Andersson}
\affiliation{Stratospheric Observatory for Infrared Astronomy, Universities Space Research Association, NASA Ames Research Center, MS 232-11, Moffett Field, 94035 CA, USA}

\author{Simon Coud\'e}
\affiliation{Stratospheric Observatory for Infrared Astronomy, Universities Space Research Association, NASA Ames Research Center, MS 232-11, Moffett Field, 94035 CA, USA}

\author{Archana Soam}
\affiliation{Stratospheric Observatory for Infrared Astronomy, Universities Space Research Association, NASA Ames Research Center, MS 232-11, Moffett Field, 94035 CA, USA}

\author{William D. Vacca}
\affiliation{Stratospheric Observatory for Infrared Astronomy, Universities Space Research Association, NASA Ames Research Center, MS 232-11, Moffett Field, 94035 CA, USA}

\author{Hyeseung Lee}
\affiliation{Korea Astronomy and Space Science Institute, Daejeon 34055, South Korea}

\author{Michael Gordon}
\affiliation{Ball Aerospace, Boulder, 80301 CO, USA}

\begin{abstract}
The heart of the Large Magellanic Cloud, 30 Doradus, is a complex region with a clear core-halo structure. Feedback from the stellar cluster R$\,$136 has been shown to be the main source of energy creating multiple pc-scale expanding-shells in the outer region, and carving a nebula core in the proximity of the ionization source. We present the morphology and strength of the magnetic fields (B-fields) of 30 Doradus inferred from the far-infrared polarimetric observations by SOFIA/HAWC+ at 89, 154, and 214$\,\mu$m. The B-field morphology is complex, showing bending structures around R$\,$136. In addition, we use high spectral and angular resolution [\textsc{CII}] observations from SOFIA/GREAT and CO(2-1) from APEX. The kinematic structure of the region correlates with the B-field morphology and shows evidence of multiple expanding shells. Our B-field strength maps, estimated using the Davis-Chandrasekhar-Fermi method and structure-function, show variations across the cloud within a maximum of 600, 450, and 350$\,\mu$G at 89, 154, and 214$\,\mu$m, respectively. We estimated that the majority of the 30 Doradus clouds are sub-critical and sub-Alfv\'enic. The probability distribution function of the gas density shows that the turbulence is mainly compressively driven, while the plasma beta parameter indicates supersonic turbulence. We show that the B-field is sufficient to hold the cloud structure integrity under feedback from R$\,$136. We suggest that supersonic compressive turbulence enables the local gravitational collapse and triggers a new generation of stars to form. The velocity gradient technique (VGT) using [\textsc{CII}] and CO(2-1) is likely to confirm these results.
\end{abstract}

\section{Introduction\label{sec:intro}}
In modern Astrophysics, magnetic fields (B-fields) and turbulence are believed to affect the star-formation process. The B-fields support against gravitational collapse, while turbulence plays a dual role. Turbulence can against cloud global collapse, but can also produce local compression (\citealt{2004RvMP...76..125M}) with the compressible and solenoidal motions acting in the opposite directions (\citealt{2003MNRAS.345..325C}). The role of B-fields can be different depending on whether B-fields are dynamically important or subdominant (see a review in \citealt{2012ARA&A..50...29C}). For weak B-fields, the cloud is super-critical (the mass-to-flux ratio is greater than unity), and the B-fields are insufficient to prevent gravitational collapse. For strong B-fields, the fields are strong enough to counteract the collapse. In this case, other mechanisms must be invoked for star formation to occur in the sub-critical cloud (the mass-to-flux ratio is lower than unity). There are two candidates for such mechanisms: (1) ambipolar diffusion (e.g., \citealt{1966MNRAS.133..265M}) can increase the mass faster than the B-field strength, enhancing the gravitational counterpart, and (2) fast turbulent reconnection (\citealt{1999ApJ...517..700L}) removes the magnetic flux, weakening the magnetic support. These two mechanisms are able to increase the mass-to-flux ratio, which can lead clouds to collapse and possibly to co-exist (\citealt{2014SSRv..181....1L}).

The fast turbulent reconnection induces a turbulence cascade perpendicular to the ambient B-fields. As a result, turbulent eddies can freely mix the B-fields parallel to the rotation axes, where the velocity gradients (VGs) are perpendicular to the local direction of B-fields. This is the basis of the velocity gradient technique (VGT; \citealt{2017ApJ...835...41G}) to study B-fields in diffuse gas (\citealt{2017ApJ...837L..24Y}; \citealt{2018ApJ...853...96L}; \citealt{2018ApJ...865...46L}; \citealt{2018MNRAS.480.1333H,2019ApJ...886...17H}), in molecular clouds (\citealt{2019NatAs...3..776H}; \citealt{2019ApJ...878...10T}; \citealt{2021ApJ...912....2H, 2021arXiv210503605H}; \citealt{2022A&A...658A..90A}), and in the atomic-molecular transition (\citealt{2021arXiv211011878S}). Nevertheless, self-gravity is able to break that relationship. The gravitational forces pull the gas in the direction along the B-fields, so the VGs are dominated by the infall acceleration. In this case, the VGs are parallel to the ambient B-fields. The misalignment between VGT and B-fields becomes a proxy for the gravitational collapse (\citealt{2017arXiv170303026Y};\citealt{2018ApJ...853...96L};\citealt{2019ApJ...878...10T}; \citealt{2021ApJ...912....2H}). Therefore, VGT is a promising tool to probe the B-fields morphology and the local gravitational collapse.

The B-fields, turbulence, and stellar feedback shape the cloud and regulate the star-formation processes. The simulations with uniform B-fields from \cite{2009MNRAS.398..157H} and \cite{2011MNRAS.412.2079M} showed that the B-fields have a significant contribution in shaping the cloud. This result depends on the orientation between the initial B-fields and the radiation from the source. Specifically, these authors found that clouds are flatter (broad head) if the B-fields are parallel to the radiation direction, while the cloud becomes a more elongated structure (tail-like structure) if the B-fields are perpendicular to the radiation direction. These features appear to be confirmed from observations, e.g., IC 1396 (\citealt{2018MNRAS.476.4782S}), M16 (\citealt{2018ApJ...860L...6P}), Ophiuchus-A (\citealt{2019ApJ...882..113S}). The simulations of the feedback in a turbulent magnetized cloud by \cite{2011MNRAS.414.1747A} showed that the B-fields tend to be amplified and slow down the formation of stars.

The morphology of the B-fields is affected by gravity (resulting in a well-known hour-glass shape; \citealt{2013ApJ...767...33E}), and regulated by the supersonic gas motion as proposed by \cite{2013ApJ...774L..31I}. The latter seems to be frequently observed, e.g., deformation of B-field geometry in M16 (\citealt{2018ApJ...860L...6P}), Orion-A (\citealt{2019A&A...632A..68T}), Musca filament (\citealt{2020A&A...641A..17B,2020A&A...644A..27B}) and BHR 71 bipolar outflow system (\citealt{2020ApJ...892..128K}). Using simulations, \cite{2021ApJ...916...83A} demonstrated that a shock with a velocity of $\sim 7\,\rm km\, s^{-1}$ is able to wrap the B-fields.

Even though B-fields may affect the star formation processes, the direct measurement of B-fields is difficult. Alternatively, the B-fields are inferred using several data analysis methods. One of the methods consists in using dust polarization (see e.g., \citealt{2007JQSRT.106..225L} and \citealt{2015ARA&A..53..501A} for reviews). The basic idea of this technique relies on the fact that irregular dust grains tend to align with their shortest axis parallel to the local B-fields due to various physical effects (see e.g., \citealt{2022arXiv220502334H} for details) so that their thermal emission is polarized with the polarization vector perpendicular to the B-fields (see \citealt{2022arXiv220813195T}). The measured position angle of thermal dust polarization is then perpendicular to the local B-fields in the plane of the sky. Hence, the polarimetric data allows us to map the B-field geometry by rotating the polarization angles (E-vectors) by $90^{o}$. The polarized thermal dust emission is feasible at long wavelengths, i.e., FIR to submm. The strength of the B-fields on the plane of the sky ($B_{\rm POS}$) can be estimated using the Davis-Chandrasekhar-Fermi (DCF; \citealt{1951PhRv...81..890D}; \citealt{1953ApJ...118..113C}) method. This method is commonly used, although some modifications need to be taken into account as a function of the object to be analyzed (\citealt{2022ApJ...925...30L}). Another approach (namely Differential Measure Approach or DMA) is recently proposed by \cite{2020arXiv200207996L,2022ApJ...935...77L}, which is suggested to be able to measure the B-field strength locally. 

\begin{figure}
    \centering
    \includegraphics[width=0.5\textwidth]{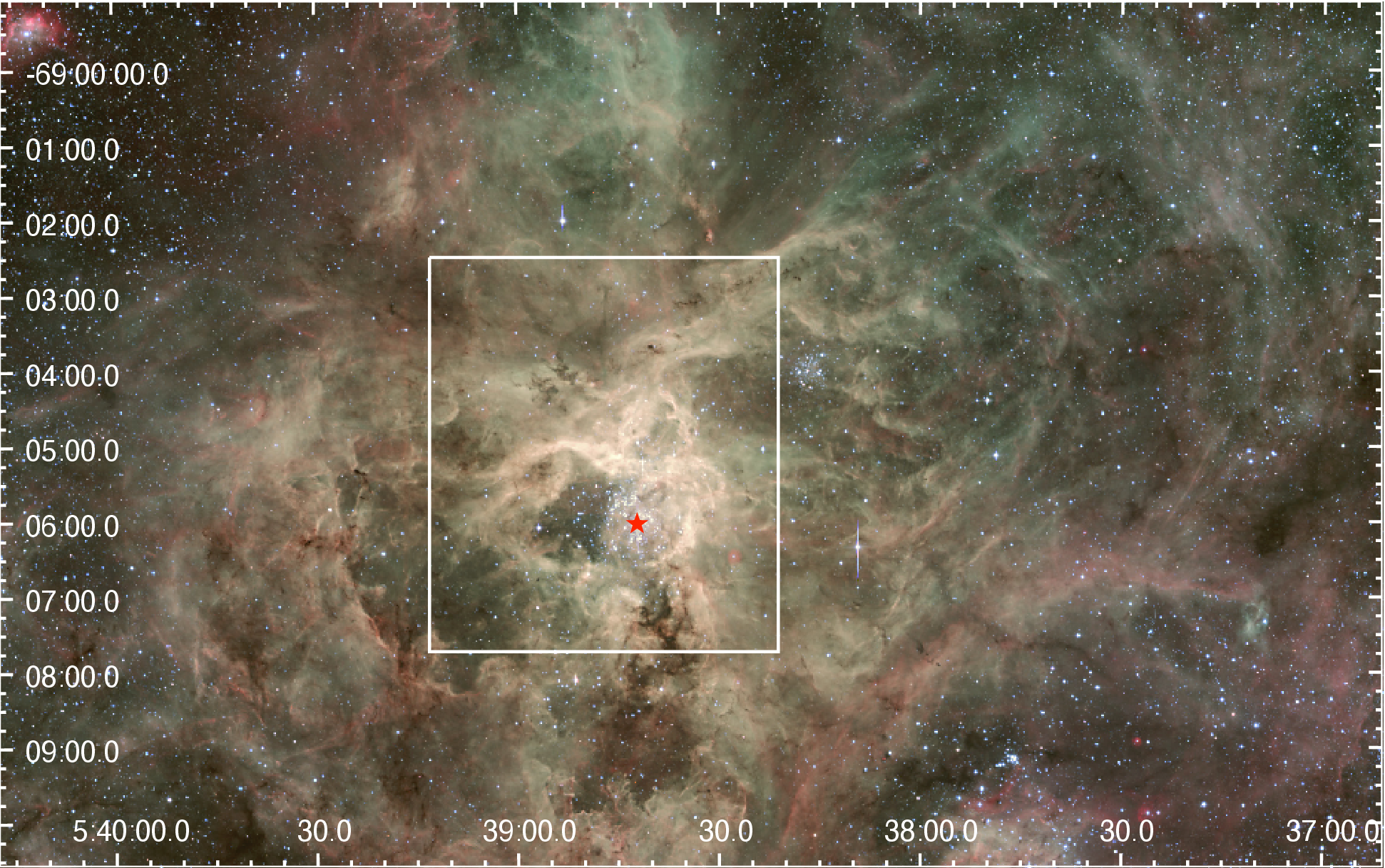}
    \caption{Public composite image of 30 Dor observed by La Silla 2.2$\,$m telescope with $H_{\alpha}$-658.827$\,$nm (red), a combination of V-539.562$\,$nm and [\textsc{OIII}]-502.393$\,$nm (green), and B-451.100$\,$nm (blue). This image shows a complex structure of the region with multiple large expanding shells produced by the hot cluster-wind from R$\,$136 (indicated by a red star), and a slow expanding-shell from the supernova remnant 30DorB (lower right). The white box shows the region covered by SOFIA/HAWC+ that we analyze in this work.}
    \label{fig:RGB_image}
\end{figure}

Our target is the star-formation region, 30 Doradus (hereafter 30 Dor) in the Large Magellanic Cloud (LMC). With a distance of $\simeq 50\,$kpc away from Earth (\citealt{2008AJ....135..112S}), it is close enough to obtain parsec-scale resolutions to study the impact of the feedback and turbulence on the surrounding molecular cloud.
30 Dor hosts a massive star cluster, R$\,$136, which is associated to the HII giant expanding-shells (\citealt{1984ApJ...287..116K}; \citealt{1994ApJ...425..720C}; \citealt{2005ASSL..329...49B}; \citealt{2006AJ....131.2140T}; \citealt{2011ApJ...731...91L}), and a nearby Supernova Remnant (\citealt{2006AJ....131.2140T}). Figure \ref{fig:RGB_image}\footnote{public data at https://archive.eso.org/cms/eso-data/data-packages/30-doradus.html} shows a composite image of the 30 Dor region with an overlaid of the field-of-view covered in our study. This complex system is embedded by multiple HI giant-shells (\citealt{1999AJ....118.2797K}). A combination of stellar winds and supernovae (\citealt{1994ApJ...425..720C}) or only the cluster-wind (not the stellar wind of individual stars) from R$\,$136 (\citealt{2021A&A...649A.175M}) are demonstrated to be the main sources to create these giant \textsc{HII} expanding-structures. The authors also clearly unveiled two structures in 30 Dor. For the nebula's core (within a distance of $25\,\rm pc$ proximity to R$\,$136), surprisingly, the thermal gas pressure is lower than that of the stellar radiation (see figure 18 in \citealt{2011ApJ...738...34P}), and the mass is lower than the virial mass (\citealt{2021A&A...649A.175M}). Hence, the important questions remaining are: how can this structure survive? and how can stars form? (the locations of protostar candidates are shown in e.g., \citealt{2019A&A...628A.113L}; \citealt{2009ApJ...694...84I}). Here, we focus on the closest region to R$\,$136, which is indicated by the white box in Figure \ref{fig:RGB_image}. For the sake of simplicity, we refer to this region as 30 Dor in this work.

Our goals are to 
\begin{itemize}
    \item[1.] map the morphology and strength of the B-fields in 30 Dor using FIR polarimetric observations with SOFIA/HAWC+ as introduced in \cite{2021arXiv210509530T} (hereafter Paper I); 
    \item[2.] examine the gas kinematic in 30 Dor by making use of [CII] and CO(2-1) data acquired by SOFIA/GREAT and APEX (see \citealt{2019A&A...621A..62O}); 
    \item[3.] make use of the VGT to probe the local gravitational collapse in 30 Dor; 
    \item[4.] quantify the effect of B-fields on supporting the cloud integrity; and 
    \item[5.] perform an energy budget to quantify the effect of gravity, B-fields, and turbulence on the star-forming processes of 30 Dor.
\end{itemize}

This paper is structured as follows. We analyze the gas kinematics in Section \ref{sec:gas_kinematic}. The analysis of the B-field orientations and strengths are shown in Section \ref{sec:Bfields}. Our discussions are presented in Section \ref{sec:discussions}. The conclusions are presented in Section \ref{sec:conclusions}. 

\section{Large scale kinematics in 30 Doradus}\label{sec:gas_kinematic}
\begin{figure}
    \centering
    \includegraphics[width=0.4\textwidth]{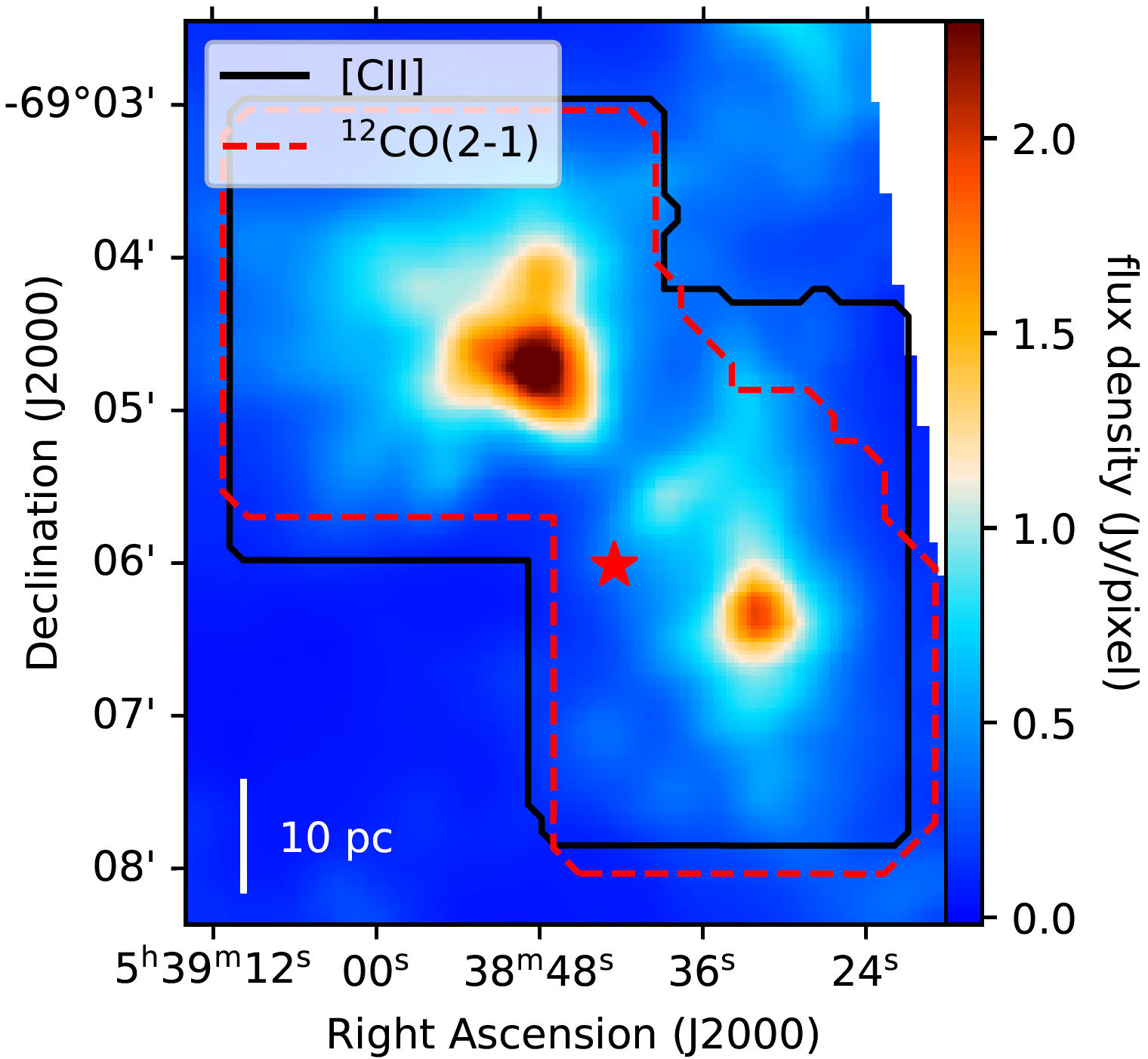}
    \caption{HAWC+ dust continuum flux density map at 214$\,\mu$m. The black full line indicates the region that is covered by the [CII] observations and the red dashed line outlines the region covered by the $^{12}$CO(2-1) observations (\citealt{2019A&A...621A..62O}). The red star indicates the location of R136.}
    \label{fig:covered}
\end{figure}

To analyze the pc scale kinematics of 30 Dor at the location of the HAWC+ observations, we make use of the spectral lines presented in \citet{2019A&A...621A..62O} that were observed with SOFIA \citep{2012ApJ...749L..17Y} \& the APEX telescope \citep{2006A&A...454L..13G}. For details on the data reduction, we refer to \citet{2019A&A...621A..62O}. The SOFIA observations of the [CII] line have a spatial resolution of 16$^{\prime\prime}$ and the APEX CO(2-1) observations have a spatial resolution of 30$^{\prime\prime}$. The region of the HAWC+ map and of the [CII] and $^{12}$CO(2-1) is presented in Figure \ref{fig:covered}. Both lines cover a similar area of 30 Dor where the dust polarization is detected.

\begin{figure*}
    \centering
    \includegraphics[width=0.43\hsize]{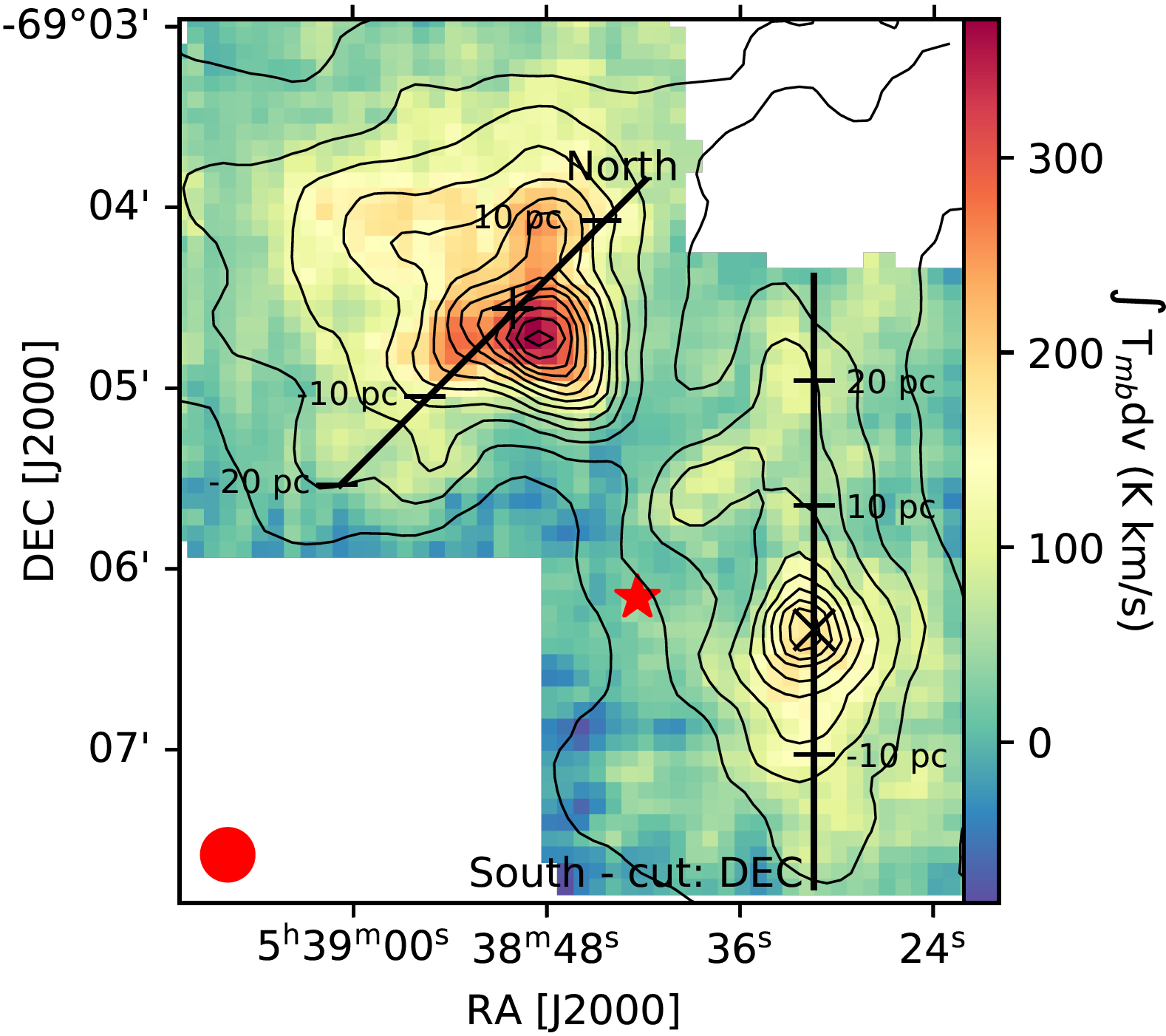}
    \includegraphics[width=0.43\hsize]{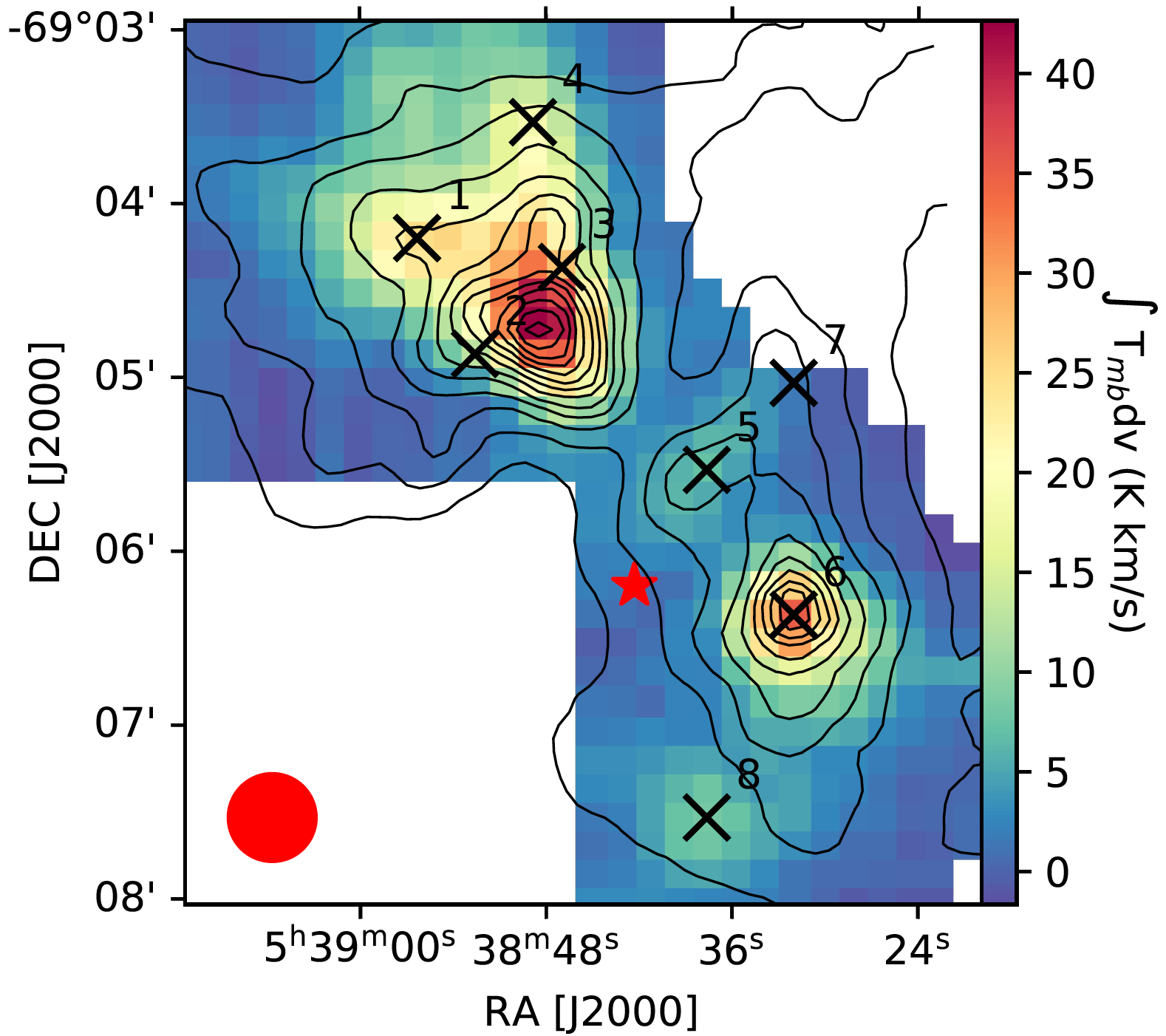}
    \includegraphics[width=0.43\hsize]{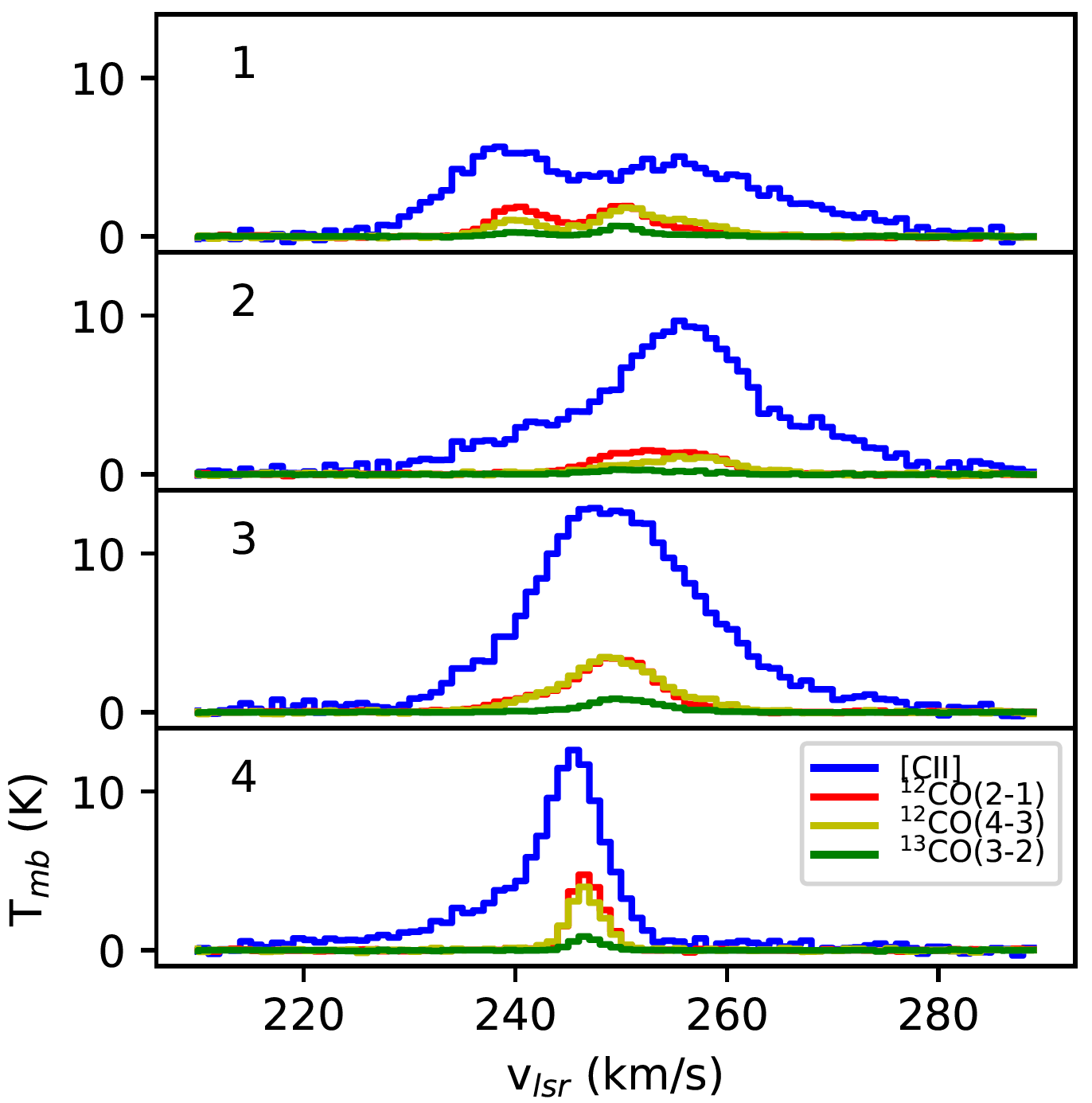}
    \includegraphics[width=0.43\hsize]{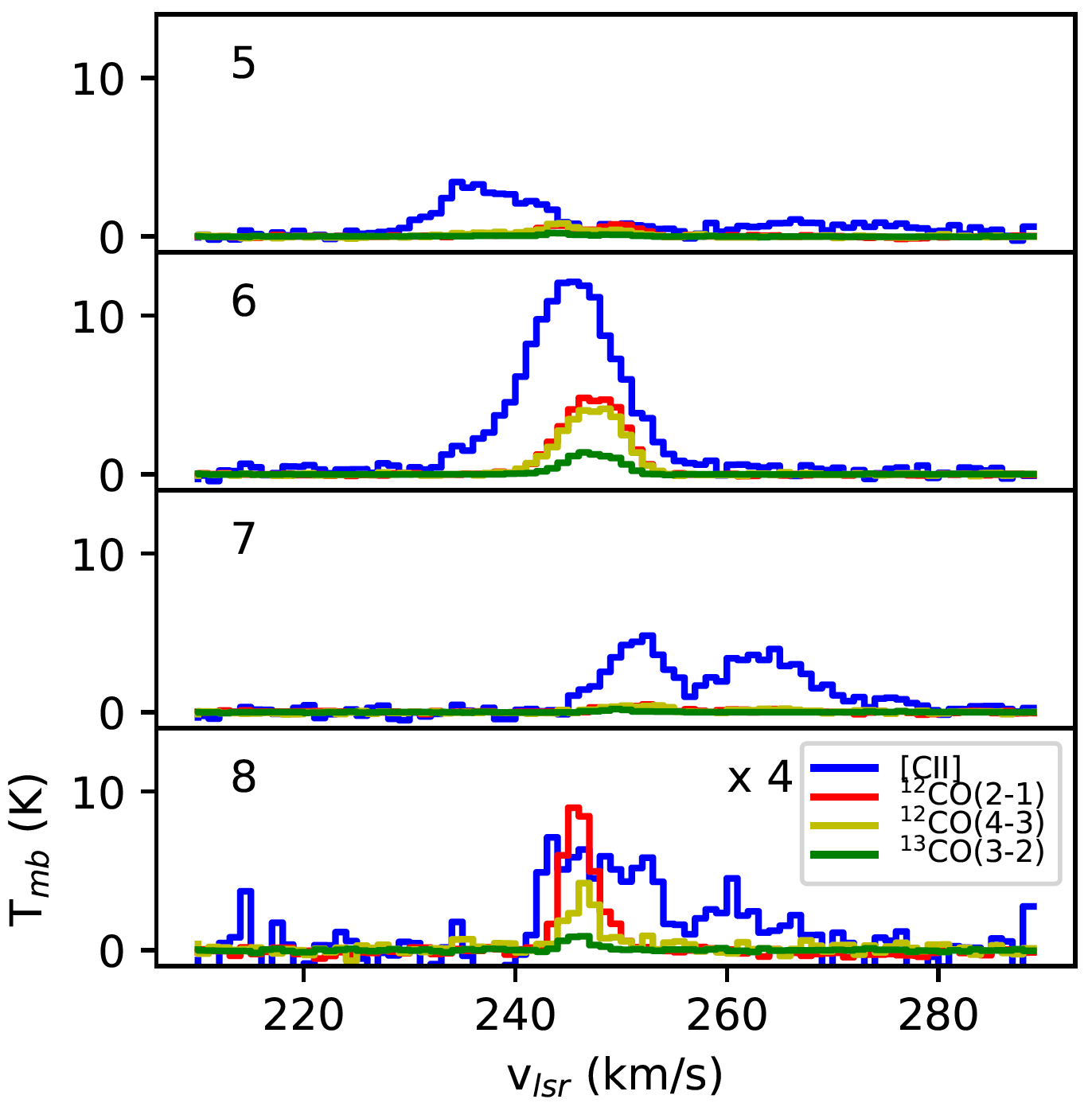}
    \caption{{\bf Left}: [CII] integrated intensity map of 30 Dor. The black contours show the HAWC+ dust continuum emission at 214$\,\mu$m starting at 0.1 Jy pixel$^{-1}$ with increments of 0.1$\,$Jy$\,$pixel$^{-1}$. The two lines indicate the spatial axes that were used to construct the position-velocity (PV) diagrams presented in Figure \ref{fig:pvDiagrams}. The crosses indicate the center, i.e. 0 pc, of the PV diagrams, and the horizontal locations indicate the physical distance along the axis at the location of 30 Dor. The red star indicates the location of R136. {\bf Right}: The integrated intensity map of $^{12}$CO(2-1), overlaid with the same HAWC+ contours shown on the left. The crosses indicate the spatial locations of the spectra shown below. {\bf Bottom}: The extracted [CII], $^{12}$CO(2-1), $^{12}$CO(4-3) and $^{13}$CO(3-2) spectra at the indicated locations over the 30 Dor region, displaying multiple components and several wings in [CII].}
    \label{fig:integratedMaps}
\end{figure*}

\subsection{The integrated intensity maps}

In Figure \ref{fig:integratedMaps}, the integrated intensity maps of [CII] and $^{12}$CO(2-1) are compared with the dust continuum maps from HAWC+ at 214$\,\mu$m. This figure shows that the [CII] peak intensities are closely correlated with the dust continuum peak and that it also traces the extended emission. We find that $^{12}$CO(2-1) is more sensitive to the gas located in the high column density regions, which can be expected for this lower metallicity region. As a result, the [CII] emission will be more sensitive to the kinematics of the ambient gas, whereas CO is more sensitive to the dense clumps in the cloud.

\subsection{The velocity structure of 30 Doradus}

\begin{figure*}
    \centering
    \includegraphics[width=0.48\hsize]{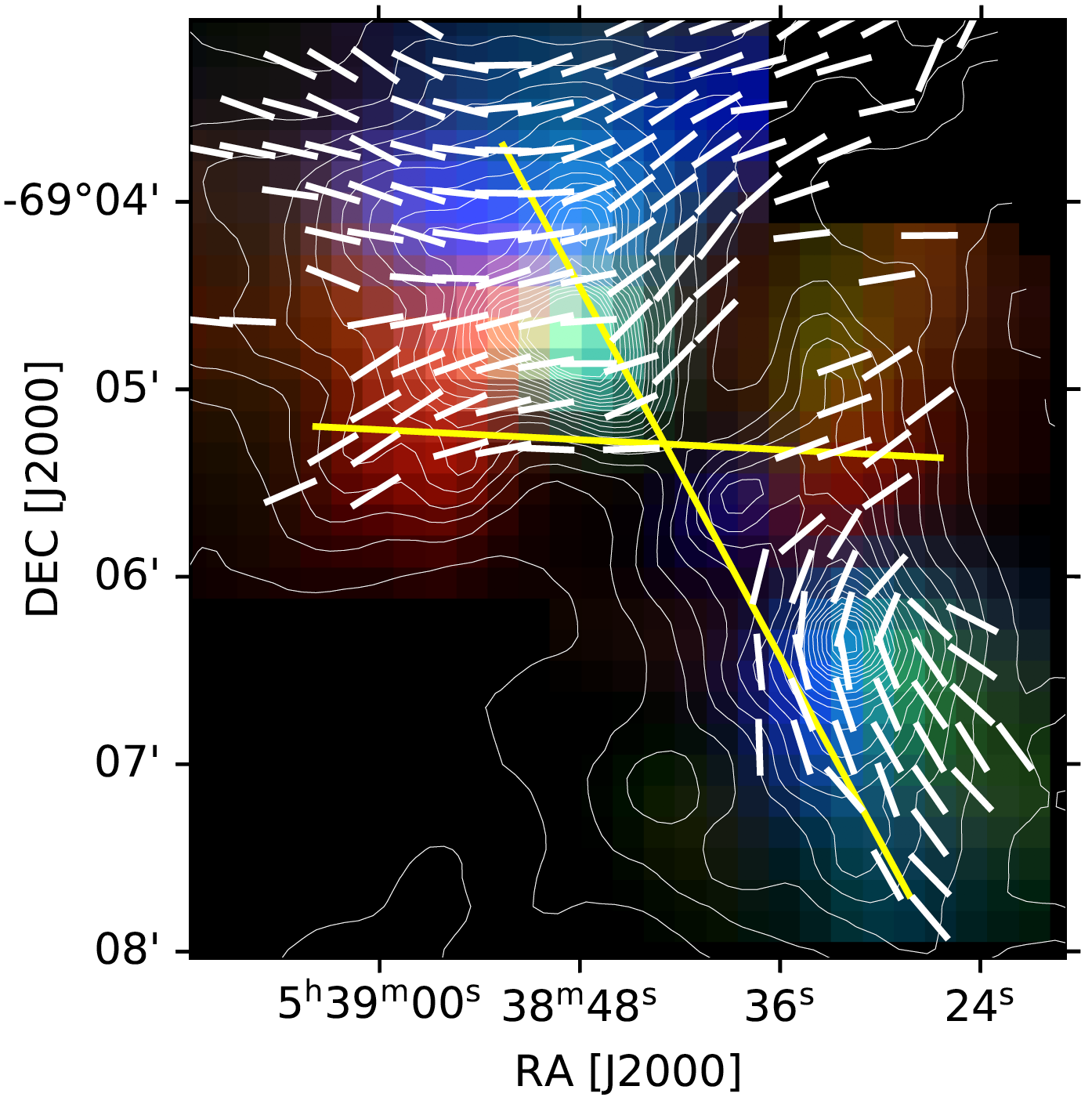}
    \includegraphics[width=0.48\hsize]{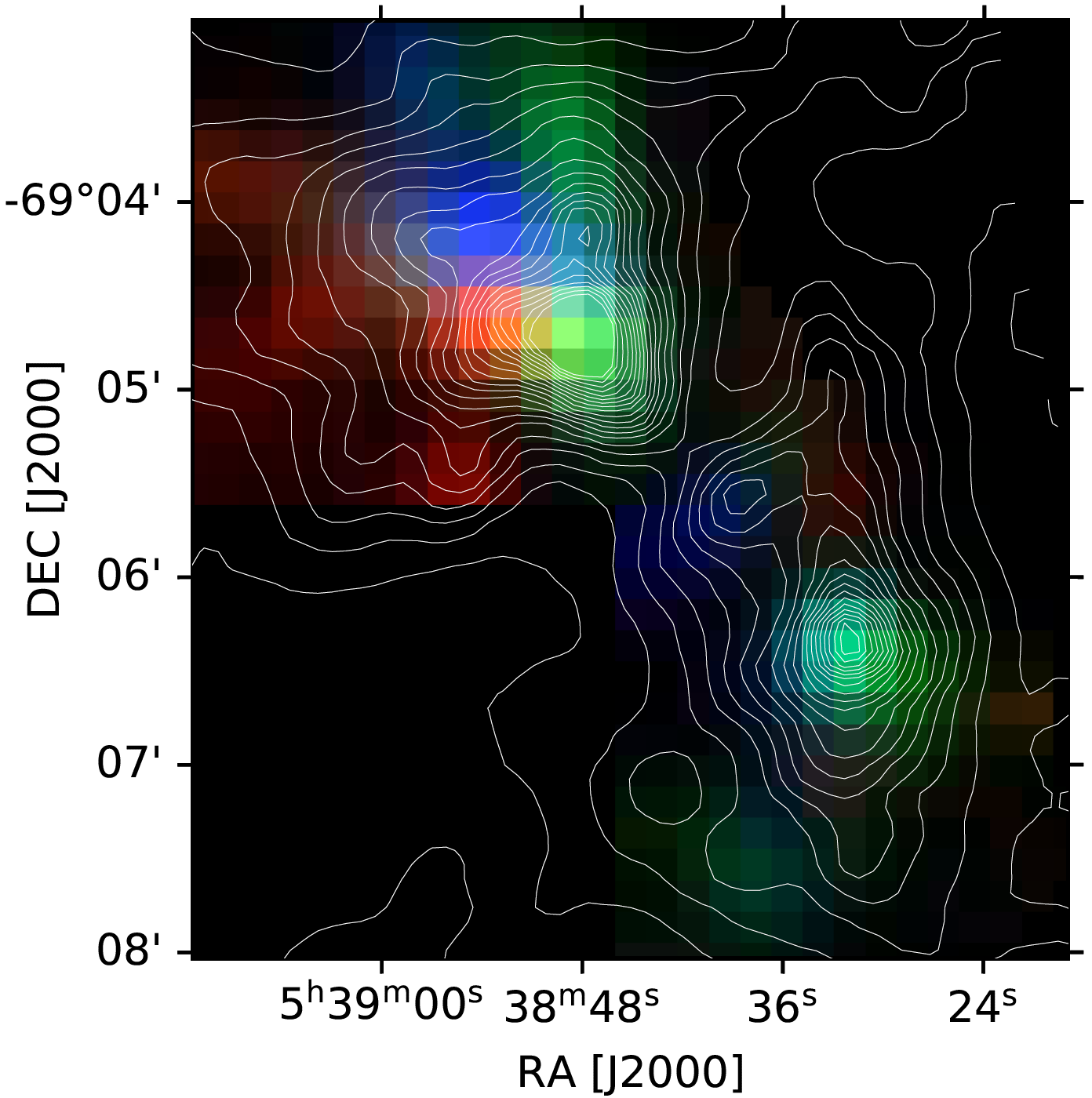}
    \caption{{\bf Left}: RGB image of 30 Dor for [CII] with blue: 235-245$\,\rm km\,s^{-1}$, green: 245-255$\,\rm km\,s^{-1}$ and red: 255-270$\,\rm km\,s^{-1}$. The contours indicate the HAWC+ 214$\,\mu$m dust continuum emission starting at 0.1$\,$Jy$\,$pixel$^{-1}$ with increments of 0.1$\,$Jy$\,$pixel$^{-1}$. The yellow cross is added to guide the eye at the axes of what appears to be dominantly blue- and red-shifted gas. The white segments indicate the magnetic field morphology on the RGB image. {\bf Right}: The same for $^{12}$CO(2-1).}
    \label{fig:RGBims}
\end{figure*}

\begin{figure*}
    \centering
    \includegraphics[width=0.49\hsize]{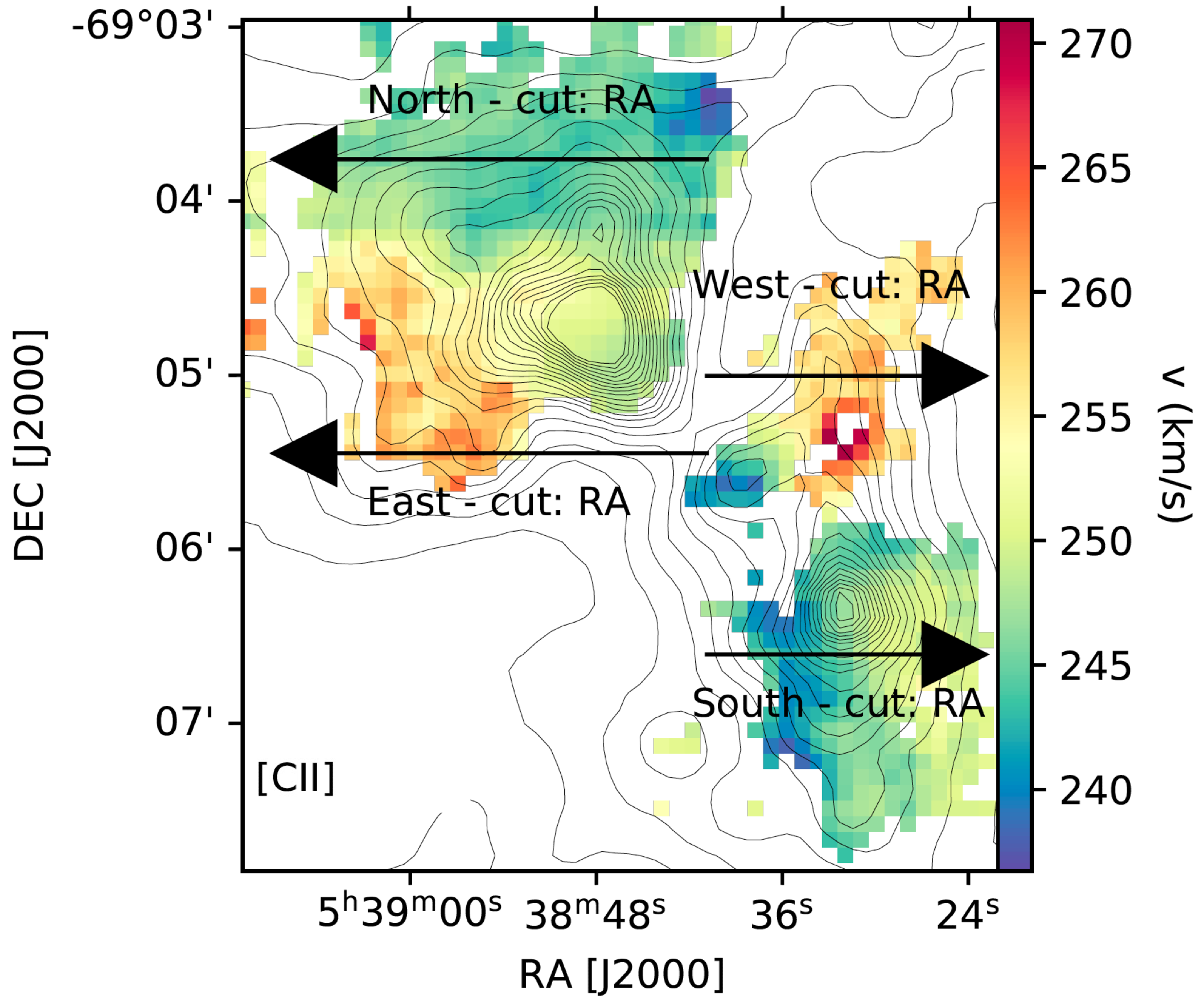}
    \includegraphics[width=0.49\hsize]{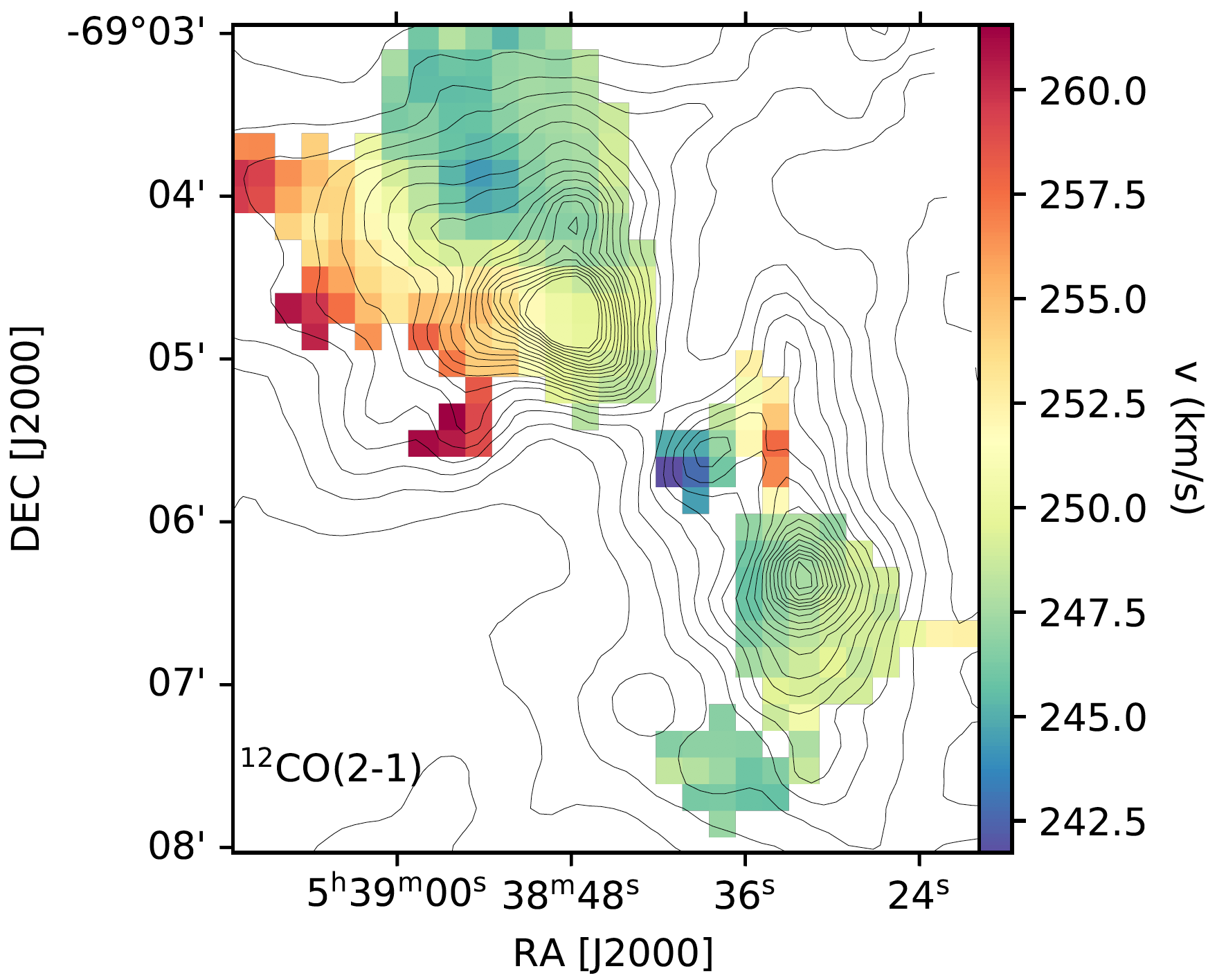}
    \includegraphics[width=0.49\hsize]{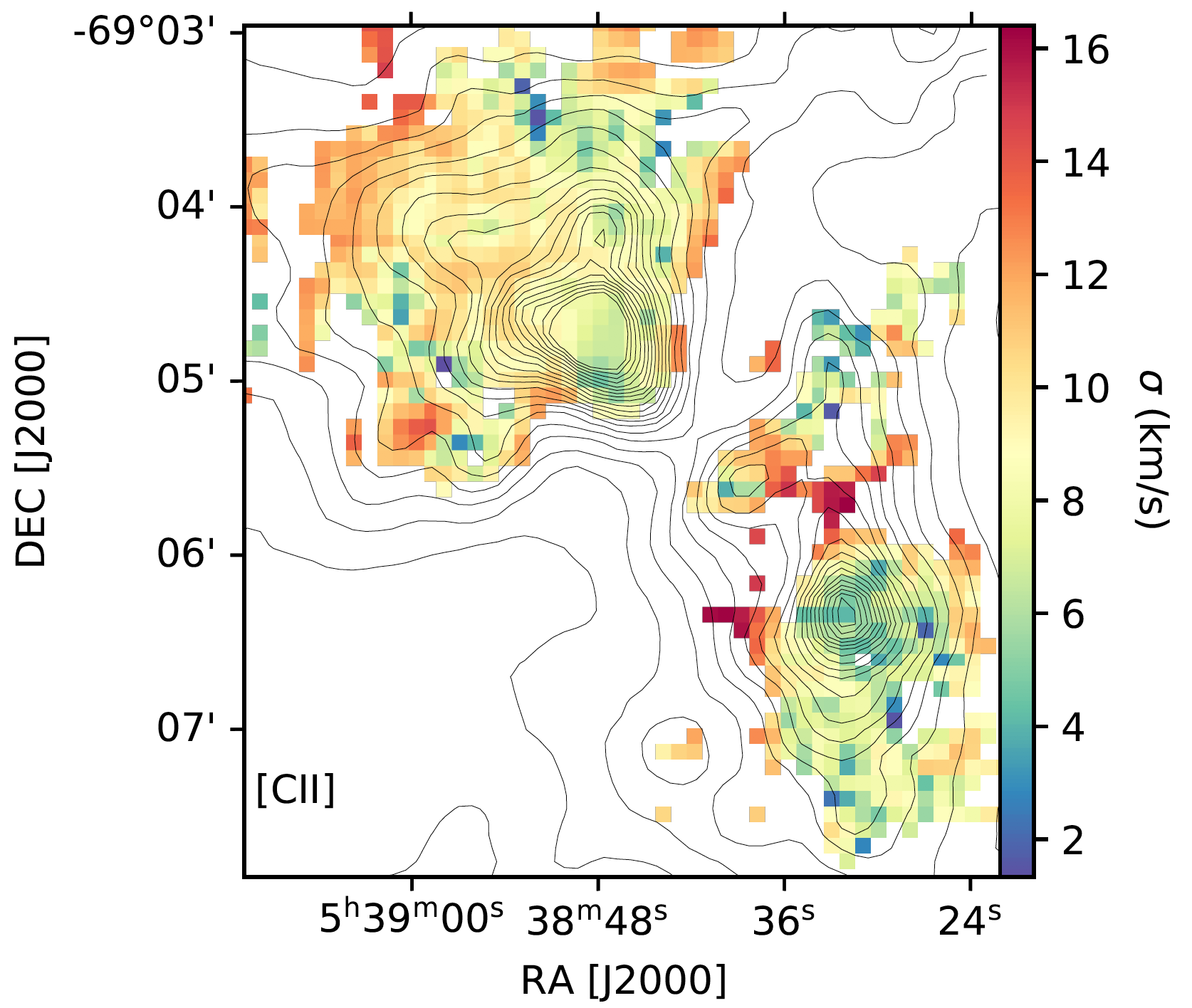}
    \includegraphics[width=0.49\hsize]{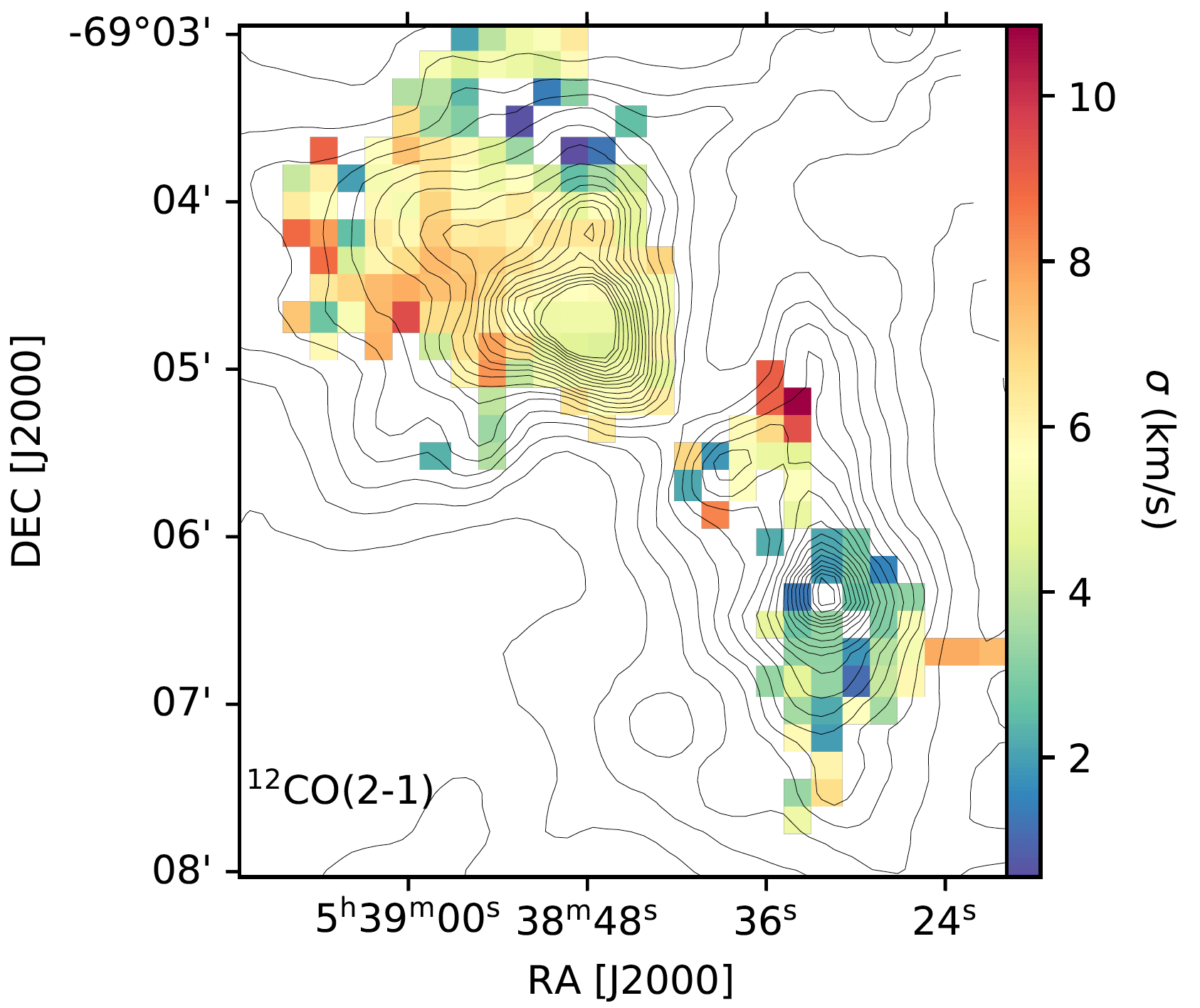}
    \caption{{\bf Upper Left}: First-moment map obtained with the [CII] line. The black contours indicate the HAWC+ emission at 214 $\mu$m starting at 0.1$\,$Jy$\,$pixel$^{-1}$ with increments of 0.1$\,$Jy$\,$pixel$^{-1}$. The black arrows indicate the regions, with their name, for the position-velocity (PV) diagrams presented in Figure \ref{fig:pvDiagrams}. The arrows indicate the increasing distance (d) in the PV diagrams. {\bf Lower left}: The second-moment map of [CII]. {\bf Right}: The same for $^{12}$CO(2-1).}
    \label{fig:momentMaps}
\end{figure*}

In \citet{2019A&A...621A..62O} and Figure \ref{fig:integratedMaps} several [CII] and CO spectra of 30 Dor are presented. These observations show that most of the emission is found between 220 and 280$\,$km$\,$s$^{-1}$. In this velocity range, multiple velocity components and high-velocity wings are observed. In Figure \ref{fig:integratedMaps}, the wings are most clear in spectra 1-4 and specifically in the velocity range between 220-240 km s$^{-1}$ and between 255-280$\,$km$\,$s$^{-1}$. Such wing-like structures were recently also found in spectrally resolved [CII] observations toward galactic HII regions \citep[e.g.][]{2018A&A...617A..45S,2020PASP..132j4301S,2021ApJ...914..117T,2022ApJ...935..171B}. The 30 Dor region has a complex dynamical structure. This is illustrated in more detail for [CII] and $^{12}$CO(2-1) in Figures \ref{fig:RGBims} and \ref{fig:chanMapsFull} (in App. \ref{sec:appChanMaps}). Despite the complexity of the kinematics and the fact that they trace different densities, Figure \ref{fig:RGBims} shows that both lines have a broadly similar velocity structure on large scales. Inspecting the velocity structure in Figure \ref{fig:RGBims}, it can be noted in particular with [CII], that the blue- and red-shifted gas appear to create intersecting axes toward the middle of the SOFIA map. This intersection is shown in Figure \ref{fig:RGBims} by the north-east to south-west axis for blue-shifted velocities and the east-to-west axis for red-shifted velocities. This intersecting structure formed by the blue and red axis is also observed in the channel maps of the region that are presented in Figure \ref{fig:chanMapsFull} of App. \ref{sec:appChanMaps}. 
The channel maps also confirm that the southern part of the maps consists of two subregions, which are discontinuous in velocity. This is consistent with the HAWC+ integrated maps, which show that this western region contains two clumps. Furthermore, these two subregions have a noticeably different B-field structure in Figure \ref{fig:LIC_maps}. In the northern region of the map it is also observed that there is a noteworthy change in B-field orientation at the transition from blue- to red-shifted gas which confirms that the B-field morphology is directly related to the large-scale kinematics of the 30 Dor region, see Figure \ref{fig:RGBims}. \\\\
From the data cubes, the moment maps can be calculated. However, before presenting these moment maps, it has to be noted that there are limitations to the moment maps due to the inherent complexity (i.e., multiple components and high-velocity wings) in the spectra. The resulting moment maps are presented in Figure \ref{fig:momentMaps}. The first-moment map (or the velocity field) has an organized gradient that looks similar in [CII] and $^{12}$CO(2-1) and fits with the velocity field obtained at smaller scales with ALMA observations \citep{2013ApJ...774...73I}. Even though both lines show a similar morphology, it is found that the [CII] velocity field covers a significantly larger velocity range. The CO kinematics thus tend to follow the [CII] kinematics, but less drastically.\\
Both lines are expected to trace different regions in the cloud because of the critical density\footnote{The critical density is defined by the balance between collisional de-excitation and spontaneous decay. The collisional rate and Einstein coefficients are adopted from https://home.strw.leidenuniv.nl/$\sim$moldata/. For [CII] and CO(2-1) this gives values of the order of $10^{3}$ and $10^{4}\,\rm cm^{-3}$, respectively.} and the fact that FUV radiation propagates further into the cloud in low-metallicity regions. 
From the second moment maps, presented in Figure \ref{fig:momentMaps}, it is observed that the linewidth significantly depends on the observed line. For this, it has to be taken into account that there are indications of multiple components, also reported in \cite{2021A&A...649A.175M}, which are particularly clear in [CII]. For example, two velocity components are prominent in the most north-eastern part of the map (see Figure \ref{fig:integratedMaps}), which is translated in a large second moment in that region. From the maps, it is also observed that the northern region has a higher second moment than the southern region, which could be related to the smaller velocity range covered by the velocity field in that region.  Both lines do show a common behavior for the second moment over the map. Specifically, they show a reduction in linewidth towards the densest gas of the northern and southern regions.

\begin{figure*}
    \centering
    \includegraphics[width=0.495\hsize]{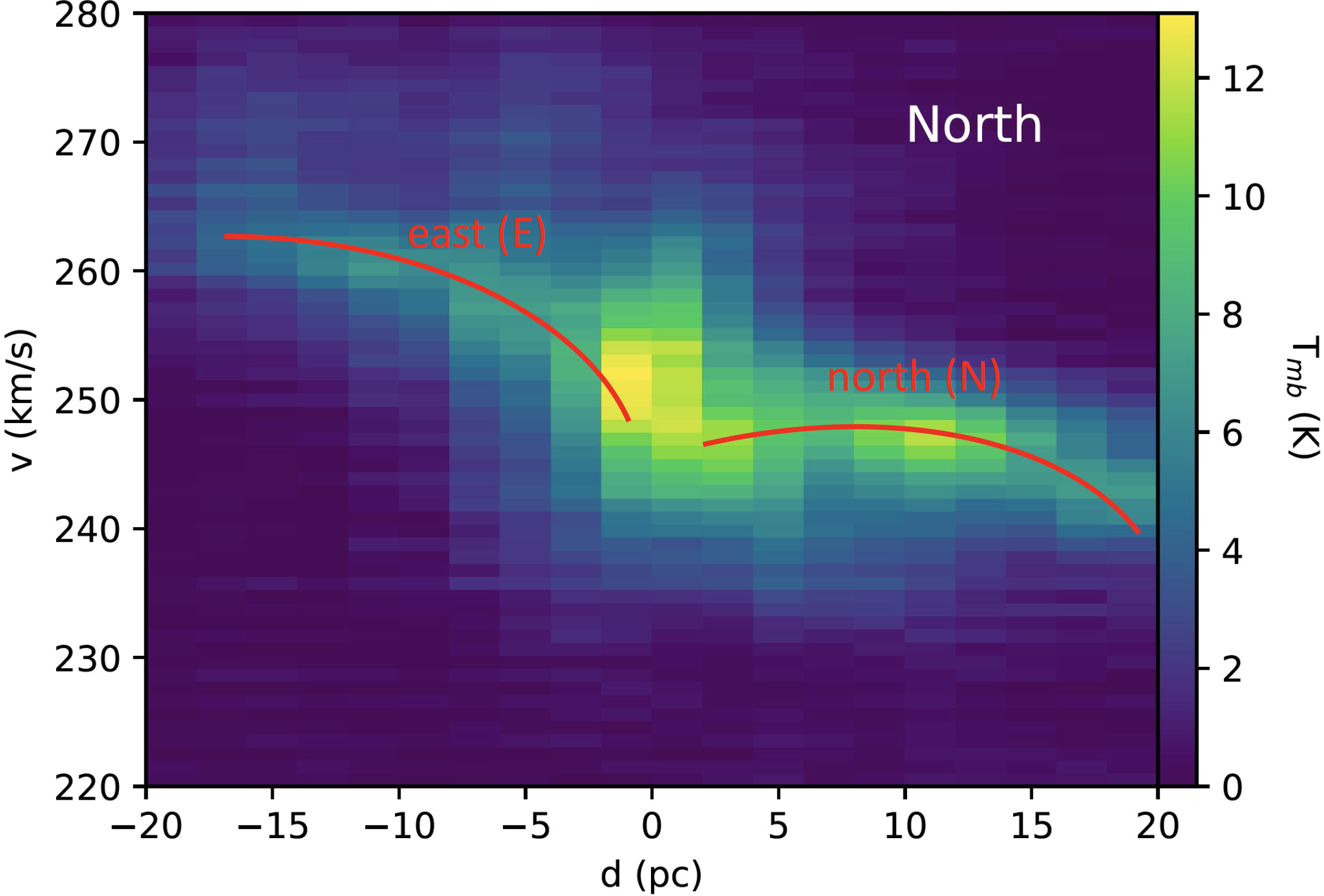}
    \includegraphics[width=0.495\hsize]{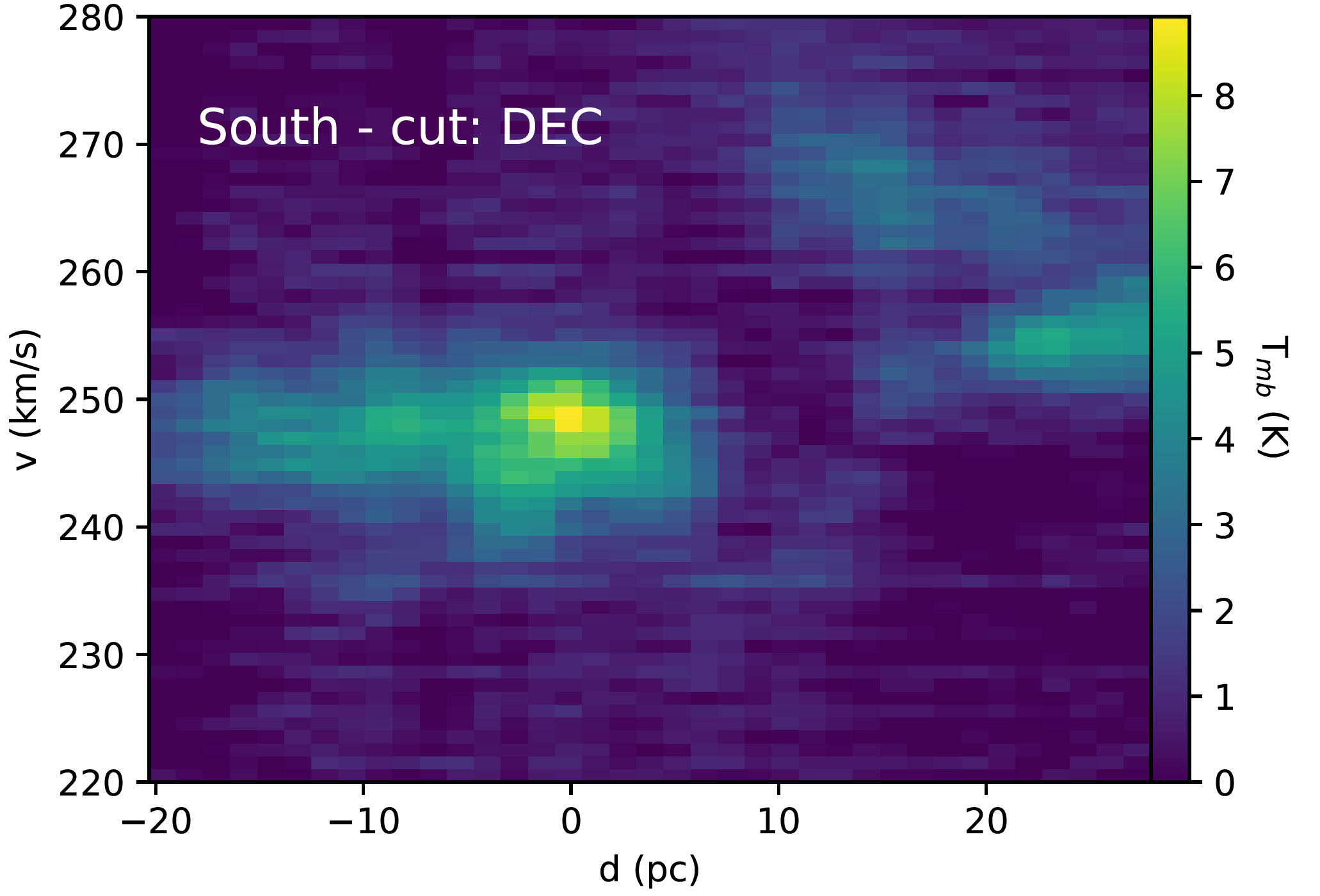}
    \includegraphics[width=0.495\hsize]{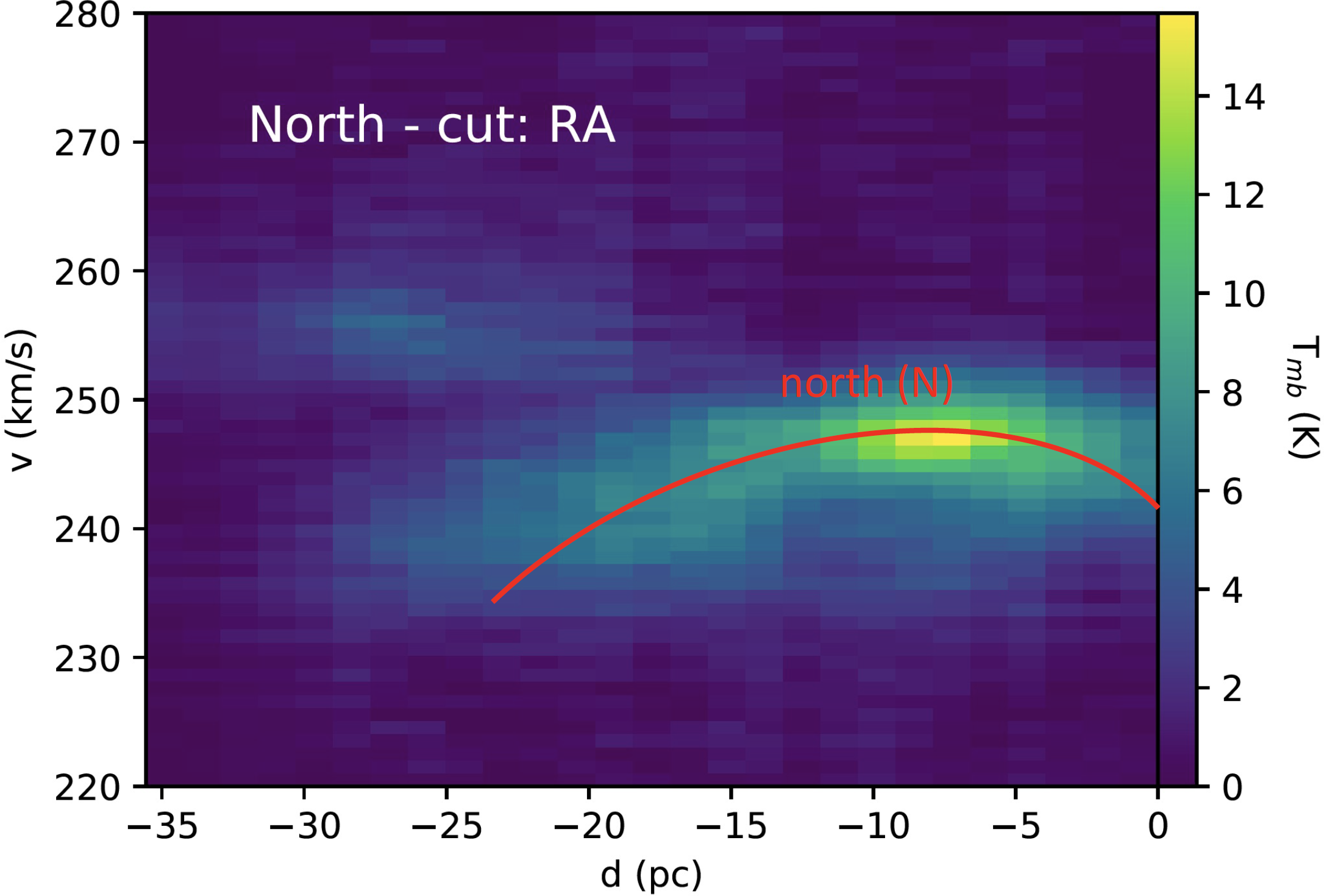}
    \includegraphics[width=0.495\hsize]{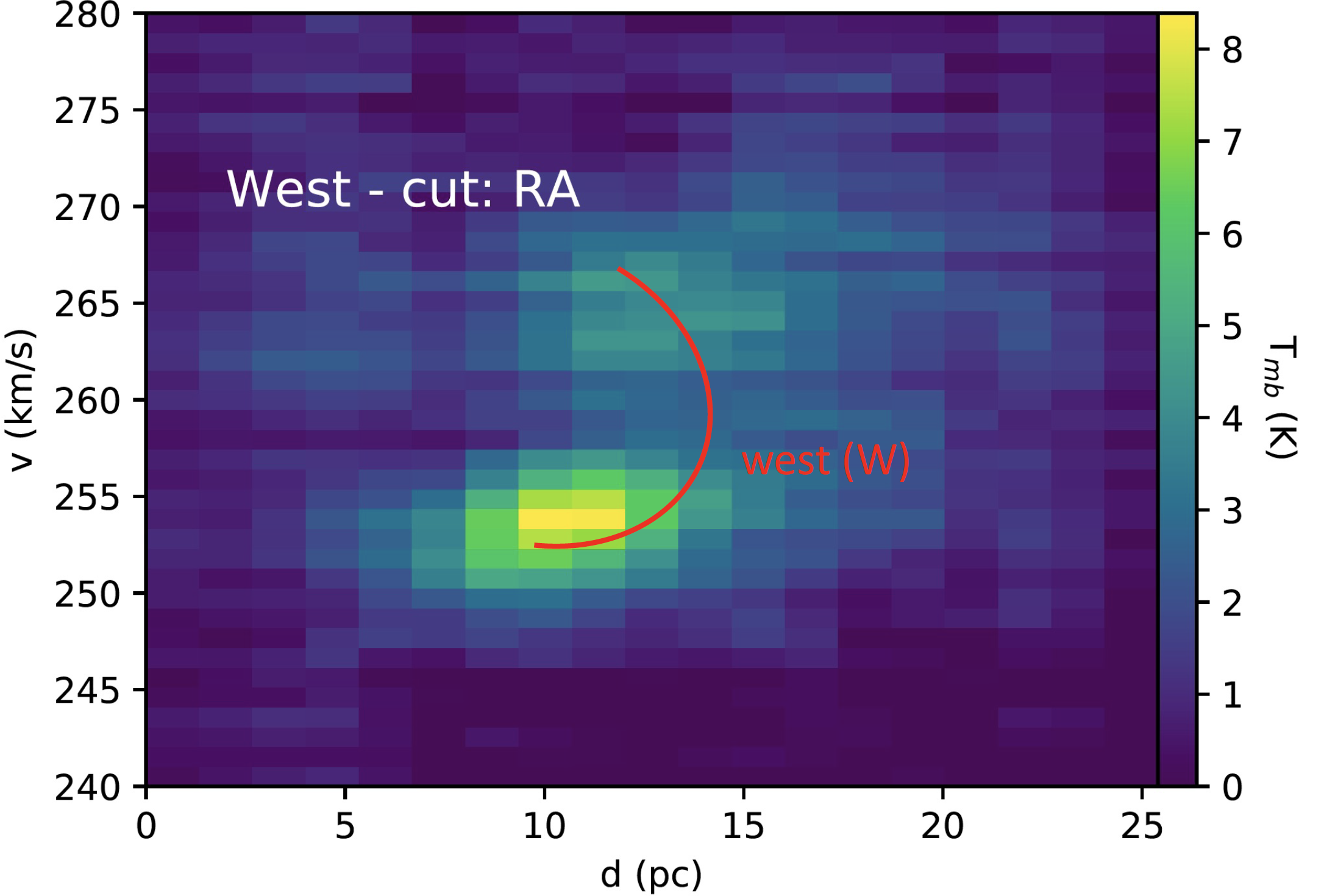}
    \includegraphics[width=0.495\hsize]{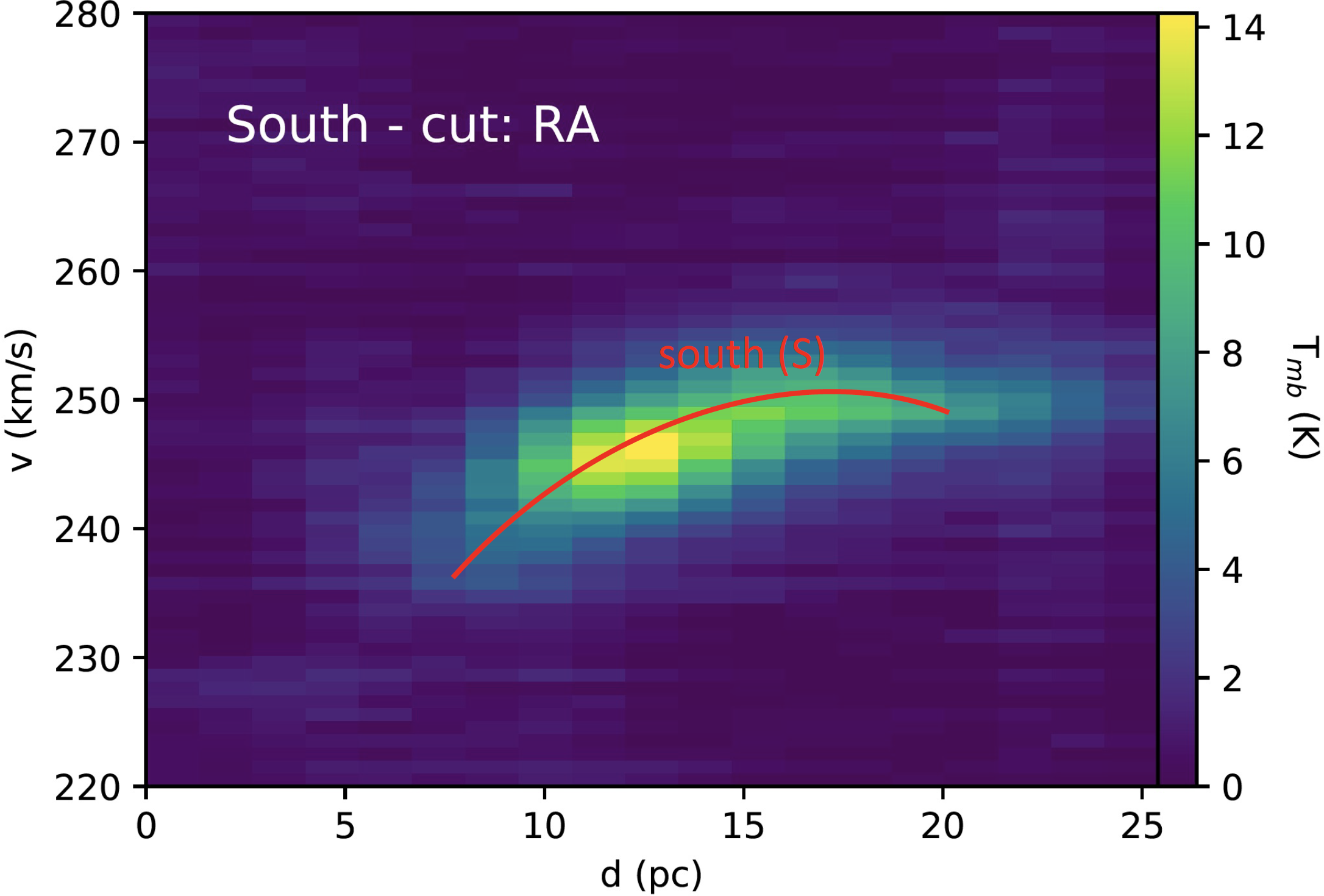}
    \includegraphics[width=0.495\hsize]{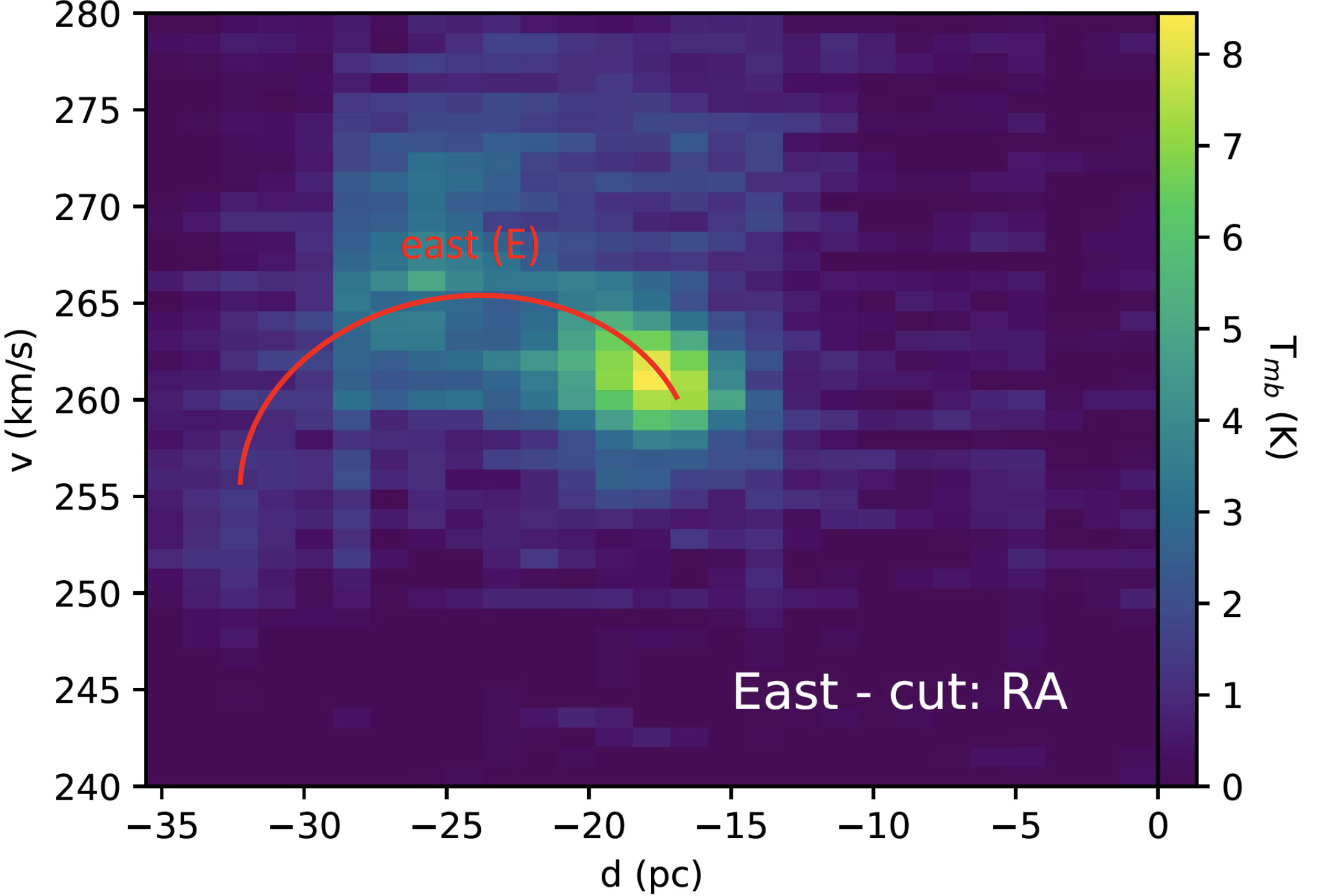}
    \caption{[CII] position-velocity diagrams, with a spatial resolution of $\sim$ 4 pc and a spectral resolution of 0.5 km s$^{-1}$, over multiple regions of the 30 Dor cloud at the locations indicated by the name in white. The cuts used to produce the PV diagrams in the top row are indicated in Figure \ref{fig:integratedMaps} and the cuts used to produce the PV diagrams in the two bottom rows are indicated in Figure \ref{fig:momentMaps}. All cuts show large gradients, often with a curve like nature, over 5-15 km s$^{-1}$ which could be associated with the presence of multiple expanding shells in this cloud. These candidate expanding shells, with their name, are indicated in red.}
    \label{fig:pvDiagrams}
\end{figure*}
\subsection{Position-velocity diagrams of 30 Doradus} \label{sec:pv_diagram}
The magnetic field observations of 30 Dor, presented in Figure \ref{fig:LIC_maps} and \citet{2021arXiv210509530T}, unveil a relatively organized structure in most regions of the cloud. As a result, it defines the main direction for both the northern clump and the southern regions of the map. This direction of the magnetic field in both regions will be used to study the [CII] kinematics of 30 Dor using position-velocity (PV) diagrams. These two directions, which are aligned with the mean field lines, are presented in Figure \ref{fig:integratedMaps}. Additionally, cuts through 4 regions in the RA direction were made for PV diagrams of [CII]. The location of these cuts is presented in Figure \ref{fig:momentMaps}. All the resulting PV diagrams are presented in Figure \ref{fig:pvDiagrams}.\\
These PV diagrams confirm that there are several organized velocity gradients in the region. These gradients cover a velocity interval of 5 - 15$\,$km$\,$s$^{-1}$ in most PV diagrams and come in the form of curves/half-elliptical features that have been associated with expanding shells \citep{2019Natur.565..618P,2020A&A...639A...2P,2021SciA....7.9511L, 2021ApJ...914..117T,2022A&A...659A..77B,2022ApJ...935..171B}. Based on the presence of these curved features, we identify 4 expanding shell candidates over the full map using the PV diagrams. The curved velocity structures associated with the candidate expanding shells are indicated in Figure \ref{fig:pvDiagrams}. With the channel maps in Figure \ref{fig:chanMapsFull}, we confirm that these candidates have a spatially curved/ring-like morphology as expected for expanding shell features. We name these candidate expanding shells: north, east, west, and south after their location on the map. The location and velocity range of these expanding shell candidates are presented in Table \ref{tab:shellTable}. The location of these shell candidates is also indicated over their velocity range in Figure \ref{fig:chanMapsFull}. Two expanding shell candidates are found in the range of 230 to 250$\,$km$\,$s$^{-1}$ and two are found in the range of 250 to 265$\,$km$\,$s$^{-1}$. As some of these shells appear to be only partly covered, larger maps will be required to: study their full extent, better characterize their properties and investigate whether some of the identified shells have the same expansion origin or not. This will also help to study, in combination with data at different wavelengths, whether radiation or stellar winds drive these expanding shell motions. As the expanding shells are directly identified by the two blue and red regions in the RGB image in Figure \ref{fig:RGBims}, this shows that the magnetic field curvature is associated with these expanding [CII] shell candidates. The highest-density region in the northern region is located at the intersection of two expanding shells. It might be possible that this region is formed by the collision of the two expanding features. This could result in additional magnetic field bending through oblique shocks \citep[e.g.][]{2013ApJ...774L..31I,2020A&A...644A..27B}. However, in the complex dynamical structure, and with the limited spectral [CII] resolution, it is not easy to establish indications of such magnetic field bending with confidence. Additionally, the increasing role of gravity might also alter the magnetic field morphology in this high-density region.

\begin{table}[]
    \centering
    \begin{tabular}{cccc}
    \hline
    \hline
    \multicolumn{4}{c}{Expanding shell candidates}\\
    \hline
        location & $\alpha_{J2000}$ & $\delta_{J2000}$ & velocities (km$\,$s$^{-1}$) \\
    \hline
        north (N) & 05:38:55 & -69:04:00 & 230-250 \\
        east (E) & 05:38:55 & -69:05:30 & 250-265 \\
        west (W) & 05:38:32 & -69:05:00 & 250-265 \\
        south (S) & 05:38:32 & -69:06:30 & 235-250 \\
    \hline
    \end{tabular}
    \caption{The location and velocity range for the identified expanding shell features in the 30 Dor region.}
    \label{tab:shellTable}
\end{table}

\begin{figure*}
    \centering
    \includegraphics[width=0.9\textwidth]{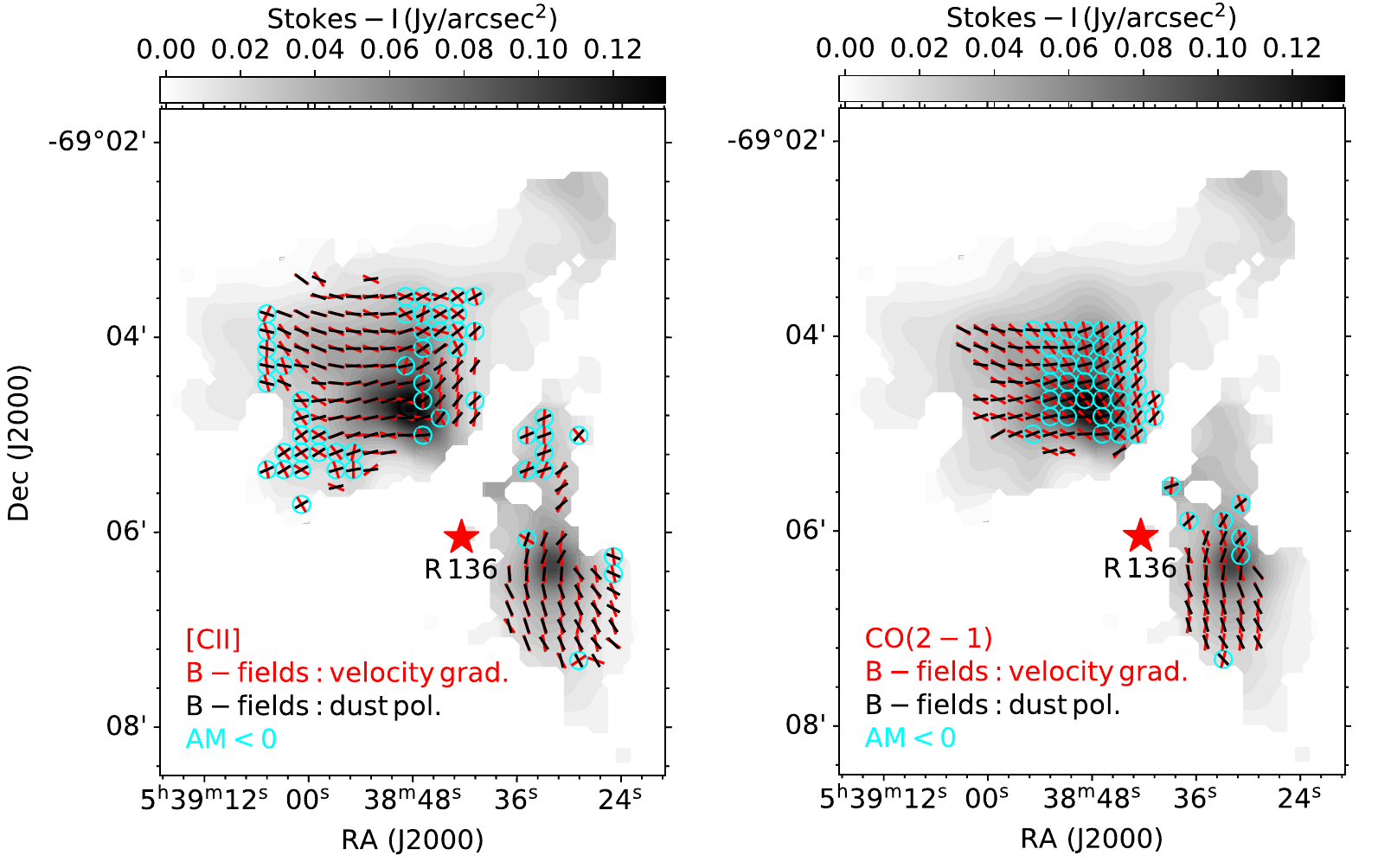}
    \caption{Comparison of magnetic fields inferred from the VGTs (red) and from the thermal dust polarization (black). The VGs estimated from [CII] are shown on the left, and from CO(2-1) on the right. The cyan circles indicate where these inferred magnetic fields are misaligned (i.e., the velocity gradient is parallel to the field lines). The background is the total dust intensity at $214\,\mu$m.}
    \label{fig:VGT}
    \includegraphics[width=0.9\textwidth]{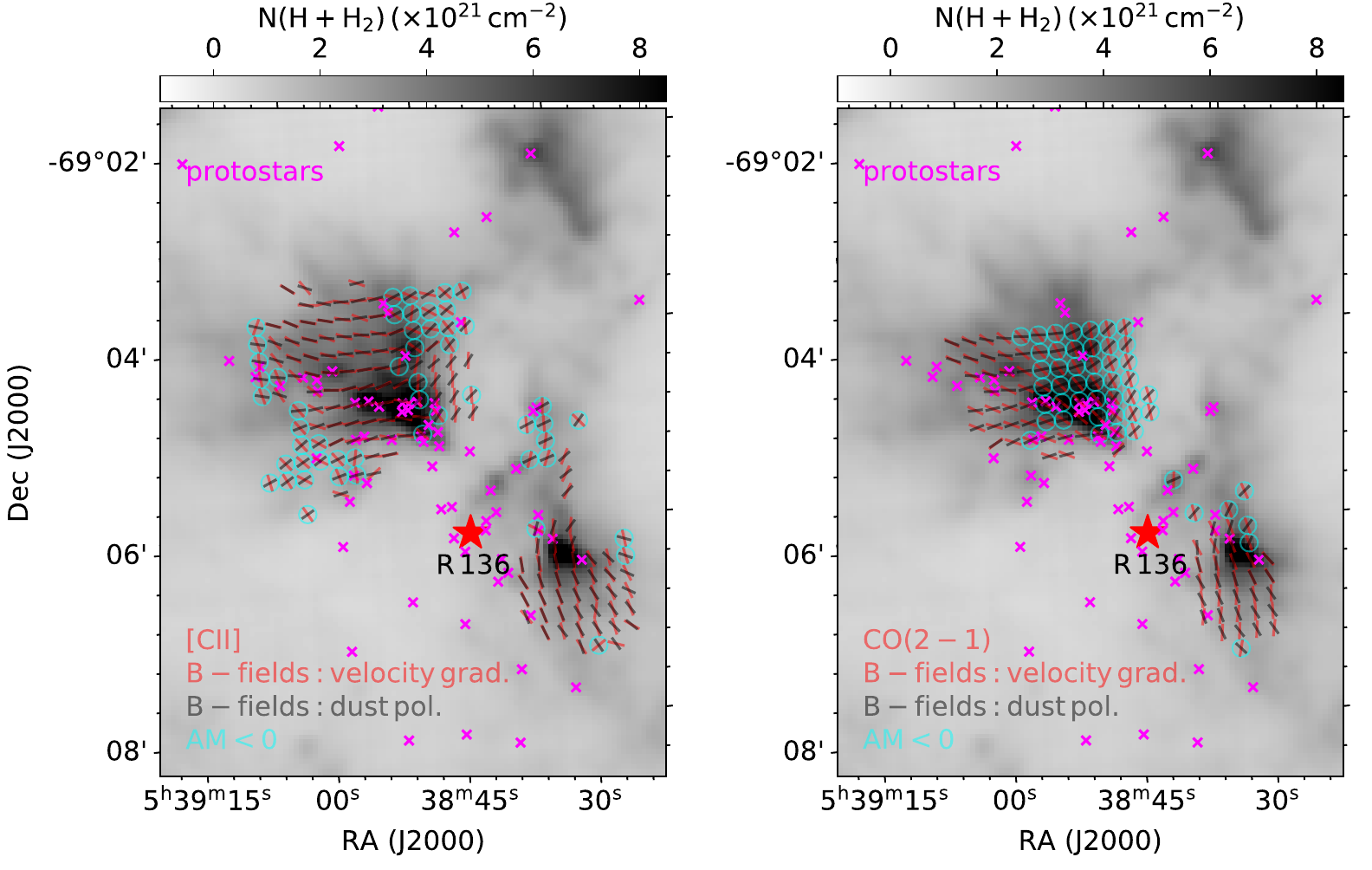}
    \caption{Same as Figure \ref{fig:VGT} but overlaid with the protostars adopted from \cite{2009ApJ...694...84I}. Interestingly, the protostars likely locate at which VGTs misalign with B-fields (i.e., VGs are parallel to the field lines).}
    \label{fig:VGT_protostars}
\end{figure*}
\subsection{Velocity gradient technique}
In this section, we computed the velocity gradients of both [CII] and CO(2-1) from their channel maps adopted from \cite{2021ApJ...912....2H}, which is referred to as VChGs (\citealt{2018ApJ...853...96L}). Here we recall the principle of this method. The methodology follows these steps: (1) the initial Position-Position-Velocity datacube was pre-processed using principle component analysis by splitting a very large number of velocity channels along the LOS as long as the thin channel map\footnote{a thin velocity channel does have information of the density and velocity fluctuation, but it is more sensitive to the velocity as the channel gets thinner so that the channel map will be used to compute the velocity gradient.} criteria is satisfied (\citealt{2000ApJ...537..720L}); (2) the product was convoluted with a $3\times 3$ Sobel Kernel to create a raw gradient map (pixels are blanked out if their intensity is less than three times root-mean-square level); (3) the gradient angle at each pixel was statistically computed from an adaptive sub-block\footnote{The size of the sub-block was chosen such that the histogram of the gradient orientation within this block follows a Gaussian distribution.} average method (\citealt{2017ApJ...837L..24Y}), in which all single gradient orientation within a rectangle sub-block in the raw gradient map was taken into account; (4) the pseudo-Stokes Q and U of the gradient were created (\citealt{2020MNRAS.496.2868L}), which allows computing the velocity gradient morphology. As the principle of VGT, the magnetic fields are inferred from this technique by rotating the velocity gradient orientation by 90 degrees (\citealt{2017ApJ...835...41G}). However, please note that the zeros-angle of the velocity gradient is defined along the East-West direction, which is an offset angle of the polarization angle defined by the SOFIA/HAWC+. Therefore, the velocity gradient orientation naturally infers the magnetic fields in the frame of the SOFIA/HAWC+ thermal dust polarization. 

Figure \ref{fig:VGT} shows the magnetic fields inferred by the VGTs (red vectors) from both [CII] (left panel) and CO(2-1) (right panel). To measure the correlation between B-fields inferred from VGTs and SOFIA/HAWC+ (black vectors), we computed the local alignment measurement (AM) as (\citealt{2017ApJ...835...41G}) 
\bea
    AM = 2\left(\cos^{2}\theta_{r} - \frac{1}{2}\right)
\ena
with $\theta_{r} = |\theta_{\rm VG} - \theta_{\rm HAWC+}|$ the angle differences between VGTs and HAWC+. These two vectors are parallel when $AM=1$, while they become perpendicular
once $AM=-1$. $AM<0$ indicates where these two vectors are misaligned (i.e., VGs are parallel to B-field lines). One can see that the VGT agrees well with the dust polarization in the South (i.e., VGs are perpendicular to B-field lines), but only fairly in the North. The VGT computed by [\textsc{CII}] shows that the misalignment occurs mostly at the edge of the regions where the gas density is not peaked (see the paper I). The VGT from CO(2-1) further shows that the misalignment occurs at the denser regions. The reason is that [CII] traces for more diffuse gas than CO so that it is less affected by gravity.

\section{Magnetic fields in 30 Doradus} \label{sec:Bfields}
To map the B-fields of 30 Dor, we make use of the SOFIA/HAWC+ polarimetric data at 89, 154, and 214$\,\mu$m observed under the Strategic Director's Discretionary Time (S-DDT) program (PI: Yorke, H., ID: 76$\_$0001) during the SOFIA New Zealand deployment in July 2018. The reduction of data was introduced in \cite{2018arXiv181103100G}. 
\begin{figure}
    \centering
    \includegraphics[width=0.4\textwidth]{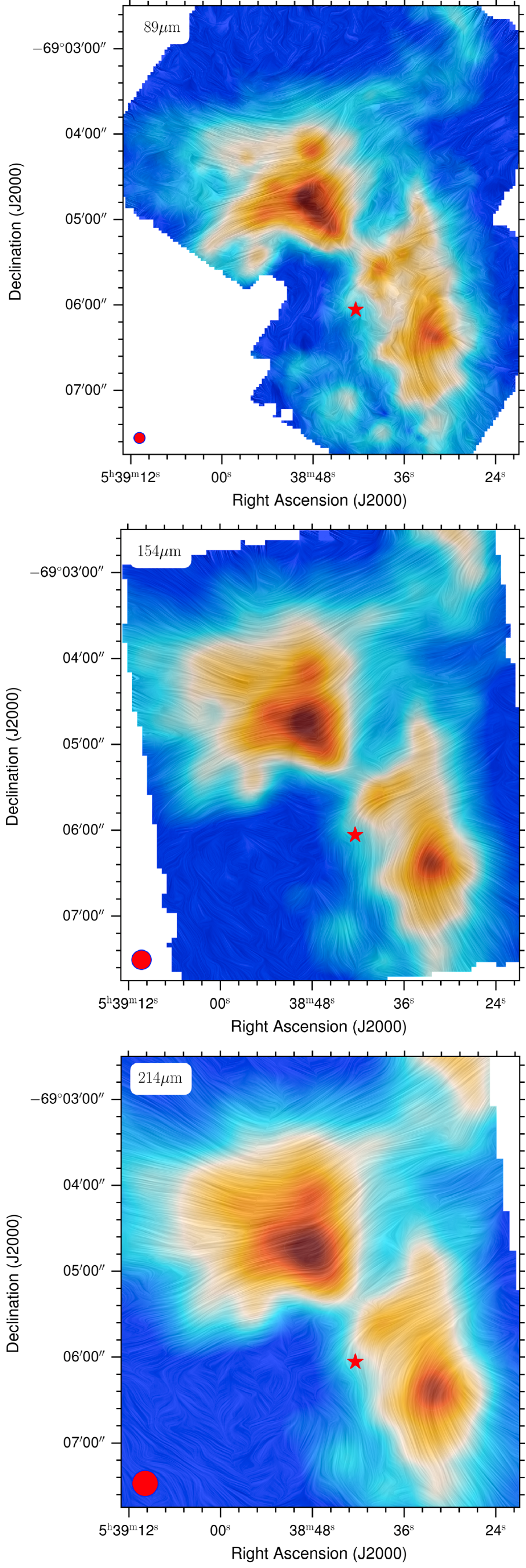}
    \caption{Magnetic field morphology in 30 Dor for all three bands (from top to bottom). The background is the total intensity stokes I. Streamlines are the LIC (\citealt{10.1145/166117.166151}). The Star symbol indicates the R$\,$136 location.}
    \label{fig:LIC_maps}
\end{figure}

\subsection{Magnetic field morphology}
{Assuming grains are perfectly aligned with the B-fields}\footnote{We validate this assumption in the upcoming work (Tram et al. in prep.)}, Figure \ref{fig:LIC_maps} shows the geometry of the fields for all three bands (top to bottom), which is inferred from the polarization vectors. The background color is the original total intensity (Stokes I). One can see that there are two main regions in 30 Dor, these are North and South regions relative to the massive star cluster R$\,$136, whose center is located by the red star. The B-fields are complex but highly ordered, and show a curved feature around the peak intensity (or the peak gas density) in both the North and the South regions. The field lines are curved toward the central cluster. In addition, the field structure at the East-side of the peaked intensity in the North region looks like an hourglass. These behaviors are observed similarly in all three bands. The discrepancy among the three panels shown in Figure \ref{fig:LIC_maps} is mainly due to the fact that shorter wavelength has higher spatial resolution than longer ones.

\begin{figure}
    \centering
    \includegraphics[width=0.49\textwidth]{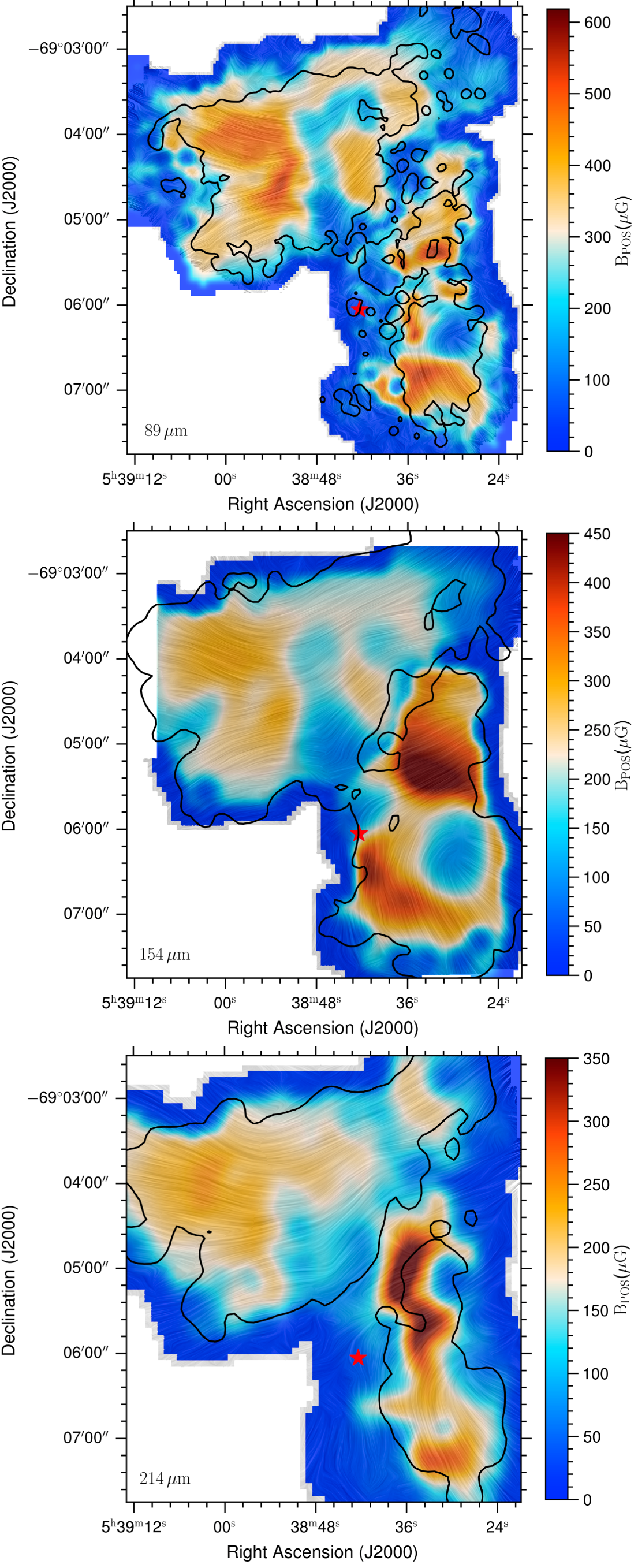}
    \caption{Maps of $B_{POS}$ strength for all three bands (from top to bottom) using [\textsc{CII}] overlaid with the B-field orientation. The black contours indicate the region where $I/\sigma_I\geq 100$ and $p/\sigma_p\geq 3$. The star symbol indicates the R$\,$136 location. The B-field strength is weaker at longer wavelengths with peaks offset from the total intensity.}
    \label{fig:Bpos}
\end{figure}

\subsection{Magnetic field strength}

The distortion of the B-field morphology could result from the local conditions. For example, this could possibly be explained by the compression of the magnetic field by turbulence, shock (\citealt{2013ApJ...774L..31I}), or gravitational contraction (\citealt{2013ApJ...767...33E}). To address which is a likely cause of this interesting situation, we thus estimate the map of the B-field strength. 

We determine the spatial distribution of the POS magnetic field strength in 30 Dor according to the strategy developed by \cite{2021ApJ...908...98G}:

\begin{itemize}
    \item A map of the strength of the magnetic field in the POS can be determined using the Davis-Chandrasekhar-Fermi (DCF; \citealt{1951PhRv...81..890D}; \citealt{1953ApJ...118..113C}), based on the equipartition of the turbulent magnetic energy and the turbulent kinematic energy, approximation:
    
    \begin{equation}
        B_{\rm POS} \simeq \sqrt{4\pi \rho} \frac{\delta V_{\rm t}}{\delta \phi}
        \label{eq:dcf}
    \end{equation}
    
    \noindent
    where $\delta V_{\rm t}$ is a map of the velocity dispersion of the gas, $\rho$ is the map of the gas density and $\delta \phi$ is a map of the angle dispersion of the polarization orientation. Following a modification in \citet{2009ApJ...706.1504H}, the angle dispersion is replaced by the angular dispersion (or structure-function) proportional to the ratio of large-to-small scales magnetic energies as $\delta\phi \simeq [\langle B^{2}_{t}\rangle/\langle B_{0}^{2}\rangle]^{1/2}$. This improvement is able to avoid the spatial change in the magnetic field morphology with $\langle B_{t}^{2}\rangle/\langle B_{0}^{2}\rangle$.
    
    \item The structure-function dispersion analysis \citep{2009ApJ...706.1504H} was applied to every map of $\vec{B}$-field angles ($\phi$) in a pixel-by-pixel fashion. This means, for a given pixel $i$, a circular kernel of radius $w$-pixel is defined around it. All the pixels inside this kernel are then used to calculate the dispersion function $1-\langle\cos[\Delta\phi(\ell)]\rangle$ for pixel $i$. This dispersion function can be described by the two-scale model:
    
    \begin{equation}
        1-\langle\cos[\Delta\phi(\ell)]\rangle = \frac{1-e^{ -\ell^{2}/2(\delta^{2} + 2W^{2})}}{1 + \mathcal{N}\left[\frac{\langle B_{t}^{2}\rangle}{\langle B_{0}^{2}\rangle}\right]^{-1}} + a_{2}\ell^{2}.
        \label{eq:disp_funct}
    \end{equation}
    
    \noindent
    where $\Delta\phi(\ell)$ is the difference between a pair of $\phi$ separated by the distance $\ell$. The first term in Equation \ref{eq:disp_funct} describes the small-scale, turbulent $\vec{B}$ contribution to the dispersion. In this term, $\delta$  is the turbulence correlation length, $\mathcal{N}(=\Delta^{\prime}(\delta^{2}+2W^{2})/\sqrt{2\pi\delta^{3}})$ is the number of turbulent cells along the LOS, and $W$ is beam's radius (e.g. the $\sigma$ value) of the polarimetric observations. $\Delta^{\prime}$--the cloud's effective depth--is used as a proxy for the depth along the LOS (see \citet{2009ApJ...706.1504H} for definition). The second term in Equation \ref{eq:disp_funct} quantifies the contribution from the large-scale, ordered magnetic field to the dispersion function. 
    
    \item The dispersion function in every pixel is fitted with Equation \ref{eq:disp_funct} using a MCMC solver. This fitting process can determine values of $\langle B_{t}^{2}\rangle/\langle B_{0}^{2}\rangle$, $a_{2}$, and $\delta$ in every pixel. The map of $\langle B_{t}^{2}\rangle/\langle B_{0}^{2}\rangle$ is required to estimate the POS magnetic field strength.
    
    \item Before combining all three maps through Equation \ref{eq:dcf}, they must have the same angular resolution. In order to do that, the lowest angular resolution (target resolution;$\sigma_{T}$) among $\delta V$, $\rho$, and $\langle B^{2}_{t}\rangle/\langle B_{0}^{2}\rangle$ is chosen and the other maps are smoothed using a Gaussian kernel with width $\sigma^{2} \xrightarrow{} \sigma_{T}^{2} - \sigma^{2}$. 
    
\end{itemize}

The choice of kernel size, $w$, is important for the fitted values of $\ratio$, $a_{2}$, $\delta$(if fitted), and therefore the $B_{\rm POS}$ maps. According to \cite{2021ApJ...908...98G}, the kernel size needs to be large enough to result in statistically-significant dispersion functions but small enough for any resulting map to have meaningful angular resolution and avoid smoothing them over large areas. The optimal kernel size, $w_{\rm opt}$, can be found by monitoring the distribution of $\rho$\footnote{Spearman, non-linear, rank correlation. Values range between 1 and -1, with the former signaling a perfect correlation.} values -- between the constructed dispersion function and its fit in every pixel. We performed tests using the 214-$\micron$ map and values of $w$ = 7, 9, 11 pixels and found that $w$ = 9 pixels produced the largest fraction of pixels with $\rho\sim$1. Table \ref{tab:Delta_prime_val} shows the values of $w_{\rm opt}$, global $\delta$, and correlated beam size ($\sqrt{2}W$) for all three bands. From these values, it is clear that the 9-pixel-radius circular kernel contains 7.5 correlated beams and 2 to 3 turbulence correlation lengths.

Using kernels with $w > w_{\rm opt}$ will not result in better-determined fitting parameters since the addition of more pairs of vectors will not modify the small-scale portion of the dispersion function. Instead, it will result in the contribution of scales larger than those described by the term $\propto \ell^{2}$ which are not described by the model in Equation \ref{eq:disp_funct}.

In order to evaluate Eq. \ref{eq:dcf}, we used the $N(\rm H_{2})$ map from Paper I and for $\delta V$ we used the second moment map of either CO or [\textsc{CII}] (Figure \ref{fig:momentMaps}). To transform column density to mass density, the cloud's depth must be used. In this work, we used the $\Delta^{\prime}$ defined above, which is calculated as the width of the auto-correlation function of the polarized intensity. However, taking into consideration that the 30 Dor complex is distinctly separated into two clumps, assuming one single value for $\Delta^{\prime}$ can result in less accurate values of $B_{\rm POS}$. Therefore, we created a map of $\Delta^{\prime}$ that consists of two areas (north and south) each with a constant, different value. These values are displayed in Table \ref{tab:Delta_prime_val}. On the other hand, the second moment ($\delta V$) calculated from CO and [CII] observations, contains contributions from thermal motions of the molecules as well as turbulent motions. Therefore, the velocity dispersion values were corrected as $(\delta V_{\rm t})^{2} = (\delta V)^{2} - k_{\rm B}T_{\rm gas}/m$, where $k_{\rm B}$, $T_{\rm gas}$, and $m$ are the Boltzmann constant, the gas temperature, and the molecule's mass. In this work, we assume 30 Dor is in LTE condition and adopt $T_{\rm gas}=T_{\rm ex}$. We computed the excitation temperature $T_{\rm ex}$ at each pixel using its maximum main-beam temperature as in \cite{2008ApJ...679..481P}.

\begin{table}[]
    {\it 30 Dor}
    \begin{center}
    \begin{tabular}{c|c|c|c}
        $\lambda$ [$\micron$] & $\delta$ [$^{\prime\prime}$] & $\sqrt{2}W$ [$^{\prime\prime}$] & $w_{\rm opt}$ [$^{\prime\prime}$] \\
        \hline
        89 & 16.00 & 4.68 & 17.55 \\
        154 & 21.56 & 8.17 & 30.51 \\
        214 & 23.68 & 10.93 & 40.95 \\
        \hline
    \end{tabular}\\
    \end{center}
    {\it North}
    \begin{center}
    \begin{tabular}{c|c|c}
        $\lambda$ [$\micron$] & $\Delta^{\prime}$ [arcmin] & $\Delta^{\prime}$ [pc] \\
        \hline
        89 & 0.55 & 7.84 \\
        154 & 0.74 & 10.53 \\
        214 & 0.76 & 10.82 \\
        \hline
    \end{tabular}\\
    \end{center}
    {\it South}
    \begin{center}
    \begin{tabular}{c|c|c}
        $\lambda$ [$\micron$] & $\Delta^{\prime}$ [arcmin] & $\Delta^{\prime}$ [pc] \\
        \hline
        89 & 0.33 & 4.70 \\
        154 & 0.40 & 5.70 \\
        214 & 0.38 & 5.41 \\
        \hline
    \end{tabular}
    \end{center}
    \caption{Values of turbulence correlation length ($\delta$), correlated beam size ($\sqrt{2}W$), and optimal kernel size ($w_{\rm opt}$) for the map-making procedure (Top). Cloud's effective thickness ($\Delta^{\prime}$) for the north (middle) and south (bottom) regions of 30 Dor. These values were calculated as the half-width half-max of the one-dimensional auto-corrrelation of the polarized flux at each wavelength. Linear depths are calculated from angular depths assuming a distance of 49 kpc.}
    \label{tab:Delta_prime_val}
\end{table}

According to \cite{2021ApJ...908...98G}, there are two alternatives to fitting the dispersion function (see their Appendix A): 1) solving for all three parameters ($a_{2},\ratio,\delta$) simultaneously. If this provides the majority of pixels in the map with values of $\delta$ greater than $\sqrt{2}W$, then the turbulent contribution is well resolved by the polarimetric observations and the values of $\ratio$ are well-defined. 2) If the condition $\delta>\sqrt{2}W$ is not met for the majority of pixels, the parameter $\delta$ can be fixed to a prescribed value and the MCMC solver computes only two parameters, $a_{2}, \ratio$. For this investigation, the first approach was tried with all polarimetric maps, but only the 214-$\micron$ satisfied the condition mentioned above. For the results here presented we used the second approach, with a prescribed $\delta$ value for each pixel equal to that from the global analysis (e.g. when only one dispersion function is calculated for the entire observation). Using the $\delta$-fixed approach results in underestimated values of $\ratio$ and overestimated values of $B_{\rm POS}$. However, a test performed with the 214-$\micron$ data (the only map for which $\delta>\sqrt{2}W$ was met in the majority of pixels) showed that differences in $B_{\rm POS}$ values using these two approaches were small, with an average value of $\sim$16\%.

The resulting maps of $B_{\rm POS}$ strength, calculated using the [\textsc{CII}] data, are presented in Figure \ref{fig:Bpos} for all three bands 89, 154, 214 $\micron$ (from the top to bottom). These maps have angular resolutions of 33$^{\prime\prime}$, 58$^{\prime\prime}$, 77$^{\prime\prime}$, respectively. The estimated strengths show a dependency with the wavelength of the polarimetric data. Values range between few and $\sim$350, $\sim$450, and $\sim$600 $\mu$G for 214, 154, and 89 $\micron$. This wavelength discrepancy should be mainly because of the angular resolution of the map -- shorter wavelengths have smaller resolution. However, there is also a possibility that such differences are the result of each individual wavelength tracing different layers of the cloud. At the same wavelength, the strength structure using CO is similar to the one using [\textsc{CII}], but the amplitudes are slightly lower (see Appendix \ref{sec:B_CO}). The uncertainty of our calculation is discussed in Section \ref{sec:uncertainties}. The black contours in Figure \ref{fig:Bpos} indicate the spatial area in which $I/\sigma_I\geq 100$ and $p/\sigma_{p}\geq 3$ (e.g., the polarization measurement is statistically significant and intrinsically associated with the source as discussed in Paper I).

\subsection{Magnetic fields vs. gravity}
An important factor indicating the influence of the magnetic field is the mass-to-flux ratio. We adopt the ratio as in \cite{2004ApJ...600..279C} as
\bea \label{eq:lambda}
    \lambda = 7.6\times 10^{-21}\frac{N(\H_{2})}{B_{\rm tot}}
\ena
where $B_{\rm tot}=\sqrt{B^{2}_{\rm POS}+B^{2}_{\rm LOS}}$ in $\mu$G is the strength of the magnetic field in three-dimension, which is missing in this work because we do not have the LOS component of the field, and $N_{\rm H_{2}}$ in $\rm cm^{-2}$ is the gas column density. Statistically, $\overline{B}_{\rm tot}$ could be approximated as $4/\pi\times \overline{B}_{\rm POS}$ (\citealt{2004ApJ...600..279C}). Using the measured Zeeman LOS component of the field, \cite{2021ApJ...908...98G} showed that $B_{\rm tot} \simeq B_{\rm POS}$ in the specific case of OMC-A (see their Equation 11 and Table 1). Lacking information on the LOS component, we simply adopt $B_{\rm tot} = B_{\rm POS}$ with the same caveat as many other studies (e.g., \citealt{2018ApJ...861...65S}; \citealt{2021ApJ...912L..27E}; \citealt{2021ApJ...908...10N}; \citealt{2021arXiv210810045H}). The gas column density is derived by a modified black-body fitting from Herschel data at 100, 160, 250, 350, and 500$\,\mu$m (\citealt{2013AJ....146...62M}) as in Paper I. The cloud is called supercitical for $\lambda>1$, in which the magnetic field is insufficient to provide support against the gravitational potential. Otherwise, the cloud is called subcritical for $\lambda<1$ in which the magnetic field prevents the cloud from collapsing.
\begin{figure*}
    \centering
    \includegraphics[width=0.9\textwidth]{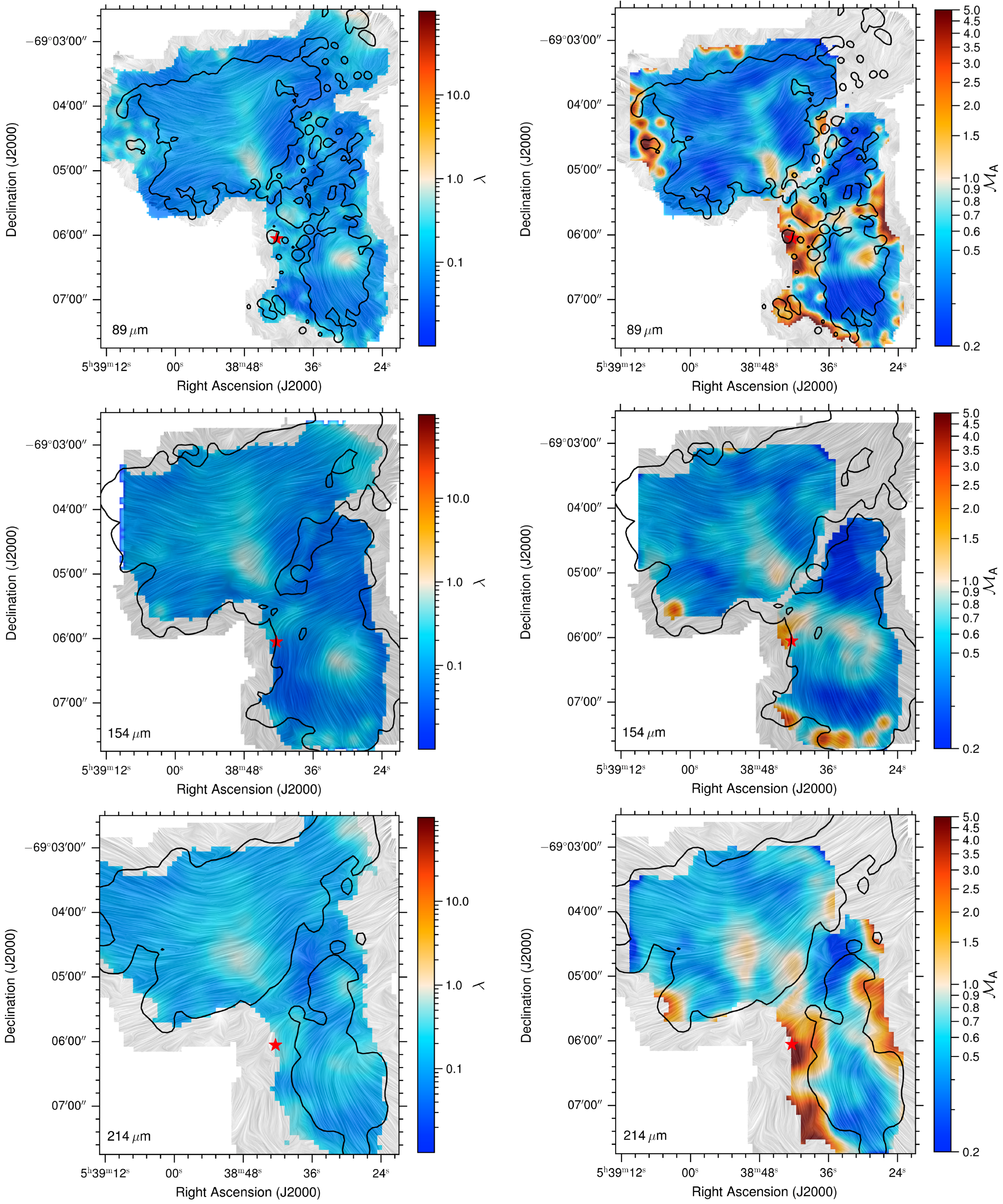}
    \caption{Maps of the mass-to-flux ratio (left panels), and the Alfv\'enic Mach number (right panels) for all three bands (wavelength increases from top to bottom) overlaid with the fields morphology. The black contours border is defined where $I/\sigma_I\geq 100$ and $p/\sigma_p\geq 3$. The red star indicates the R$\,$136 location.}
    \label{fig:maps_lamMA_CII}
\end{figure*}
\begin{figure*}
    \centering
    \includegraphics[width=0.9\textwidth]{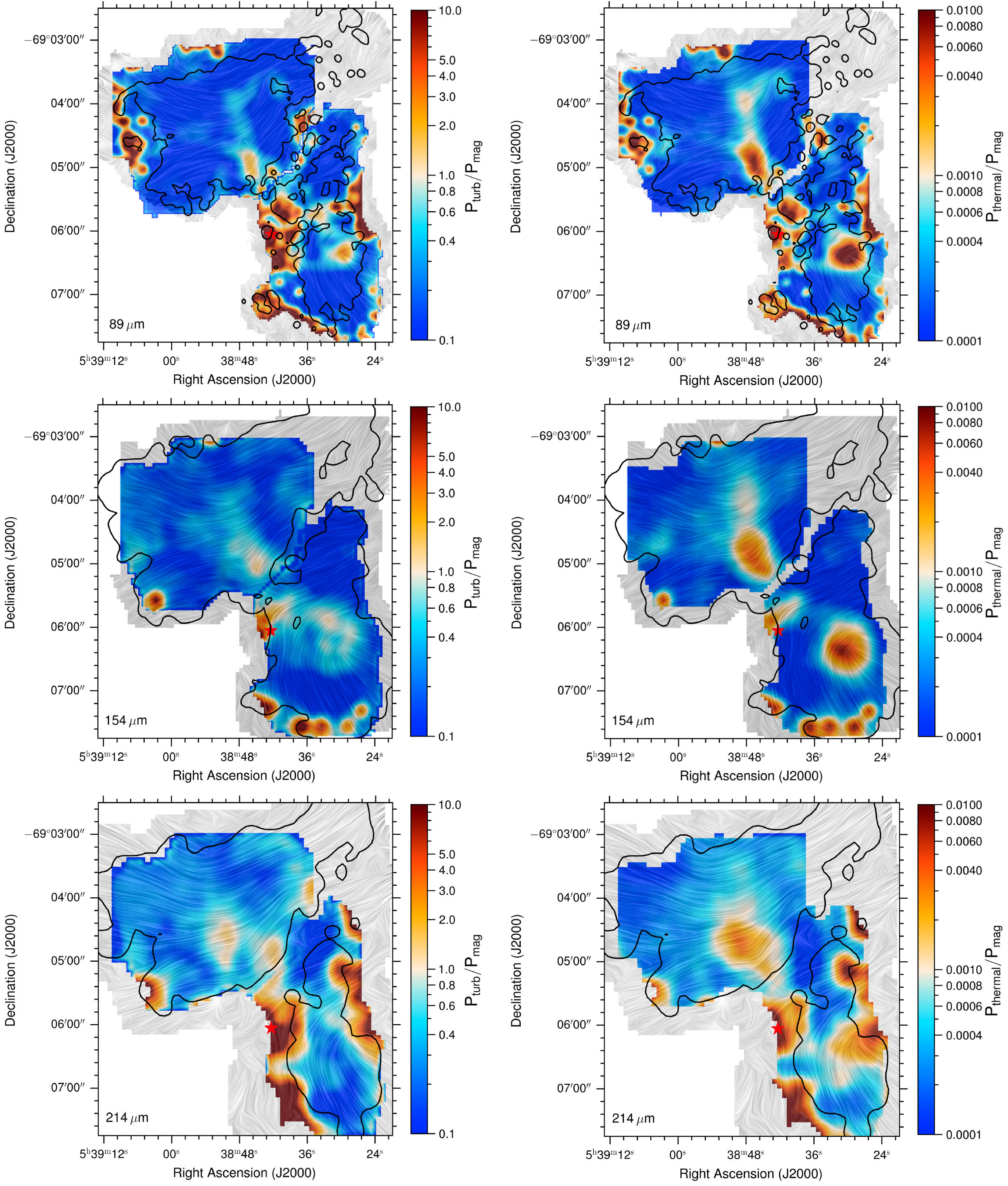}
    \caption{Turbulent-to-magnetic pressure ratio (left panels) and thermal-to-magnetic pressure ratio (right panels). The symbols are the same as in Figure \ref{fig:maps_lamMA_CII}. The magnetic pressure is higher than that of the turbulence everywhere across the cloud except in the peak intensity and the Eastern edge in the North where the turbulent pressure is higher or comparable to the magnetic pressure at the peak intensity in the North and South regions. The magnetic pressure is dominating the thermal one over the cloud.}
    \label{fig:maps_pratio}
\end{figure*}

Figure \ref{fig:maps_lamMA_CII}, left panels show the maps of the mass-to-flux ratio for all three bands (from top to bottom) overlaid with the field lines. Generally, the cloud is sub-critical ($\lambda<1$) for most of the region, except at the peak intensity in both the North and South ($\lambda \geq 1$), and the regions where $\lambda \sim 1$ spreads larger in area at longer wavelengths. Since $\lambda \sim N(\rm H_{2})$, and $B_{\rm pos}\sim \sqrt{N(\rm H_{2})}$, the gas density is the main uncertainty factor, which is discussed in Section \ref{sec:uncertainties}. Our derivation of $N(\rm H_{2})$ agrees quite well with the one from PDR models (\citealt{2020MNRAS.494.5279C}). As the maximum of gas column density is about $10^{22}\,\rm cm^{-2}$, a hundred $\mu$G B-field can make $\lambda<1$ (see Equation \ref{eq:lambda}).

\subsection{Magnetic field vs. Turbulence}
The interplay between the B-field and turbulence can be defined through the Alfv\'enic Mach number. Because the DCF assumes that the turbulent magnetic energy balances the turbulent kinematic energy, the total magnetic energy estimated by the DCF method always greater than the turbulent kinematic energy, which leads to an Alfv\'enic Mach number is always lower than 1. Therefore,  we use the 3D Alfv\'enic Mach number to assess the relative importance between the magnetic field and turbulence as
\bea
    \mathcal{M}_{A} = \sqrt{3}\frac{\sigma_{\rm NT}}{\mathcal{V}_{A}}  
\ena
where $\sigma_{\rm NT}$ is the non-thermal velocity dispersion, and $\mathcal{V}_{A}=B_{\rm tot}/\sqrt{4\pi \rho}$ is the Alfv\'enic velocity. $\mathcal{M_{A}}<1$ stands for the sub-Alfv\'enic, the impact of the gas turbulence is minimal and the magnetic fields are able to regulate the gas motion. $\mathcal{M_{A}}>1$ stands for the super-Alfv\'enic, the gas turbulence drives the magnetic fields to be random fields.

The right panel in Figure \ref{fig:maps_lamMA_CII} shows the map of the Alfv\'enic Mach number in three bands (from top to bottom). These maps are color scaled such that the blue-color represents $\mathcal{M}_{A}<1$ and the red-color is for $\mathcal{M}_{A}>1$. One can see that the entire cloud is made up of sub-Alfv\'enic motions, while it is trans- or super-Alfv\'enic at around the peaked intensity both in North and South, except at 214$\,\mu$m where the the South peak becomes sub-Alfv\'enic. Similar to mass-to-flux ratio maps, the trans-Alfv\'enic area is more widespread at longer wavelength.

The magnetic and isotropic turbulent pressures in units of $\rm dyn\, cm^{-2}$ are computed as
\bea
    P_{\rm mag} &=& \frac{B_{\rm tot}^{2}}{8\pi} \\ \nonumber
    P_{\rm turb} &=& \frac{3}{2}\rho (\sigma_{\rm NT})^{2}
\ena

If $P_{\rm turb}/P_{\rm mag}$ is greater than 1, the turbulence pressure is higher than the magnetic pressure, otherwise the magnetic pressure is dominant. The maps in all three bands are shown in the left panels of Figure \ref{fig:maps_pratio}, indicating that the turbulent pressure is comparable and higher than that of the B-field at which $\mathcal{M}_{A}\gtrsim 1$, elsewhere it becomes lower (see Figure \ref{fig:maps_lamMA_CII}, right panels). 

\subsection{Magnetic vs. thermal pressure}
The relative importance between the magnetic and thermal pressure could be quantified through a so-called plasma beta parameter, which is a ratio of these two pressures as
\bea
	\beta = \frac{P_{\rm thermal}}{P_{\rm mag}} = \frac{2c^{2}_{s}}{\mathcal{V}^{2}_{\rm A}}
\ena
where $c_{s}=\sqrt{k_{B} T_{\rm gas}/\mu_{\rm H_{2}} m_{\rm H}}$ is the thermal sound speed. Figure \ref{fig:maps_pratio}, right panels show the maps of the thermal to magnetic pressure ratio. The magnetic pressure is much stronger than the thermal one and varies across the cloud, which is confirmed by the constraints observed by \cite{2019A&A...628A.113L} and \cite{2020MNRAS.494.5279C} using the PDR Meudon code. Indeed, the authors constrained that $P_{\rm thermal} \sim 10^{4}-10^{6}\,\rm K\, cm^{-3}$, while we showed that $P_{\rm mag} \sim 10^{7} - 10^{8}\,\rm K\, cm^{-3}$ with $B = 200-500\,\mu$G.

\section{Discussions}\label{sec:discussions}
\subsection{Gas column density probability distribution and turbulent driving force}\label{sec:grav_sf}
The interplay between the gravity, magnetic fields and turbulence in the star-formation has been raised and discussed in numerous studies. The probability density function (PDF) of the gas column density was demonstrated to be a key to study the dynamics of molecular clouds. The PDF density of the isothermal and non-self-gravity gas follows closely a log-normal distribution; while this PDF develops a power-law tail once the self-gravity dominates and the collapse is significant (e.g., \citealt{2000ApJ...535..869K}; \citealt{2013ApJ...763...51F}; \citealt{2014Sci...344..183K}; \citealt{2014ApJ...781...91G}; \citealt{2013ApJ...766L..17S,2015MNRAS.453L..41S,2022arXiv220714604S}). 
\begin{figure}
    \centering
    \includegraphics[width=0.5\textwidth]{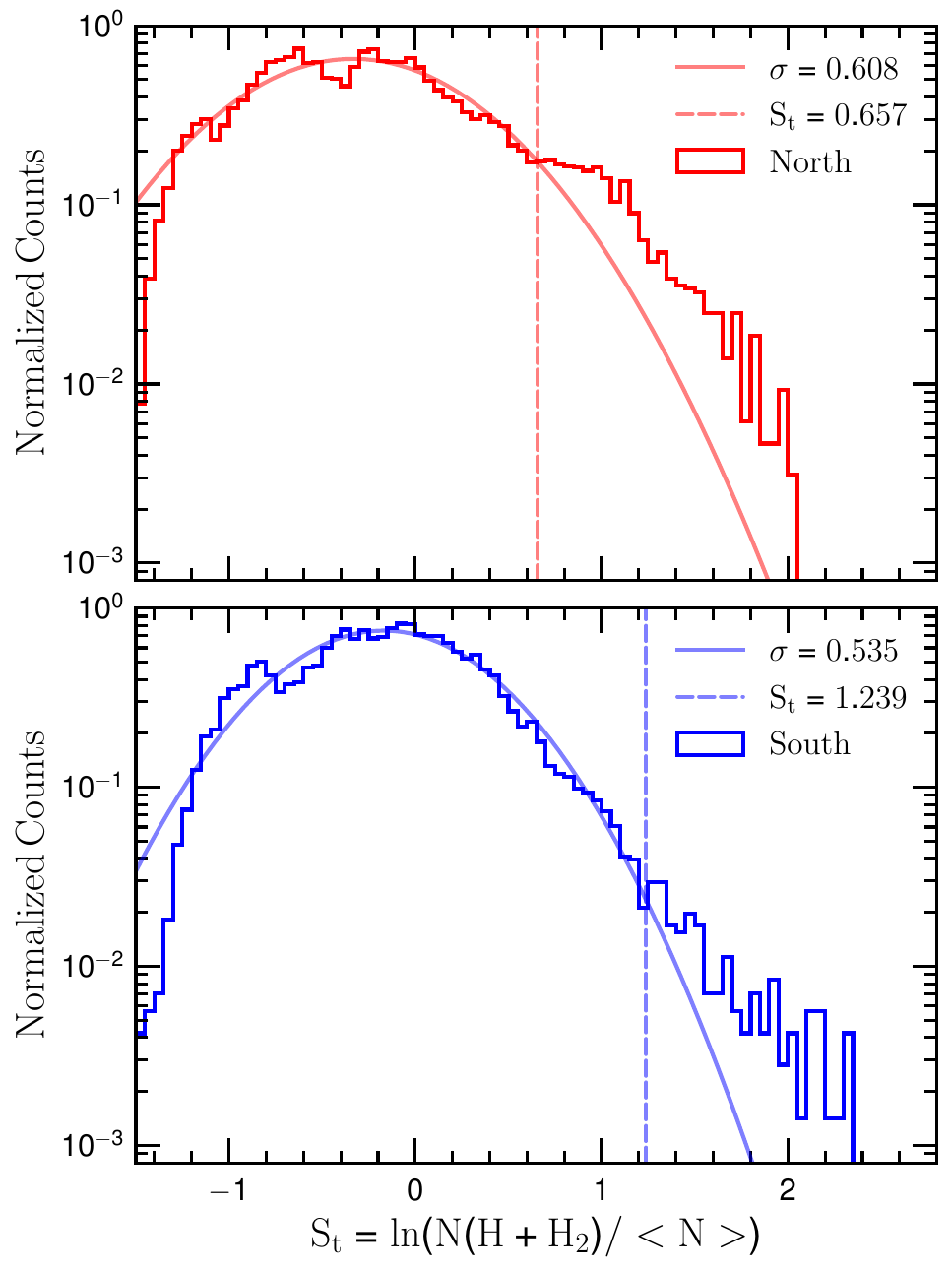}
    \caption{PDF distribution of the logarithm of the gas column density in the North (upper) and in the South (lower). A log-normal just fits up to $S_{t}\simeq 0.657$ (North) or $S_{t}\simeq 1.239$ (South), which are followed by a power-law tail. Apparently, the tail-structure in the North appears at lower gas density than in the South.}
    \label{fig:pdf_NH}
\end{figure}

However, the shape of the PDF strongly depends on the way observations are performed as studied in \cite{2019MNRAS.482.5233K}. The authors demonstrated that small or insufficient field-of-view will result in a cut-off in the log-normal part of the distribution, and argued that a sufficiently large field-of-view is needed for the PDF to be built completely. Thus, we used the whole column density map to build the PDF in Figure \ref{fig:pdf_NH}. Then, we fit a log-normal to the histogram as shown by the solid lines. One can see that a log-normal could not fit entirely the PDFs of both North's and South's gas density, but up to $S_{t} \simeq 0.657$ or $N \geq 1.8\times \bar{N}_{\rm North} = 3\times 10^{21}\,\cm^{-2}$ in the North (upper panel), and $S_{t}\simeq 1.239$ or $N\geq 3.5\times \bar{N}_{\rm South} = 3.8\times 10^{21}\,\cm^{-2}$ in the South (lower panel). Above these values, power-law tails can be seen in both regions. This reveals that the gravitational instability likely sets in relatively low gas density in 30 Dor cloud in general, and the North experiences it at a slightly lower density than that of the South because its power-law tail pops up earlier. The gravitational collapse in parsec-scale has been proposed in various studies (e.g., \citealt{2007ApJ...654..988H}; \citealt{2006A&A...445..979P,2013A&A...555A.112P}; \citealt{2010A&A...520A..49S}; \citealt{2019ApJ...870....5J}; \citealt{2022arXiv220610889B}). Another note is that the gravitational collapse in the South is much more compact than in the North. 

Figure \ref{fig:VGT_protostars} shows the location of the protostars candidates adopted from \cite{2009ApJ...694...84I}. 
These locations span from low to the peak gas column density. Interestingly, the protostars locations are coincident with the misaligned vectors between the rotated VGs by [CII] and CO and the B-fields, or at which these gas motions are parallel to the field lines.

The power-law tails have been seen to develop even in low column density gas, and we suspect that the turbulence could be the key gradient to make that happened. There are three main turbulent types, i.e., compressive (curl-free), solenoidal (divergence-free) and the mixing, in which the compressive turbulence results in stronger compression and thus larger deviation of the PDF (\citealt{2010A&A...512A..81F}). These three types of turbulence could be identified through the turbulence driving parameter $b$ defined as
\bea
    b^{2} = \frac{e^{\sigma^{2}} -1 }{M^{2}_{S}}\frac{\beta + 1}{\beta}
\ena
where $M_{S}$ is the sonic Mach number, $\sigma$ is the standard deviation of the log-normal distribution in Figure \ref{fig:pdf_NH}, $\beta = 2M^{2}_{A}/M^{2}_{S}$ is the compressibility (or the plasma beta parameter in the right panel of Figure \ref{fig:maps_pratio}) with $M_{A}$ the Alfvenic Mach number. Purely solenoidal driving has $b=1/3$, the compressive driving turbulence yields $b=1$, and a stochastic forcing mixture has $b\sim 0.4$ (\citealt{2010A&A...512A..81F}). Table \ref{tab:driving_force} shows the mean values in the North and South regions. One can see that $b \simeq 1$ in both regions, which indicates that the driving force of turbulence in 30 Dor is mainly compressive. This is consistent with the existence of the supersonic expanding shells within the cloud discussed in Section \ref{sec:pv_diagram}, and similar to other star-forming regions; such as Serpens cloud (\citealt{2021ApJ...912....2H}). 
\begin{table}[]
    \centering
    \begin{tabular}{cccccc}
    \hline
    \hline
    \multicolumn{6}{c}{turbulent driving force} \\
    \multicolumn{6}{c}{North region} \\
    \hline
        band & $\bar{T}_{\rm gas}$ & $\bar{\mathcal{M}}_{\rm A}$ & $\bar{\sigma}_{v}$ & $\sigma$ & $\bar{b}$  \\
    \hline
	C & 31.68       & 0.41 &  8.46      &  0.608       & 1.15 \\
	D & 31.68       & 0.46 &  8.46      &  0.608       & 1.03 \\
	E & 31.68       & 0.61 &  8.46      &  0.608       & 0.78 \\
    \hline
    \hline
    \multicolumn{6}{c}{South region} \\
    \hline
	C & 30.68     & 0.56 &   7.60    &  0.535    & 0.73 \\
	D & 30.68     & 0.52 &   7.60    &  0.535    & 0.78 \\
	E & 30.68     & 0.62 &   7.60    &  0.535    & 0.66 \\    
    \end{tabular}
    \caption{Mean values of $\bar{b}\simeq 1$ in both North and South regions indicates that the turbulence in 30 Dor is likely the compressive driven.}
    \label{tab:driving_force}
\end{table} 

\subsection{Can Star-formation occur even in the strongly magnetized environment of 30 Doradus?}
\begin{figure}
    \centering
    \includegraphics[width=0.5\textwidth]{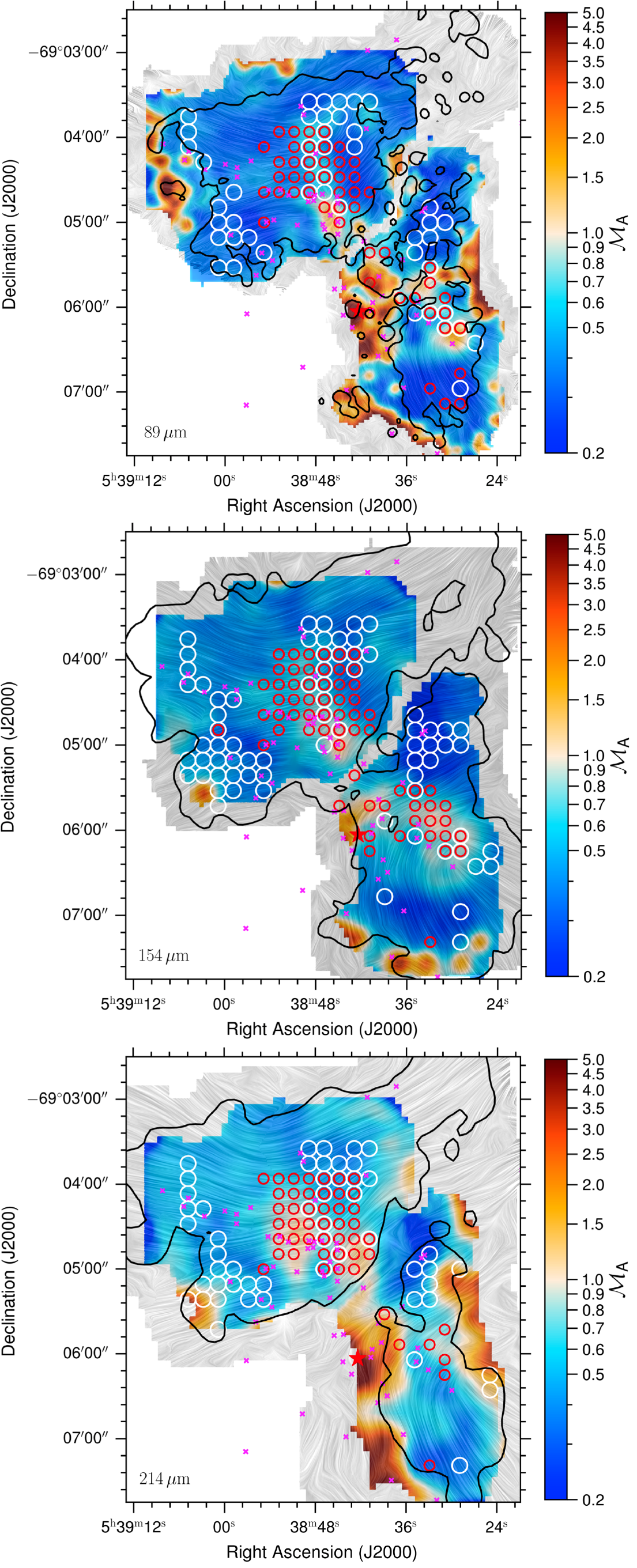}
    \caption{Alfv\'enic Mach number as shown in right panel of Figure \ref{fig:maps_lamMA_CII}. The white and red circles mark the positions where VGs from \textsc{CII} and CO are parallel to the fields (i.e., misaligned between VGT and magnetic field lines as shown in Figure \ref{fig:VGT}), respectively. The magenta symbols show the locations of the protostars.}
    \label{fig:MA_VGTs}
\end{figure}
Figures \ref{fig:VGT_protostars} and \ref{fig:MA_VGTs} show that the velocity gradients of both \textsc{CII} and CO misalign with the B-fields at certain locations in both North and South regions, which indicates that (1) gas moves parallel to the field lines and (2) there is gravitational collapse. However, these positions distribute from low to maximum gas density, and mostly in  magnetically sub-critical, sub-Alfv\'enic regions where the magnetic pressure is higher than that of the turbulence and thermal. The PDF of gas density in Figure \ref{fig:pdf_NH} illustrates the power-law tail, an indicator of star-formation activities, at not very dense gas (column density few times higher than $10^{21}\,\rm cm^{-2}$) supporting the local collapse scenario, where protostars candidates are likely to locate. How can we explain the on-going star-formations in strong B-fields?

The turbulence driving parameters $b \simeq 1$ (see Table \ref{tab:driving_force}) reveals that the turbulence is mainly driven by compressive forcing in 30 Dor. The plasma parameter $\beta\ll 1$ (see Figure \ref{fig:maps_pratio}, right panel) shows that the sonic Mach number $\mathcal{M}_{s}\gg 1$, indicating the presence of supersonic turbulence. The supersonic compressive turbulence could create overdense gas in which the gravity gets unstable and stars can form by directly compressing the gas and/or accumulating the gas along the B-fields line. The latter will not be affected by the magnetic pressure because this pressure acts preferentially perpendicular to the field lines. As a result, the supersonic compressive turbulence could help to trigger the gravitational collapse. That turbulence could result from the cluster-wind from R$\,$136 (\citealt{2021A&A...649A.175M}) or probably the interaction with much larger giant HI-bubbles in the LMC (\citealt{1999AJ....118.2797K}). After that, the B-fields get amplified and slow down the star-formation activities, as we observe today.

\subsection{Role of magnetic fields in holding 30 Dor integrity?}\label{sec:Bfield_role}
30 Dor contains a spectacular feature with multiple large expanding-shells (see Figure \ref{fig:RGB_image}). The kinematic of these structures is just about $10^{50}-10^{51}\,\rm ergs$ (\citealt{1994ApJ...425..720C}), which could be generated by a cluster-wind (\citealt{2021A&A...649A.175M}) or a combination with supernovae (\citealt{1994ApJ...425..720C}) from R$\,$136. 
\cite{2021A&A...649A.175M} estimated the virial mass within a distance of 25$\,$pc from R$\,$136 as a few times of $10^{6}\,M_{\odot}$ which is larger than the mass of 30 Dor gas (a few times of $10^{5}\,M_{\odot}$). \cite{2011ApJ...738...34P} demonstrated that the thermal pressure (in order of $10^{-9}\,\rm dyn\,cm^{-2}$) is about 1.6 times lower than that of the radiation pressure (see their Figure 18), and is about 3 times higher than the hot gas pressure (i.e., the hot gas pressure is higher elsewhere, see their Figure 21). Thus, how can the 30 Dor cloud survive? 

We suspect that B-fields play a crucial role here in holding the cloud integrity. The B-field morphology orients perpendicular to the radiation direction, so that the magnetic pressure could resist pressure coming from this direction. Indeed, the magnetic pressure is $P_{\rm mag}\simeq 1.6\times 10^{-9}-10^{-8}\,\rm dyn\,cm^{-2}$ for $B=200-500\,\mu$G, which is absolutely higher than both thermal and radiation pressure and thus is capable to resist against the radiative impact. This calculation is in agreement with the right panel in Figure \ref{fig:maps_pratio}. The role of B-fields could be deduced from a kinematic point of view. \cite{2021A&A...649A.175M} showed that the kinematic energy of the region within 25pc from R$\,$136 is about 24$\%$ the total kinematic energy (which is equivalent to $\simeq 5\times 10^{50}\,\rm ergs$\footnote{numberic values are adopted from Table 3 in \cite{2021A&A...649A.175M}}). For $B=200-500\,\mu$G, the magnetic energy is orders of magnitude larger as $\simeq 10^{51} - 2\times 10^{52}\,\rm ergs$ within the same spherical volume. In addition, \cite{2021A&A...649A.175M} showed the turbulent kinematic energy could be up to $10^{51}\,\rm ergs$; thus, the turbulence only dominates over the B-field for $B<200\,\mu$G which is consistent with the maps of $\mathcal{M}_{A}$ (right panel in Figure \ref{fig:maps_lamMA_CII}) and $P_{\rm turb}/P_{\rm mag}$ (left panel in Figure \ref{fig:maps_pratio}).

The next question could be how large external shells developed around our cloud? 30 Dor opens to the Eastern direction from R$\,$136, which helps hot gas (traced by X-ray, see Figure 14 in \citealt{2006AJ....131.2140T}) easily escape and creates the large expanding-shell along this direction. The mechanism could otherwise be complicated along other directions, especially Northern and Western. There are multiple expanding-shells on a smaller scale within the cloud (see Figure \ref{fig:pvDiagrams}) referring to the fact that the cloud is being carved by the cluster feedback. \cite{2021A&A...649A.175M} proposed that this impact could lead the certain cloud's structure to break, which enables hot gas to go through and "inflate" the external expanding-shells along these directions. Interestingly, Figure \ref{fig:Bpos} shows that the B-field strength is relatively weak (i.e., lower than 200$\,\mu$G) at certain regions in 30 Dor along directions that point to or adjoin to these external structures. Along these directions, the field is insufficient to resist feedback; thus, there are nozzles where the hot gas can escape through the cloud.

\subsection{Comparison with the larger scale magnetic field of the hosting LMC}
The larger-scale B-fields have been measured using both near-infrared (NIR) (e.g., \citealt{2007PASJ...59..519N,2016ApJS..222....2K}) and radio polarimetric observations (e.g., \citealt{1993A&A...271..402K, 2012ApJ...759...25M}). 

The absorption polarization at NIR probed the B-field structure outside our region (see Figure 3 in \citealt{2007PASJ...59..519N}). These large field lines are seen to associate with the dust cloud in LMC and the large expanding structures around 30 Dor. The B-field strength from this more diffuse region was estimated of $3-25\,\mu$G using DCF. 

The B-fields measured by synchrotron polarized emission inferred the larger scale of LMC. The low spatial resolution (e.g., $14'$ in \citealt{2012ApJ...759...25M}) was unable to resolve our region. The total B-field strength is measured of $<7\,\mu$G, assuming equipartition with cosmic-ray electrons, with an upper-limit B-field on POS of $11\,\mu$G in the HI southwestern filaments up to the 30 Dor region. 
The higher B-field strength in the filaments may be due to anisotropic turbulent B-fields mostly located in POS. 

Our B-field strengths of $350-600\,\mu$G at pc-scales using thermal dust polarization are much higher than the NIR and radio polarimetric observations in more diffuse media, which may be enhanced due to turbulent dynamo mechanism and star formation activity. The additional larger FIR and higher spatial resolution radio polarimetric observations of the LMC are required to better understand the connection between galactic B-fields and those B-fields responsible for the star-formation activity and molecular cloud formation.

\subsection{Uncertainties}\label{sec:uncertainties}
In this section, we discuss several possible uncertainties. Our first concern lies in the estimate of the column density, which is constructed by the modified SED fitting from Herschel data (see Paper I). Our derived column density agrees well in a range of $1.9\times 10^{21}-1.3\times 10^{22}\,\rm cm^{-2}$ estimation from PDR modelling (\citealt{2020MNRAS.494.5279C}), but $1.3\times 10^{22}\,\cm^{-2}$ is 1.12 times higher than our maximum derivation. Hence if $B_{\rm pos} \sim N(\rm H_{2})^{0.5}$, the mass-to-flux ratio $\lambda$ increases by a factor 1.06, and the Alfv\'enic Mach number correspondingly decreases by a factor of 0.94, the magnetic pressure increases by a factor of 1.12. The values are slightly changed but do not affect our conclusions. 

Our second concern is related to the volume density derived from the column density. The depth of the cloud is fixed and estimated by the auto-correlation function from the polarized intensity. Thus, our derived volume density is insensitive to the local cloud depth, which could be either over-/under-estimated.

The third concern is the non-thermal velocity dispersion. We computed this term by using [\textsc{CII}] and $^{12}$CO emission lines. These lines are not necessarily optically thin and show multiple components; hence it is not ideal for probing the turbulence component of the gas motion even though it shows no evident signs of absorption. If these emissions are optically thick, the non-thermal velocity dispersion might be overestimated because ($\delta V_{\rm turb})^{2} \sim (\delta V)^{2}$, (e.g., \citealt{1979ApJ...231..720P}; \citealt{2016A&A...591A.104H}).

The fourth concern is the gas temperature, because ($\delta V_{\rm turb})^{2} \sim (\delta V)^{2} - k_{\rm B}T_{\rm gas}/m_{\rm gas}$. We assume the gas is in the LTE condition, so we adopt $T_{\rm gas}=T_{\rm ex}$ across the entire cloud. This assumption will fail at the locations where the LTE is simply incorrect. However, we expect that $T_{\rm gas}$ has a little impact on the final result because (1) the thermal contribution is relatively small ($\simeq 2.6\,\rm km\,s^{-1}$ for $T_{\rm gas}=10^{4}\K$) compared to the total velocity dispersion (Figure \ref{fig:momentMaps}), and (2) this velocity dispersion is supersonic for [\textsc{CII}].

The fifth concern is the strength of the B-field in 3D space. In this work, we adopt the value of the POS component of B-field since its LOS component is unknown in 30 Dor. The higher value of the total B-field strength will result in lower values of $\lambda$, $\mathcal{M}_{A}$, while increased in $P_{\rm mag}$. \cite{2019MNRAS.485.3499C} proposed a method to estimate the inclination of B-fields by assuming a uniform B-fields (i.e., no variation in position angle and inclination of the field lines) and a constant polarization coefficient (i.e., $p_{0}$ in their Equation 10) throughout a cloud. These assumptions are unlikely to be satisfied in the case of 30 Dor given a variation of the field lines and the polarization degree across the cloud. 

The sixth concern is the method to estimate the B-field strength. In this work, we use an improved version of the DCF method from \cite{2009ApJ...706.1504H}, in which the structure-function of polarization angle is used rather than the dispersion function of this quantity. \cite{2021A&A...647A.186S} and \cite{2021A&A...656A.118S} showed that the strength of the mean B-field component is proportional to $1/\sqrt{\delta \phi}$ which is different from Equation \ref{eq:dcf}. However, this approach assumed an equipartition between turbulent motions and the parallel component of turbulent B-fields whose physical basics behind is not clear (see Appendix A3 in \citealt{2022MNRAS.510.6085L} and Section 10.3 in \citealt{2022ApJ...935...77L}). Recently, at the time of this work, \cite{2022ApJ...935...77L} proposed a new method to estimate the field strength, namely differential measure approach (DMA) as $B \simeq f\sqrt{4\pi \rho}D^{1/2}_{v}/D^{1/2}_{\phi}$ where $D_{v}$, $D_{\phi}$ are the structure functions of the velocity centroids and polarization angle, and $f$ is the factor based on the anisotropic turbulence. The authors showed that the DMA method could estimate the field strength at the smallest scale (lower than the turbulence injection scale) and with anisotropic turbulence properties. These two modifications contrast the DCF method and could increase the estimation accuracy. A comparison between DCF and DMA is beyond the scope of this work, and it hence serves as an important improvement for future works (not only for 30 Dor but also other star-formation regions).

To summarize, the quantitative value of B-field strengths could be different from what is presented in this work once using a proper value of the non-thermal velocity dispersion and the gas temperature for the DCF or different methods. However, it could not change by orders of magnitude, and we do not expect it will change the effect of the B-field. Indeed, when the strength is as high as hundreds $\,\mu$G, the field is strong enough to support against the gravitational potential (Equation \ref{eq:lambda}) and is sufficient to resist the hot gas and radiation pressures (Section \ref{sec:Bfield_role}).

\section{Conclusions}\label{sec:conclusions}
The entire 30 Dor is a complex star-forming region, which clearly shows a core-halo structure, in which there are multiple pc-scale expanding-shells structures in the outer region and a cloud in the inner region. Cluster-wind or a combination of supernovae from the star cluster R$\,$136 is demonstrated to be the main source of energy for these giant shells (\citealt{1994ApJ...425..720C}; \citealt{2021A&A...649A.175M}). However, it is not very clear how an inner structure is able to remain close to this source of energy. 
Thus, we study the gas kinematics and B-fields in this region in this work, for which we use the same name as 30 Dor for the sake of convenience. Our main findings are summarized as follows.

\begin{itemize}
    \item[1.] We derived the B-field morphology from thermal dust polarization observed by SOFIA/HAWC+, which shows the highly ordered yet complex B-field structure. The field lines are wrapped and bent around the peak intensity (where the gas is densest, Figure \ref{fig:LIC_maps}). The tip of the bending fields points towards the star cluster, R$\,$136.
    
    \item[2.] We performed velocity field analysis on two different gas tracers. One is expected to trace warmer and more diffuse gas ([\textsc{CII}]), and the other cold and dense gas (CO). Our analysis indicates a complex dynamic structure but organized with multiple gas components and wing-like structures. The change from the blue- to red-shifted gas is directly related to the B-field morphology change. Furthermore, the position-velocity diagrams showed possible multi expanding-shells in 30 Dor.

    \item[3.] We compared the B-fields obtained using the VGT of both [CII] and CO and showed that the B-fields mostly align with the one inferred from thermal dust polarization observed by SOFIA/HAWC+ in the South, while in the North, numerous misalignment vectors are seen (Figure \ref{fig:VGT}). CO line shows anti-alignment at dense gas, and C[II] complementary illustrates the miss-alignment at more diffuse regions. 
    
    \item[4.] The miss-alignment between the B-fields induced by the VGT and the thermal dust polarization evidences the local gravitational collapse because it could pull the gas along the fields; the VGs are thus parallel rather than perpendicular to the field lines. 
 
    \item[5.] We showed that the probability distribution function (PDF) of the gas column density (Figure \ref{fig:pdf_NH}) exhibits a power-law tail, which is a sign of gravitational collapse. This is consistent with the constraints of the velocity gradients.

    \item[6.] We estimated the $B_{\rm pos}$ using the DCF method. Our estimation shows that the B-fields are quite strong in 30 Dor with maxima of 600, 450, and 350$\,\mu$G at 89, 154, and 214$\,\mu$m, respectively (Figure \ref{fig:Bpos} ). Thus, 30 Dor is mostly sub-critical ($\lambda<1$, see Figure \ref{fig:maps_lamMA_CII}, left panel) and sub-Alfv\'enic ($\mathcal{M}_{A}<1$, see Figure \ref{fig:maps_lamMA_CII}, right panels), except at the peak intensity where the gas density is densest. The magnetic field is also dominant over the turbulent (Figure \ref{fig:maps_pratio}, left panels), and substantially higher than the thermal contribution (Figure \ref{fig:maps_pratio}, right panels).

    \item[7.] We found that turbulence is mainly driven by compressive forcing and supersonic. We proposed that the supersonic compressive turbulence helps form a new generation of stars (as shown in \citealt{2009ApJ...694...84I}) in strong B-fields in 30 Dor by compressing and/or accumulating the gas along the B-field lines. As the latter is not influenced by magnetic pressure, this process could happen even in the regions where the B-fields are strong enough to act against the gravitational collapse. 
    
\end{itemize}

Our considered region is being carved by the R$\,$136 feedback resulting in multiple expanding-shells. The B-fields are sufficiently strong and seem to play a key role in supporting the cloud structure against this stellar feedback. However, there are certain regions at which the B-fields are relatively weak, where the gas can escape and inflate the external giant shells. Inside 30 Dor, supersonic and compressive turbulence accumulates gas along the field lines, enabling gravitational collapse and triggering stars' formation in strong B-fields environments. We argue that future polarimetric observations covering a large area in 30 Dor will be necessary to better understand the role of B-fields in the kinematical evolution of the entire 30 Dor region. In addition, the B-field strength estimated from the thermal dust polarization of 30 Dor is much higher than the average derived from the diffuse radio polarimetric observations of the hosting LMC. More sensitivity and resolution of polarimetric observations at radio wavelengths are needed to better understand the link from the galactic-scale to cloud-scale B-fields. 

However, we would like to warn the reader that our estimation of B-field strengths suffers from numerous uncertainties, as discussed in Section \ref{sec:uncertainties}. Consequently, the values of B-field strengths and other related parameters could vary. Nevertheless, we do not expect our main conclusions on the role of B-fields will be changed significantly or be reversed.

\textit{Acknowledgments}: This research is based on observations made with the NASA/DLR Stratospheric Observatory for Infrared Astronomy (SOFIA). SOFIA is jointly operated by the Universities Space Research Association, Inc. (USRA), under NASA contract NNA17BF53C, and the Deutsches SOFIA Institut (DSI) under DLR contract 50 OK 0901 to the University of Stuttgart. Financial support for this work was provided by NASA through award 4$\_$0152 issued by USRA. T.H is funded by the National Research Foundation of Korea (NRF) grant funded by the Korea government (MSIT) through the Mid-career Research Program (2019R1A2C1087045). AL and HY acknowledge the support of NSF AST 1816234, NASA TCAN 144AAG1967 and NASA ATP AAH7546.

\textit{Facility}: SOFIA

\begin{appendix}
\section{Channel and moment maps of [\textsc{CII}]}\label{sec:appChanMaps}
\begin{figure*}
    \centering
    \includegraphics[width=0.8\hsize]{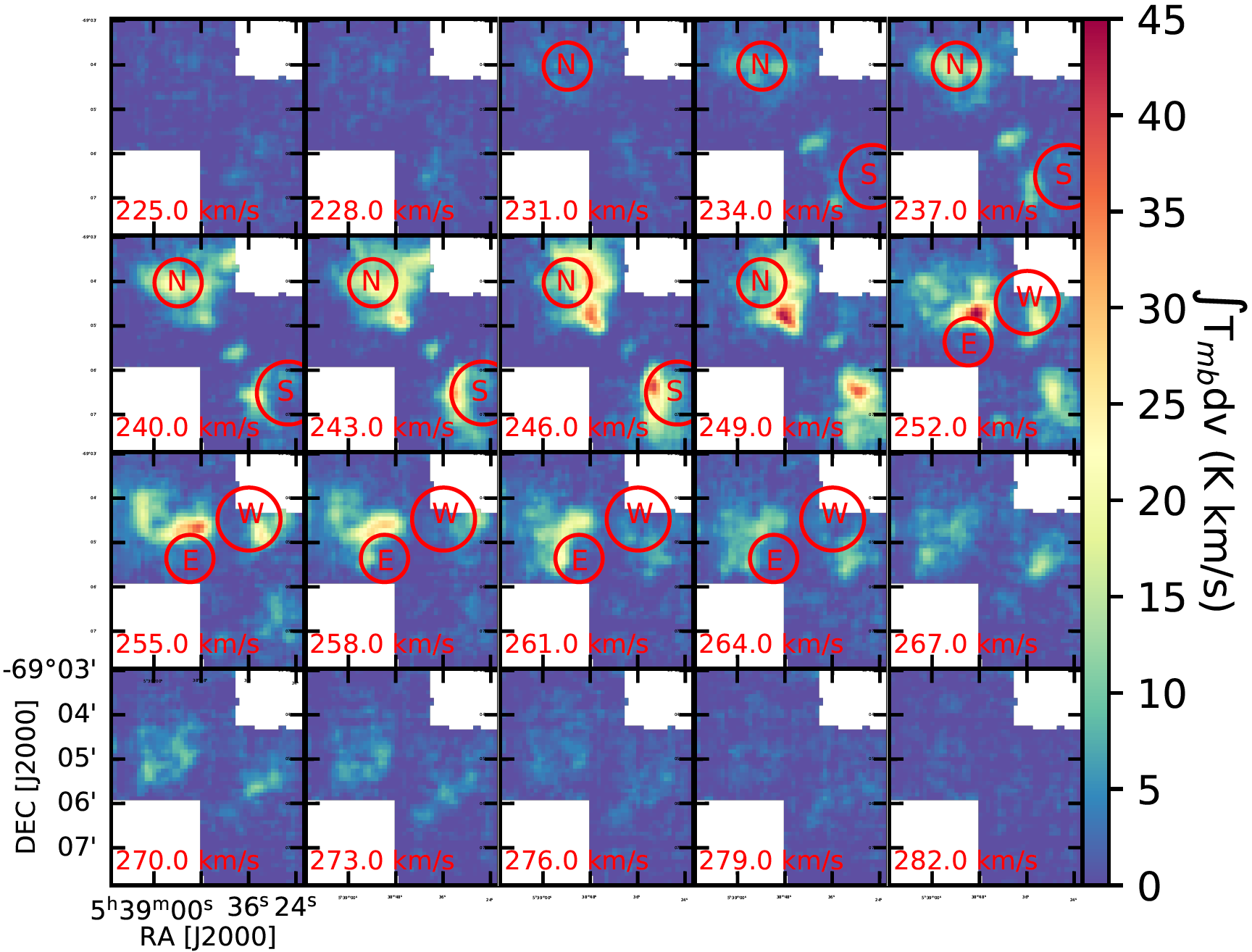}
    \caption{Channel maps of the 30 Dor region, in steps of 3 km s$^{-1}$, displaying the [CII] dynamics of the region. The red circles indicate the expanding shell candidates from Table \ref{tab:shellTable} in their proposed velocity with their name.}
    \label{fig:chanMapsFull}
\end{figure*}

Figure \ref{fig:chanMapsFull} shows the channel maps of \textsc{CII}. It demonstrates the presence of relatively complex kinematics in the region. Expanding shell candidates, established based on the PV diagrams of Figure \ref{fig:pvDiagrams}, are indicated with red circles in their respective velocity ranges. The channel maps indeed show several curved velocity structures that could be associated with the expanding shells.

\section{Magnetic fields using CO(2-1)}\label{sec:B_CO}
\begin{figure*}
    \centering
    \includegraphics[width=0.9\textwidth]{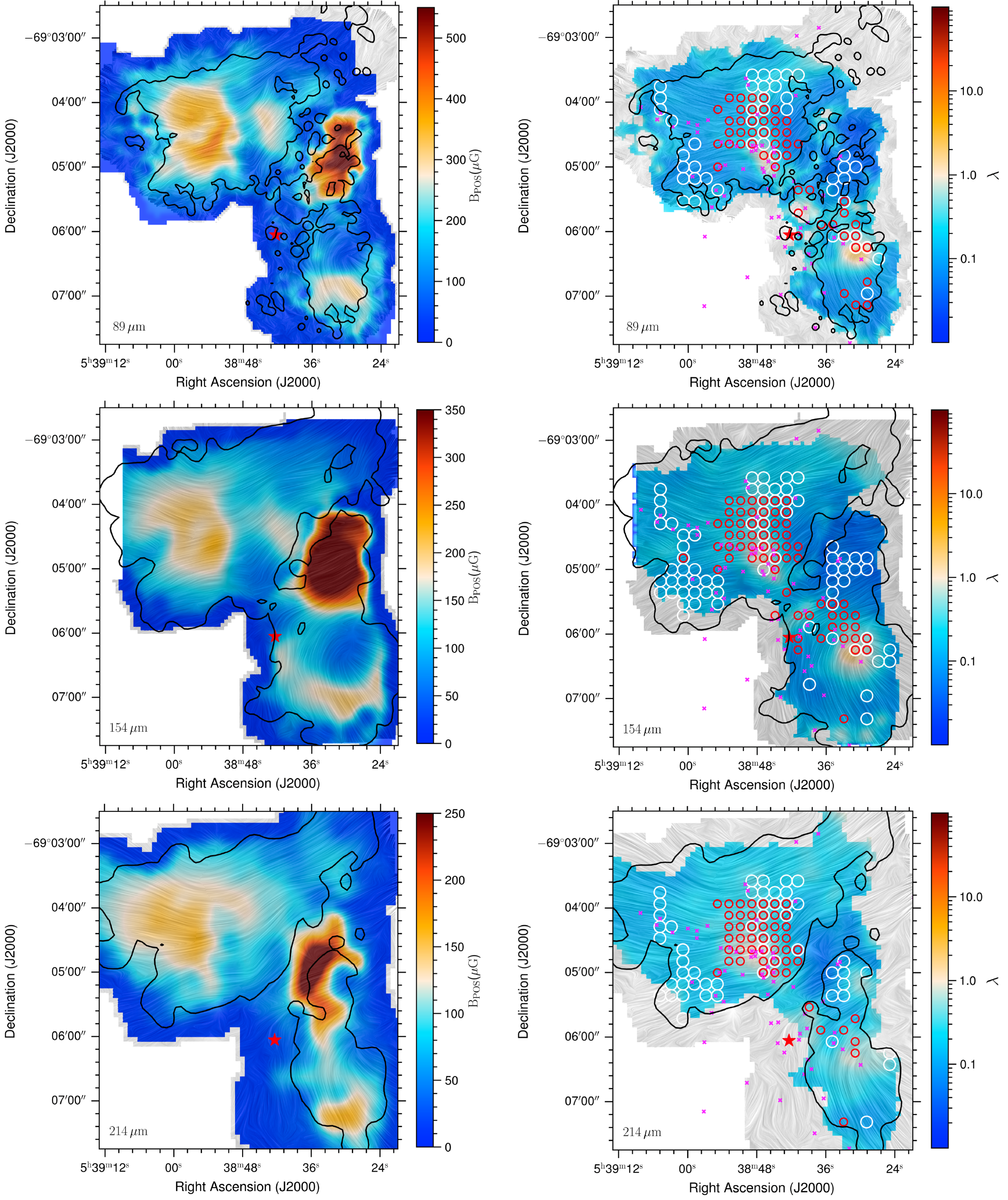}
    \caption{{\bf Left}: Magnetic field strength estimated by DCF method similar using CO(2-1) data. {\bf Right}: Mass-to-flux ratio derived by the field strength on the left panel. Symbols are the same as in Figure \ref{fig:maps_lamMA_CII}.}
    \label{fig:CO_Blam}
\end{figure*}

\begin{figure*}
    \centering
    \includegraphics[width=0.9\textwidth]{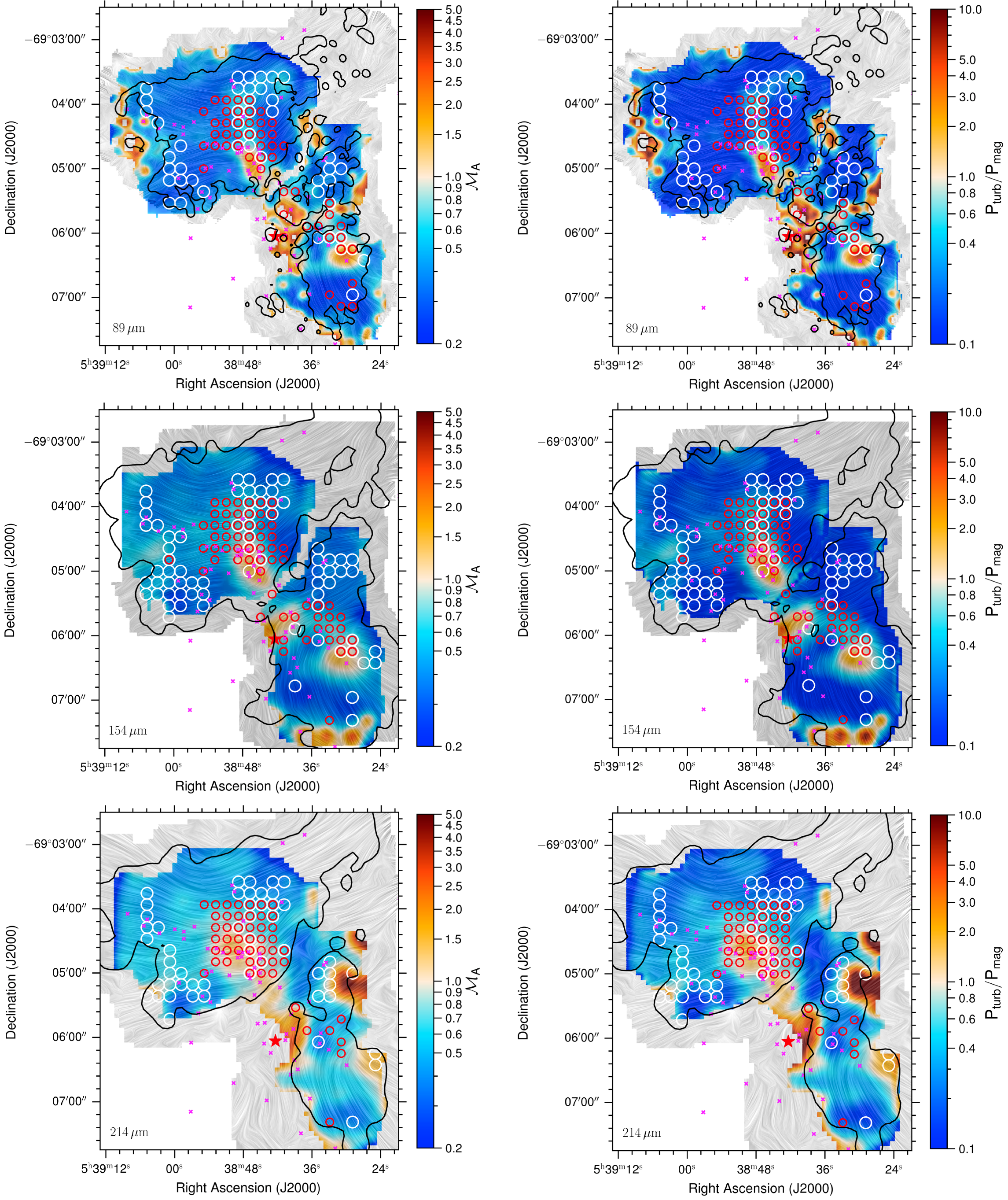}
    \caption{Afv\'enic Mach number (left panel) and turbulent-to-magnetic pressure ratio (right panel) derived by the field strength in Figure \ref{fig:CO_Blam}. Symbols are the same as in Figure \ref{fig:maps_lamMA_CII}.}
    \label{fig:CO_ma_p}
\end{figure*}

Figure \ref{fig:CO_Blam}, left panel shows maps of the strength of B-field estimated using the velocity dispersion from CO(2-1). A similar structure as in Figure \ref{fig:Bpos} is seen but the amplitude is slightly lower because of the narrower velocity dispersion of CO(2-1) (see lower panels in Figure \ref{fig:momentMaps}). Right panels of Figure \ref{fig:CO_Blam} and Figure \ref{fig:CO_ma_p} show the mass-to-flux ratio, Afv\'enic Mach number and turbulence-to-magnetic pressure ratio. As seen in [\textsc{CII}], magnetic field is sufficient strong to take over the gravitational potential and turbulence pressure except at the peak intensity. 

\end{appendix}


\newpage
\begin{thebibliography}{}
\expandafter\ifx\csname natexlab\endcsname\relax\def\natexlab#1{#1}\fi
\providecommand{\url}[1]{\href{#1}{#1}}
\providecommand{\dodoi}[1]{doi:~\href{http://doi.org/#1}{\nolinkurl{#1}}}
\providecommand{\doeprint}[1]{\href{http://ascl.net/#1}{\nolinkurl{http://ascl.net/#1}}}
\providecommand{\doarXiv}[1]{\href{https://arxiv.org/abs/#1}{\nolinkurl{https://arxiv.org/abs/#1}}}

\bibitem[{{Abe} {et~al.}(2021){Abe}, {Inoue}, {Inutsuka}, \&
  {Matsumoto}}]{2021ApJ...916...83A}
{Abe}, D., {Inoue}, T., {Inutsuka}, S.-i., \& {Matsumoto}, T. 2021, \apj, 916, 83, \dodoi{10.3847/1538-4357/ac07a1}

\bibitem[{{Alina} {et~al.}(2022){Alina}, {Montillaud}, {Hu}, {Lazarian},
  {Ristorcelli}, {Abdikamalov}, {Sagynbayeva}, {Juvela}, {Liu}, \&
  {Carri{\`e}re}}]{2022A&A...658A..90A}
{Alina}, D., {Montillaud}, J., {Hu}, Y., {et~al.} 2022, \aap, 658, A90,
  \dodoi{10.1051/0004-6361/202039065}

\bibitem[{{Andersson} {et~al.}(2015){Andersson}, {Lazarian}, \&
  {Vaillancourt}}]{2015ARA&A..53..501A}
{Andersson}, B.~G., {Lazarian}, A., \& {Vaillancourt}, J.~E. 2015, \araa, 53,
  501, \dodoi{10.1146/annurev-astro-082214-122414}

\bibitem[{{Arthur} {et~al.}(2011){Arthur}, {Henney}, {Mellema}, {de Colle}, \&
  {V{\'a}zquez-Semadeni}}]{2011MNRAS.414.1747A}
{Arthur}, S.~J., {Henney}, W.~J., {Mellema}, G., {de Colle}, F., \&
  {V{\'a}zquez-Semadeni}, E. 2011, \mnras, 414, 1747,
  \dodoi{10.1111/j.1365-2966.2011.18507.x}
  
\bibitem[Beuther et al.(2022)]{2022A&A...659A..77B} Beuther, H., Schneider, N., Simon, R., et al.\ 2022, \aap, 659, A77. doi:10.1051/0004-6361/202142689

\bibitem[{{Bonne} {et~al.}(2020{\natexlab{a}}){Bonne}, {Schneider}, {Bontemps},
  {Clarke}, {Gusdorf}, {Lehmann}, {Steinke}, {Csengeri}, {Kabanovic}, {Simon},
  {Buchbender}, \& {G{\"u}sten}}]{2020A&A...641A..17B}
{Bonne}, L., {Schneider}, N., {Bontemps}, S., {et~al.} 2020{\natexlab{a}},
  \aap, 641, A17, \dodoi{10.1051/0004-6361/201937104}

\bibitem[{{Bonne} {et~al.}(2020{\natexlab{b}}){Bonne}, {Bontemps}, {Schneider},
  {Clarke}, {Arzoumanian}, {Fukui}, {Tachihara}, {Csengeri}, {Guesten},
  {Ohama}, {Okamoto}, {Simon}, {Yahia}, \& {Yamamoto}}]{2020A&A...644A..27B}
{Bonne}, L., {Bontemps}, S., {Schneider}, N., {et~al.} 2020{\natexlab{b}},
  \aap, 644, A27, \dodoi{10.1051/0004-6361/202038281}

\bibitem[Bonne et al.(2022{\natexlab{a}})]{2022ApJ...935..171B} Bonne, L., Schneider, N., Garc{\'\i}a, P., et al.\ 2022, \apj, 935, 171. doi:10.3847/1538-4357/ac8052

\bibitem[Bonne et al.(2022{\natexlab{b}})]{2022arXiv220610889B} Bonne, L., Peretto, N., Duarte-Cabral, A., et al.\ 2022, arXiv:2206.10889

\bibitem[{{Brandl}(2005)}]{2005ASSL..329...49B}
{Brandl}, B.~R. 2005, in Astrophysics and Space Science Library, Vol. 329,
  Starbursts: From 30 Doradus to Lyman Break Galaxies, ed. R.~{de Grijs} \&
  R.~M. {Gonz{\'a}lez Delgado}, 49, \dodoi{10.1007/1-4020-3539-X\_10}

\bibitem[{Cabral \& Leedom(1993)}]{10.1145/166117.166151}
Cabral, B., \& Leedom, L.~C. 1993, in Proceedings of the 20th Annual Conference
  on Computer Graphics and Interactive Techniques, SIGGRAPH '93 (New York, NY,
  USA: Association for Computing Machinery), 263–270,
  \dodoi{10.1145/166117.166151}
  
\bibitem[{{Chandrasekhar} \& {Fermi}(1953)}]{1953ApJ...118..113C}
{Chandrasekhar}, S., \& {Fermi}, E. 1953, \apj, 118, 113,
  \dodoi{10.1086/145731}

\bibitem[{{Chen} {et~al.}(2019){Chen}, {King}, {Li}, {Fissel}, \&
  {Mazzei}}]{2019MNRAS.485.3499C}
{Chen}, C.-Y., {King}, P.~K., {Li}, Z.-Y., {Fissel}, L.~M., \& {Mazzei}, R.~R.
  2019, \mnras, 485, 3499, \dodoi{10.1093/mnras/stz618}

\bibitem[{{Chevance} {et~al.}(2020){Chevance}, {Madden}, {Fischer}, {Vacca},
  {Lebouteiller}, {Fadda}, {Galliano}, {Indebetouw}, {Kruijssen}, {Lee},
  {Poglitsch}, {Polles}, {Cormier}, {Hony}, {Iserlohe}, {Krabbe}, {Meixner},
  {Sabbi}, \& {Zinnecker}}]{2020MNRAS.494.5279C}
{Chevance}, M., {Madden}, S.~C., {Fischer}, C., {et~al.} 2020, \mnras, 494,
  5279, \dodoi{10.1093/mnras/staa1106}

\bibitem[{{Cho} \& {Lazarian}(2003)}]{2003MNRAS.345..325C}
{Cho}, J., \& {Lazarian}, A. 2003, \mnras, 345, 325,
  \dodoi{10.1046/j.1365-8711.2003.06941.x}

\bibitem[{{Chu} \& {Kennicutt}(1994)}]{1994ApJ...425..720C}
{Chu}, Y.-H., \& {Kennicutt}, Robert~C., J. 1994, \apj, 425, 720,
  \dodoi{10.1086/174017}

\bibitem[{{Crutcher}(2012)}]{2012ARA&A..50...29C}
{Crutcher}, R.~M. 2012, \araa, 50, 29,
  \dodoi{10.1146/annurev-astro-081811-125514}

\bibitem[{{Crutcher} {et~al.}(2004){Crutcher}, {Nutter}, {Ward-Thompson}, \&
  {Kirk}}]{2004ApJ...600..279C}
{Crutcher}, R.~M., {Nutter}, D.~J., {Ward-Thompson}, D., \& {Kirk}, J.~M. 2004,
  \apj, 600, 279, \dodoi{10.1086/379705}

\bibitem[{{Davis}(1951)}]{1951PhRv...81..890D}
{Davis}, L. 1951, Physical Review, 81, 890, \dodoi{10.1103/PhysRev.81.890.2}

\bibitem[{{Devaraj} {et~al.}(2021){Devaraj}, {Clemens}, {Dewangan}, {Luna},
  {Ray}, \& {Mackey}}]{2021ApJ...911...81D}
{Devaraj}, R., {Clemens}, D.~P., {Dewangan}, L.~K., {et~al.} 2021, \apj, 911,
  81, \dodoi{10.3847/1538-4357/abe9b1}

\bibitem[{{Dolginov} \& {Mytrophanov}(1976)}]{1976Ap&SS..43..257D}
{Dolginov}, A.~Z., \& {Mytrophanov}, I.~G. 1976, \apss, 43, 257,
  \dodoi{10.1007/BF00640009}

\bibitem[{{Eswaraiah} {et~al.}(2021){Eswaraiah}, {Li}, {Furuya}, {Hasegawa},
  {Ward-Thompson}, {Qiu}, {Ohashi}, {Pattle}, {Sadavoy}, {Hull}, {Berry},
  {Doi}, {Ching}, {Lai}, {Wang}, {Koch}, {Kwon}, {Kwon}, {Bastien},
  {Arzoumanian}, {Coud{\'e}}, {Soam}, {Fanciullo}, {Yen}, {Liu}, {Hoang}, {Ping
  Chen}, {Shimajiri}, {Liu}, {Chen}, {Li}, {Lyo}, {Hwang}, {Johnstone}, {Rao},
  {Bich Ngoc}, {Ngoc Diep}, {Mairs}, {Parsons}, {Tamura}, {Tahani}, {Vivien
  Chen}, {Nakamura}, {Shinnaga}, {Tang}, {Cho}, {Won Lee}, {Inutsuka}, {Inoue},
  {Iwasaki}, {Qian}, {Xie}, {Li}, {Liu}, {Zhang}, {Chen}, {Zhang}, {Zhu},
  {Zhou}, {Andr{\'e}}, {Liu}, {Yuan}, {Lu}, {Peretto}, {Bourke}, {Byun}, {Dai},
  {Duan}, {Duan}, {Eden}, {Matthews}, {Fiege}, {Fissel}, {Kim}, {Lee}, {Kim},
  {Pyo}, {Choi}, {Choi}, {Chrysostomou}, {Jung Chung}, {Ngoc Tram},
  {Franzmann}, {Friberg}, {Friesen}, {Fuller}, {Gledhill}, {Graves}, {Greaves},
  {Griffin}, {Gu}, {Han}, {Hatchell}, {Hayashi}, {Houde}, {Kawabata}, {Jeong},
  {Kang}, {Kang}, {Kang}, {Kataoka}, {Kemper}, {Rawlings}, {Rawlings},
  {Retter}, {Richer}, {Rigby}, {Saito}, {Savini}, {Scaife}, {Seta}, {Kim}, {Hee
  Kim}, {Kim}, {Kirchschlager}, {Kirk}, {Kobayashi}, {Konyves}, {Kusune},
  {Lacaille}, {Law}, {Lee}, {Lee}, {Matsumura}, {Moriarty-Schieven}, {Nagata},
  {Nakanishi}, {Onaka}, {Park}, {Tang}, {Tomisaka}, {Tsukamoto}, {Viti},
  {Wang}, {Whitworth}, {Yoo}, {Yun}, {Zenko}, {Zhang}, {de Looze}, {Dowell},
  {Eyres}, {Falle}, {Robitaille}, \& {van Loo}}]{2021ApJ...912L..27E}
{Eswaraiah}, C., {Li}, D., {Furuya}, R.~S., {et~al.} 2021, \apjl, 912, L27,
  \dodoi{10.3847/2041-8213/abeb1c}

\bibitem[{{Ewertowski} \& {Basu}(2013)}]{2013ApJ...767...33E}
{Ewertowski}, B., \& {Basu}, S. 2013, \apj, 767, 33,
  \dodoi{10.1088/0004-637X/767/1/33}

\bibitem[{{Federrath} \& {Klessen}(2013)}]{2013ApJ...763...51F}
{Federrath}, C., \& {Klessen}, R.~S. 2013, \apj, 763, 51,
  \dodoi{10.1088/0004-637X/763/1/51}

\bibitem[{{Federrath} {et~al.}(2010){Federrath}, {Roman-Duval}, {Klessen},
  {Schmidt}, \& {Mac Low}}]{2010A&A...512A..81F}
{Federrath}, C., {Roman-Duval}, J., {Klessen}, R.~S., {Schmidt}, W., \& {Mac
  Low}, M.~M. 2010, \aap, 512, A81, \dodoi{10.1051/0004-6361/200912437}

\bibitem[{{Girichidis} {et~al.}(2014){Girichidis}, {Konstandin}, {Whitworth},
  \& {Klessen}}]{2014ApJ...781...91G}
{Girichidis}, P., {Konstandin}, L., {Whitworth}, A.~P., \& {Klessen}, R.~S.
  2014, \apj, 781, 91, \dodoi{10.1088/0004-637X/781/2/91}

\bibitem[{{Gonz{\'a}lez-Casanova} \& {Lazarian}(2017)}]{2017ApJ...835...41G}
{Gonz{\'a}lez-Casanova}, D.~F., \& {Lazarian}, A. 2017, \apj, 835, 41,
  \dodoi{10.3847/1538-4357/835/1/41}

\bibitem[{{Gordon} {et~al.}(2018){Gordon}, {Lopez-Rodriguez}, {Andersson},
  {Clarke}, {Coude}, {Moullet}, {Richards}, {Shuping}, {Vacca}, \&
  {Yorke}}]{2018arXiv181103100G}
{Gordon}, M.~S., {Lopez-Rodriguez}, E., {Andersson}, B.~G., {et~al.} 2018,
  arXiv e-prints, arXiv:1811.03100.
\newblock \doarXiv{1811.03100}

\bibitem[{{Guerra} {et~al.}(2021){Guerra}, {Chuss}, {Dowell}, {Houde},
  {Michail}, {Siah}, \& {Wollack}}]{2021ApJ...908...98G}
{Guerra}, J.~A., {Chuss}, D.~T., {Dowell}, C.~D., {et~al.} 2021, \apj, 908, 98,
  \dodoi{10.3847/1538-4357/abd6f0}

\bibitem[{{G{\"u}sten} {et~al.}(2006){G{\"u}sten}, {Nyman}, {Schilke},
  {Menten}, {Cesarsky}, \& {Booth}}]{2006A&A...454L..13G}
{G{\"u}sten}, R., {Nyman}, L.~{\r{A}}., {Schilke}, P., {et~al.} 2006, \aap,
  454, L13, \dodoi{10.1051/0004-6361:20065420}

\bibitem[{{Hacar} {et~al.}(2016){Hacar}, {Alves}, {Burkert}, \&
  {Goldsmith}}]{2016A&A...591A.104H}
{Hacar}, A., {Alves}, J., {Burkert}, A., \& {Goldsmith}, P. 2016, \aap, 591,
  A104, \dodoi{10.1051/0004-6361/201527319}

\bibitem[{{Hartmann} \& {Burkert}(2007)}]{2007ApJ...654..988H}
{Hartmann}, L., \& {Burkert}, A. 2007, \apj, 654, 988, \dodoi{10.1086/509321}

\bibitem[{{Henney} {et~al.}(2009){Henney}, {Arthur}, {de Colle}, \&
  {Mellema}}]{2009MNRAS.398..157H}
{Henney}, W.~J., {Arthur}, S.~J., {de Colle}, F., \& {Mellema}, G. 2009,
  \mnras, 398, 157, \dodoi{10.1111/j.1365-2966.2009.15153.x}

\bibitem[Hoang et al.(2022)]{2022arXiv220502334H} Hoang, T., Tram, L.~N., Phan, V.~H.~M., et al.\ 2022, arXiv:2205.02334

\bibitem[{{Hoang} {et~al.}(2021{\natexlab{b}}){Hoang}, {Bich Ngoc}, {Diep},
  {Tram}, {Hoang}, {Lim}, {Nguyen}, {Le}, {Phuong}, {Fuda}, {Van Bui},
  {Pattle}, {Truong Le}, {Phan}, \& {Chau Giang}}]{2021arXiv210810045H}
{Hoang}, T.~D., {Bich Ngoc}, N., {Diep}, P.~N., {et~al.} 2021{\natexlab{b}},
  arXiv e-prints, arXiv:2108.10045.
\newblock \doarXiv{2108.10045}

\bibitem[{{Houde} {et~al.}(2009){Houde}, {Vaillancourt}, {Hildebrand},
  {Chitsazzadeh}, \& {Kirby}}]{2009ApJ...706.1504H}
{Houde}, M., {Vaillancourt}, J.~E., {Hildebrand}, R.~H., {Chitsazzadeh}, S., \&
  {Kirby}, L. 2009, \apj, 706, 1504, \dodoi{10.1088/0004-637X/706/2/1504}

\bibitem[{{Hu} {et~al.}(2021{\natexlab{a}}){Hu}, {Lazarian}, \&
  {Stanimirovi{\'c}}}]{2021ApJ...912....2H}
{Hu}, Y., {Lazarian}, A., \& {Stanimirovi{\'c}}, S. 2021{\natexlab{a}}, \apj,
  912, 2, \dodoi{10.3847/1538-4357/abedb7}

\bibitem[{{Hu} {et~al.}(2021{\natexlab{b}}){Hu}, {Lazarian}, \&
  {Wang}}]{2021arXiv210503605H}
{Hu}, Y., {Lazarian}, A., \& {Wang}, Q.~D. 2021{\natexlab{b}}, arXiv e-prints,
  arXiv:2105.03605.
\newblock \doarXiv{2105.03605}

\bibitem[{{Hu} {et~al.}(2018){Hu}, {Yuen}, \& {Lazarian}}]{2018MNRAS.480.1333H}
{Hu}, Y., {Yuen}, K.~H., \& {Lazarian}, A. 2018, \mnras, 480, 1333,
  \dodoi{10.1093/mnras/sty1807}
  
\bibitem[{{Hu} {et~al.}(2019{\natexlab{a}}){Hu}, {Yuen}, \&
  {Lazarian}}]{2019ApJ...886...17H}
{Hu}, Y., {Yuen}, K.~H., \& {Lazarian}, A. 2019{\natexlab{a}}, \apj, 886, 17, 
\dodoi{10.3847/1538-4357/ab4b5e}

\bibitem[{{Hu} {et~al.}(2019{\natexlab{b}}){Hu}, {Yuen}, {Lazarian}, {Ho}, {Benjamin},
  {Hill}, {Lockman}, {Goldsmith}, \& {Lazarian}}]{2019NatAs...3..776H}
{Hu}, Y., {Yuen}, K.~H., {Lazarian}, V., {et~al.} 2019{\natexlab{b}}, Nature Astronomy, 3,
  776, \dodoi{10.1038/s41550-019-0769-0}

\bibitem[{{Indebetouw} {et~al.}(2009){Indebetouw}, {de Messi{\`e}res},
  {Madden}, {Engelbracht}, {Smith}, {Meixner}, {Brandl}, {Smith}, {Boulanger},
  {Galliano}, {Gordon}, {Hora}, {Sewilo}, {Tielens}, {Werner}, \&
  {Wolfire}}]{2009ApJ...694...84I}
{Indebetouw}, R., {de Messi{\`e}res}, G.~E., {Madden}, S., {et~al.} 2009, \apj,
  694, 84, \dodoi{10.1088/0004-637X/694/1/84}

\bibitem[{{Indebetouw} {et~al.}(2013){Indebetouw}, {Brogan}, {Chen}, {Leroy},
  {Johnson}, {Muller}, {Madden}, {Cormier}, {Galliano}, {Hughes}, {Hunter},
  {Kawamura}, {Kepley}, {Lebouteiller}, {Meixner}, {Oliveira}, {Onishi}, \&
  {Vasyunina}}]{2013ApJ...774...73I}
{Indebetouw}, R., {Brogan}, C., {Chen}, C. H.~R., {et~al.} 2013, \apj, 774, 73,
  \dodoi{10.1088/0004-637X/774/1/73}

\bibitem[{{Inoue} \& {Fukui}(2013)}]{2013ApJ...774L..31I}
{Inoue}, T., \& {Fukui}, Y. 2013, \apjl, 774, L31,
  \dodoi{10.1088/2041-8205/774/2/L31}

\bibitem[{{Jackson} {et~al.}(2019){Jackson}, {Whitaker}, {Rathborne}, {Foster},
  {Contreras}, {Sanhueza}, {Stephens}, {Longmore}, \&
  {Allingham}}]{2019ApJ...870....5J}
{Jackson}, J.~M., {Whitaker}, J.~S., {Rathborne}, J.~M., {et~al.} 2019, \apj,
  870, 5, \dodoi{10.3847/1538-4357/aaef84}

\bibitem[{{Kainulainen} {et~al.}(2014){Kainulainen}, {Federrath}, \&
  {Henning}}]{2014Sci...344..183K}
{Kainulainen}, J., {Federrath}, C., \& {Henning}, T. 2014, Science, 344, 183,
  \dodoi{10.1126/science.1248724}

\bibitem[{{Kandori} {et~al.}(2020){Kandori}, {Tamura}, {Saito}, {Tomisaka},
  {Matsumoto}, {Tazaki}, {Nagata}, {Kusakabe}, {Nakajima}, {Kwon}, {Nagayama},
  \& {Tatematsu}}]{2020ApJ...892..128K}
{Kandori}, R., {Tamura}, M., {Saito}, M., {et~al.} 2020, \apj, 892, 128,
  \dodoi{10.3847/1538-4357/ab7b68}

\bibitem[{{Kennicutt}(1984)}]{1984ApJ...287..116K}
{Kennicutt}, R.~C., J. 1984, \apj, 287, 116, \dodoi{10.1086/162669}

\bibitem[{{Kim} {et~al.}(2016){Kim}, {Jeong}, {Pak}, {Park}, \&
  {Tamura}}]{2016ApJS..222....2K}
{Kim}, J., {Jeong}, W.-S., {Pak}, S., {Park}, W.-K., \& {Tamura}, M. 2016,
  \apjs, 222, 2, \dodoi{10.3847/0067-0049/222/1/2}
  
\bibitem[{{Kim} {et~al.}(1999){Kim}, {Dopita}, {Staveley-Smith}, \&
  {Bessell}}]{1999AJ....118.2797K}
{Kim}, S., {Dopita}, M.~A., {Staveley-Smith}, L., \& {Bessell}, M.~S. 1999,
  \aj, 118, 2797, \dodoi{10.1086/301116}

\bibitem[{{Klein} {et~al.}(1993){Klein}, {Haynes}, {Wielebinski}, \&
  {Meinert}}]{1993A&A...271..402K}
{Klein}, U., {Haynes}, R.~F., {Wielebinski}, R., \& {Meinert}, D. 1993, \aap,
  271, 402
  
\bibitem[{{Klessen}(2000)}]{2000ApJ...535..869K}
{Klessen}, R.~S. 2000, \apj, 535, 869, \dodoi{10.1086/308854}

\bibitem[{{K{\"o}rtgen} {et~al.}(2019){K{\"o}rtgen}, {Federrath}, \&
  {Banerjee}}]{2019MNRAS.482.5233K}
{K{\"o}rtgen}, B., {Federrath}, C., \& {Banerjee}, R. 2019, \mnras, 482, 5233,
  \dodoi{10.1093/mnras/sty3071}

\bibitem[{{Lazarian}(2007)}]{2007JQSRT.106..225L}
{Lazarian}, A. 2007, \jqsrt, 106, 225, \dodoi{10.1016/j.jqsrt.2007.01.038}

\bibitem[{{Lazarian}(2014)}]{2014SSRv..181....1L}
{Lazarian}, A. 2014, \ssr, 181, 1, \dodoi{10.1007/s11214-013-0031-5}

\bibitem[{{Lazarian} \& {Pogosyan}(2000)}]{2000ApJ...537..720L}
{Lazarian}, A., \& {Pogosyan}, D. 2000, \apj, 537, 720, \dodoi{10.1086/309040}

\bibitem[{{Lazarian} \& {Vishniac}(1999)}]{1999ApJ...517..700L}
{Lazarian}, A., \& {Vishniac}, E.~T. 1999, \apj, 517, 700,
  \dodoi{10.1086/307233}

\bibitem[{{Lazarian} \& {Yuen}(2018)}]{2018ApJ...853...96L}
{Lazarian}, A., \& {Yuen}, K.~H. 2018, \apj, 853, 96,
  \dodoi{10.3847/1538-4357/aaa241}

\bibitem[{{Lazarian} {et~al.}(2018){Lazarian}, {Yuen}, {Ho}, {Chen},
  {Lazarian}, {Lu}, {Yang}, \& {Hu}}]{2018ApJ...865...46L}
{Lazarian}, A., {Yuen}, K.~H., {Ho}, K.~W., {et~al.} 2018, \apj, 865, 46,
  \dodoi{10.3847/1538-4357/aad7ff}

\bibitem[{{Lazarian} {et~al.}(2020){Lazarian}, {Yuen}, \&
  {Pogosyan}}]{2020arXiv200207996L}
{Lazarian}, A., {Yuen}, K.~H., \& {Pogosyan}, D. 2020, arXiv e-prints,
  arXiv:2002.07996.
\newblock \doarXiv{2002.07996}

\bibitem[{{Lazarian} {et~al.}(2022){Lazarian}, {Yuen}, \&
  {Pogosyan}}]{2022ApJ...935...77L}
---. 2022, \apj, 935, 77, \dodoi{10.3847/1538-4357/ac6877}

\bibitem[Schneider et al.(2018)]{2018A&A...617A..45S} Schneider, N., R{\"o}llig, M., Simon, R., et al.\ 2018, \aap, 617, A45. doi:10.1051/0004-6361/201732508

\bibitem[{{Lee} {et~al.}(2019){Lee}, {Madden}, {Le Petit}, {Gusdorf},
  {Lesaffre}, {Wu}, {Lebouteiller}, {Galliano}, \&
  {Chevance}}]{2019A&A...628A.113L}
{Lee}, M.~Y., {Madden}, S.~C., {Le Petit}, F., {et~al.} 2019, \aap, 628, A113,
  \dodoi{10.1051/0004-6361/201935215}

\bibitem[{{Li} {et~al.}(2022){Li}, {Lopez-Rodriguez}, {Ajeddig}, {Andr{\'e}},
  {McKee}, {Rho}, \& {Klein}}]{2022MNRAS.510.6085L}
{Li}, P.~S., {Lopez-Rodriguez}, E., {Ajeddig}, H., {et~al.} 2022, \mnras, 510,
  6085, \dodoi{10.1093/mnras/stab3448}
  
\bibitem[{{Liu} {et~al.}(2022){Liu}, {Qiu}, \& {Zhang}}]{2022ApJ...925...30L}
{Liu}, J., {Qiu}, K., \& {Zhang}, Q. 2022, \apj, 925, 30,
  \dodoi{10.3847/1538-4357/ac3911}

\bibitem[{{Lopez} {et~al.}(2011){Lopez}, {Krumholz}, {Bolatto}, {Prochaska}, \&
  {Ramirez-Ruiz}}]{2011ApJ...731...91L}
{Lopez}, L.~A., {Krumholz}, M.~R., {Bolatto}, A.~D., {Prochaska}, J.~X., \&
  {Ramirez-Ruiz}, E. 2011, \apj, 731, 91, \dodoi{10.1088/0004-637X/731/2/91}

\bibitem[{{Lu} {et~al.}(2020){Lu}, {Lazarian}, \&
  {Pogosyan}}]{2020MNRAS.496.2868L}
{Lu}, Z., {Lazarian}, A., \& {Pogosyan}, D. 2020, \mnras, 496, 2868,
  \dodoi{10.1093/mnras/staa1570}
  
\bibitem[{{Luisi} {et~al.}(2021){Luisi}, {Anderson}, {Schneider}, {Simon},
  {Kabanovic}, {G{\"u}sten}, {Zavagno}, {Broos}, {Buchbender}, {Guevara},
  {Jacobs}, {Justen}, {Klein}, {Linville}, {R{\"o}llig}, {Russeil}, {Stutzki},
  {Tiwari}, {Townsley}, \& {Tielens}}]{2021SciA....7.9511L}
{Luisi}, M., {Anderson}, L.~D., {Schneider}, N., {et~al.} 2021, Science
  Advances, 7, eabe9511, \dodoi{10.1126/sciadv.abe9511}

\bibitem[{{Mac Low} \& {Klessen}(2004)}]{2004RvMP...76..125M}
{Mac Low}, M.-M., \& {Klessen}, R.~S. 2004, Reviews of Modern Physics, 76, 125,
  \dodoi{10.1103/RevModPhys.76.125}

\bibitem[{{Mackey} \& {Lim}(2011)}]{2011MNRAS.412.2079M}
{Mackey}, J., \& {Lim}, A.~J. 2011, \mnras, 412, 2079,
  \dodoi{10.1111/j.1365-2966.2010.18043.x}

\bibitem[{{Mao} {et~al.}(2012){Mao}, {McClure-Griffiths}, {Gaensler},
  {Haverkorn}, {Beck}, {McConnell}, {Wolleben}, {Stanimirovi{\'c}}, {Dickey},
  \& {Staveley-Smith}}]{2012ApJ...759...25M}
{Mao}, S.~A., {McClure-Griffiths}, N.~M., {Gaensler}, B.~M., {et~al.} 2012,
  \apj, 759, 25, \dodoi{10.1088/0004-637X/759/1/25}

\bibitem[{{Meixner} {et~al.}(2013){Meixner}, {Panuzzo}, {Roman-Duval},
  {Engelbracht}, {Babler}, {Seale}, {Hony}, {Montiel}, {Sauvage}, {Gordon},
  {Misselt}, {Okumura}, {Chanial}, {Beck}, {Bernard}, {Bolatto}, {Bot},
  {Boyer}, {Carlson}, {Clayton}, {Chen}, {Cormier}, {Fukui}, {Galametz},
  {Galliano}, {Hora}, {Hughes}, {Indebetouw}, {Israel}, {Kawamura}, {Kemper},
  {Kim}, {Kwon}, {Lebouteiller}, {Li}, {Long}, {Madden}, {Matsuura}, {Muller},
  {Oliveira}, {Onishi}, {Otsuka}, {Paradis}, {Poglitsch}, {Reach},
  {Robitaille}, {Rubio}, {Sargent}, {Sewi{\l}o}, {Skibba}, {Smith},
  {Srinivasan}, {Tielens}, {van Loon}, \& {Whitney}}]{2013AJ....146...62M}
{Meixner}, M., {Panuzzo}, P., {Roman-Duval}, J., {et~al.} 2013, \aj, 146, 62,
  \dodoi{10.1088/0004-6256/146/3/62}

\bibitem[{{Melnick} {et~al.}(2021){Melnick}, {Tenorio-Tagle}, \&
  {Telles}}]{2021A&A...649A.175M}
{Melnick}, J., {Tenorio-Tagle}, G., \& {Telles}, E. 2021, \aap, 649, A175,
  \dodoi{10.1051/0004-6361/201937268}
  
\bibitem[{{Mestel}(1966)}]{1966MNRAS.133..265M}
{Mestel}, L. 1966, \mnras, 133, 265, \dodoi{10.1093/mnras/133.2.265}

\bibitem[{{Nakajima} {et~al.}(2007){Nakajima}, {Kandori}, {Tamura}, {Kusakabe},
  {Hashimoto}, {Nagayama}, {Nagata}, {Hatano}, {Kato}, \&
  {Hough}}]{2007PASJ...59..519N}
{Nakajima}, Y., {Kandori}, R., {Tamura}, M., {et~al.} 2007, \pasj, 59, 519,
  \dodoi{10.1093/pasj/59.3.519}

\bibitem[{{Ngoc} {et~al.}(2021){Ngoc}, {Diep}, {Parsons}, {Pattle}, {Hoang},
  {Ward-Thompson}, {Tram}, {Hull}, {Tahani}, {Furuya}, {Bastien}, {Qiu},
  {Hasegawa}, {Kwon}, {Doi}, {Lai}, {Coud{\'e}}, {Berry}, {Ching}, {Hwang},
  {Soam}, {Wang}, {Arzoumanian}, {Bourke}, {Byun}, {Chen}, {Chen}, {Chen},
  {Chen}, {Cho}, {Choi}, {Choi}, {Chrysostomou}, {Chung}, {Dai}, {Di
  Francesco}, {Duan}, {Duan}, {Eden}, {Eswaraiah}, {Fanciullo}, {Fiege},
  {Fissel}, {Franzmann}, {Friberg}, {Friesen}, {Fuller}, {Gledhill}, {Graves},
  {Greaves}, {Griffin}, {Gu}, {Han}, {Hatchell}, {Hayashi}, {Houde}, {Inoue},
  {Inutsuka}, {Iwasaki}, {Jeong}, {Johnstone}, {Kang}, {Kang}, {Kang},
  {Kataoka}, {Kawabata}, {Kemper}, {Kim}, {Kim}, {Pyo}, {Qian}, {Rao},
  {Rawlings}, {Rawlings}, {Retter}, {Richer}, {Rigby}, {Sadavoy}, {Saito},
  {Savini}, {Scaife}, {Seta}, {Kim}, {Kim}, {Kim}, {Kim}, {Kirchschlager},
  {Kirk}, {Kobayashi}, {Koch}, {Konyves}, {Kusune}, {Kwon}, {Lacaille}, {Law},
  {Lee}, {Lee}, {Lee}, {Lee}, {Lee}, {Lee}, {Li}, {Li}, {Li}, {Liu}, {Liu},
  {Liu}, {Liu}, {Lu}, {Lyo}, {Mairs}, {Matsumura}, {Matthews},
  {Moriarty-Schieven}, {Nagata}, {Nakamura}, {Nakanishi}, {Ohashi}, {Onaka},
  {Park}, {Peretto}, {Shimajiri}, {Shinnaga}, {Tamura}, {Tang}, {Tang},
  {Tomisaka}, {Tsukamoto}, {Viti}, {Wang}, {Whitworth}, {Xie}, {Yen}, {Yoo},
  {Yuan}, {Yun}, {Zenko}, {Zhang}, {Zhang}, {Zhang}, {Zhou}, {Zhu}, {de Looze},
  {Andr{\'e}}, {Dowell}, {Eyres}, {Falle}, {Robitaille}, \& {van
  Loo}}]{2021ApJ...908...10N}
{Ngoc}, N.~B., {Diep}, P.~N., {Parsons}, H., {et~al.} 2021, \apj, 908, 10,
  \dodoi{10.3847/1538-4357/abd0fc}

\bibitem[{{Okada} {et~al.}(2019){Okada}, {G{\"u}sten}, {Requena-Torres},
  {R{\"o}llig}, {Stutzki}, {Graf}, \& {Hughes}}]{2019A&A...621A..62O}
{Okada}, Y., {G{\"u}sten}, R., {Requena-Torres}, M.~A., {et~al.} 2019, \aap,
  621, A62, \dodoi{10.1051/0004-6361/201833398}

\bibitem[{{Pabst} {et~al.}(2019){Pabst}, {Higgins}, {Goicoechea}, {Teyssier},
  {Berne}, {Chambers}, {Wolfire}, {Suri}, {Guesten}, {Stutzki}, {Graf},
  {Risacher}, \& {Tielens}}]{2019Natur.565..618P}
{Pabst}, C., {Higgins}, R., {Goicoechea}, J.~R., {et~al.} 2019, \nat, 565, 618,
  \dodoi{10.1038/s41586-018-0844-1}

\bibitem[{{Pabst} {et~al.}(2020){Pabst}, {Goicoechea}, {Teyssier}, {Bern{\'e}},
  {Higgins}, {Chambers}, {Kabanovic}, {G{\"u}sten}, {Stutzki}, \&
  {Tielens}}]{2020A&A...639A...2P}
{Pabst}, C.~H.~M., {Goicoechea}, J.~R., {Teyssier}, D., {et~al.} 2020, \aap,
  639, A2, \dodoi{10.1051/0004-6361/202037560}

\bibitem[{{Pattle} {et~al.}(2018){Pattle}, {Ward-Thompson}, {Hasegawa},
  {Bastien}, {Kwon}, {Lai}, {Qiu}, {Furuya}, {Berry}, \& {JCMT BISTRO Survey
  Team}}]{2018ApJ...860L...6P}
{Pattle}, K., {Ward-Thompson}, D., {Hasegawa}, T., {et~al.} 2018, \apjl, 860,
  L6, \dodoi{10.3847/2041-8213/aac771}

\bibitem[{{Pellegrini} {et~al.}(2011){Pellegrini}, {Baldwin}, \&
  {Ferland}}]{2011ApJ...738...34P}
{Pellegrini}, E.~W., {Baldwin}, J.~A., \& {Ferland}, G.~J. 2011, \apj, 738, 34,
  \dodoi{10.1088/0004-637X/738/1/34}

\bibitem[{{Peretto} {et~al.}(2006){Peretto}, {Andr{\'e}}, \&
  {Belloche}}]{2006A&A...445..979P}
{Peretto}, N., {Andr{\'e}}, P., \& {Belloche}, A. 2006, \aap, 445, 979,
  \dodoi{10.1051/0004-6361:20053324}

\bibitem[{{Peretto} {et~al.}(2013){Peretto}, {Fuller}, {Duarte-Cabral},
  {Avison}, {Hennebelle}, {Pineda}, {Andr{\'e}}, {Bontemps}, {Motte},
  {Schneider}, \& {Molinari}}]{2013A&A...555A.112P}
{Peretto}, N., {Fuller}, G.~A., {Duarte-Cabral}, A., {et~al.} 2013, \aap, 555,
  A112, \dodoi{10.1051/0004-6361/201321318}

\bibitem[{{Phillips} {et~al.}(1979){Phillips}, {Huggins}, {Wannier}, \&
  {Scoville}}]{1979ApJ...231..720P}
{Phillips}, T.~G., {Huggins}, P.~J., {Wannier}, P.~G., \& {Scoville}, N.~Z.
  1979, \apj, 231, 720, \dodoi{10.1086/157237}

\bibitem[{{Pineda} {et~al.}(2008){Pineda}, {Caselli}, \&
  {Goodman}}]{2008ApJ...679..481P}
{Pineda}, J.~E., {Caselli}, P., \& {Goodman}, A.~A. 2008, \apj, 679, 481,
  \dodoi{10.1086/586883}
  
  \bibitem[Schneider et al.(2020)]{2020PASP..132j4301S} Schneider, N., Simon, R., Guevara, C., et al.\ 2020, \pasp, 132, 104301. doi:10.1088/1538-3873/aba840

\bibitem[{{Santos} {et~al.}(2019){Santos}, {Chuss}, {Dowell}, {Houde},
  {Looney}, {Lopez Rodriguez}, {Novak}, {Ward-Thompson}, {Berthoud}, {Dale},
  {Guerra}, {Hamilton}, {Hanany}, {Harper}, {Henning}, {Jones}, {Lazarian},
  {Michail}, {Morris}, {Staguhn}, {Stephens}, {Tassis}, {Trinh}, {Van Camp},
  {Volpert}, \& {Wollack}}]{2019ApJ...882..113S}
{Santos}, F.~P., {Chuss}, D.~T., {Dowell}, C.~D., {et~al.} 2019, \apj, 882,
  113, \dodoi{10.3847/1538-4357/ab3407}

\bibitem[{{Schaefer}(2008)}]{2008AJ....135..112S}
{Schaefer}, B.~E. 2008, \aj, 135, 112, \dodoi{10.1088/0004-6256/135/1/112}

\bibitem[{{Schneider} {et~al.}(2010){Schneider}, {Csengeri}, {Bontemps},
  {Motte}, {Simon}, {Hennebelle}, {Federrath}, \&
  {Klessen}}]{2010A&A...520A..49S}
{Schneider}, N., {Csengeri}, T., {Bontemps}, S., {et~al.} 2010, \aap, 520, A49,
  \dodoi{10.1051/0004-6361/201014481}

\bibitem[{{Schneider} {et~al.}(2013){Schneider}, {Andr{\'e}}, {K{\"o}nyves},
  {Bontemps}, {Motte}, {Federrath}, {Ward-Thompson}, {Arzoumanian},
  {Benedettini}, {Bressert}, {Didelon}, {Di Francesco}, {Griffin}, {Hennemann},
  {Hill}, {Palmeirim}, {Pezzuto}, {Peretto}, {Roy}, {Rygl}, {Spinoglio}, \&
  {White}}]{2013ApJ...766L..17S}
{Schneider}, N., {Andr{\'e}}, P., {K{\"o}nyves}, V., {et~al.} 2013, \apjl, 766,
  L17, \dodoi{10.1088/2041-8205/766/2/L17}

\bibitem[{{Schneider} {et~al.}(2015){Schneider}, {Bontemps}, {Girichidis},
  {Rayner}, {Motte}, {Andr{\'e}}, {Russeil}, {Abergel}, {Anderson},
  {Arzoumanian}, {Benedettini}, {Csengeri}, {Didelon}, {di}, {Griffin}, {Hill},
  {Klessen}, {Ossenkopf}, {Pezzuto}, {Rivera-Ingraham}, {Spinoglio},
  {Tremblin}, \& {Zavagno}}]{2015MNRAS.453L..41S}
{Schneider}, N., {Bontemps}, S., {Girichidis}, P., {et~al.} 2015, \mnras, 453,
  L41, \dodoi{10.1093/mnrasl/slv101}

\bibitem[Schneider et al.(2022)]{2022arXiv220714604S} Schneider, N., Ossenkopf-Okada, V., Clarke, S., et al.\ 2022, arXiv:2207.14604

\bibitem[{{Skalidis} {et~al.}(2021{\natexlab{a}}){Skalidis}, {Sternberg},
  {Beattie}, {Pavlidou}, \& {Tassis}}]{2021A&A...656A.118S}
{Skalidis}, R., {Sternberg}, J., {Beattie}, J.~R., {Pavlidou}, V., \& {Tassis},
  K. 2021{\natexlab{a}}, \aap, 656, A118, \dodoi{10.1051/0004-6361/202142045}
  
\bibitem[{{Skalidis} \& {Tassis}(2021)}]{2021A&A...647A.186S}
{Skalidis}, R., \& {Tassis}, K. 2021, \aap, 647, A186,
  \dodoi{10.1051/0004-6361/202039779}
  
\bibitem[{{Skalidis} {et~al.}(2021{\natexlab{b}}){Skalidis}, {Tassis}, {Panopoulou},
  {Pineda}, {Gong}, {Mandarakas}, {Blinov}, {Kiehlmann}, \&
  {Kypriotakis}}]{2021arXiv211011878S}
{Skalidis}, R., {Tassis}, K., {Panopoulou}, G.~V., {et~al.} 2021, arXiv
  e-prints, arXiv:2110.11878.
\newblock \doarXiv{2110.11878}

\bibitem[{{Soam} {et~al.}(2018{\natexlab{a}}){Soam}, {Maheswar}, {Lee}, {Neha},
  \& {Kim}}]{2018MNRAS.476.4782S}
{Soam}, A., {Maheswar}, G., {Lee}, C.~W., {Neha}, S., \& {Kim}, K.-T.
  2018{\natexlab{a}}, \mnras, 476, 4782, \dodoi{10.1093/mnras/sty517}

\bibitem[{{Soam} {et~al.}(2018{\natexlab{b}}){Soam}, {Pattle}, {Ward-Thompson},
  {Lee}, {Sadavoy}, {Koch}, {Kim}, {Kwon}, {Kwon}, {Arzoumanian}, {Berry},
  {Hoang}, {Tamura}, {Lee}, {Liu}, {Kim}, {Johnstone}, {Nakamura}, {Lyo},
  {Onaka}, {Kim}, {Furuya}, {Hasegawa}, {Lai}, {Bastien}, {Chung}, {Kim},
  {Parsons}, {Rawlings}, {Mairs}, {Graves}, {Robitaille}, {Liu}, {Whitworth},
  {Eswaraiah}, {Rao}, {Yoo}, {Houde}, {Kang}, {Doi}, {Choi}, {Kang},
  {Coud{\'e}}, {Li}, {Matsumura}, {Matthews}, {Pon}, {Di Francesco}, {Hayashi},
  {Kawabata}, {Inutsuka}, {Qiu}, {Franzmann}, {Friberg}, {Greaves}, {Kirk},
  {Li}, {Shinnaga}, {van Loo}, {Aso}, {Byun}, {Chen}, {Chen}, {Chen}, {Ching},
  {Cho}, {Chrysostomou}, {Drabek-Maunder}, {Eyres}, {Fiege}, {Friesen},
  {Fuller}, {Gledhill}, {Griffin}, {Gu}, {Hatchell}, {Holland}, {Inoue},
  {Iwasaki}, {Jeong}, {Kang}, {Kemper}, {Kim}, {Kim}, {Lacaille}, {Lee}, {Li},
  {Liu}, {Liu}, {Moriarty-Schieven}, {Nakanishi}, {Ohashi}, {Peretto}, {Pyo},
  {Qian}, {Retter}, {Richer}, {Rigby}, {Savini}, {Scaife}, {Tang}, {Tomisaka},
  {Wang}, {Wang}, {Yen}, {Yuan}, {Zhang}, {Zhang}, {Zhou}, {Zhu}, {Andr{\'e}},
  {Dowell}, {Falle}, {Tsukamoto}, {Kanamori}, {Kataoka}, {Kobayashi}, {Nagata},
  {Saito}, {Seta}, {Hwang}, {Han}, {Lee}, \& {Zenko}}]{2018ApJ...861...65S}
{Soam}, A., {Pattle}, K., {Ward-Thompson}, D., {et~al.} 2018{\natexlab{b}},
  \apj, 861, 65, \dodoi{10.3847/1538-4357/aac4a6}

\bibitem[{{Tahani} {et~al.}(2019){Tahani}, {Plume}, {Brown}, {Soler}, \&
  {Kainulainen}}]{2019A&A...632A..68T}
{Tahani}, M., {Plume}, R., {Brown}, J.~C., {Soler}, J.~D., \& {Kainulainen}, J.
  2019, \aap, 632, A68, \dodoi{10.1051/0004-6361/201936280}

\bibitem[{{Tang} {et~al.}(2019){Tang}, {Koch}, {Peretto}, {Novak},
  {Duarte-Cabral}, {Chapman}, {Hsieh}, \& {Yen}}]{2019ApJ...878...10T}
{Tang}, Y.-W., {Koch}, P.~M., {Peretto}, N., {et~al.} 2019, \apj, 878, 10,
  \dodoi{10.3847/1538-4357/ab1484}

\bibitem[{{Tiwari} {et~al.}(2021){Tiwari}, {Karim}, {Pound}, {Wolfire},
  {Jacob}, {Buchbender}, {G{\"u}sten}, {Guevara}, {Higgins}, {Kabanovic},
  {Pabst}, {Ricken}, {Schneider}, {Simon}, {Stutzki}, \&
  {Tielens}}]{2021ApJ...914..117T}
{Tiwari}, M., {Karim}, R., {Pound}, M.~W., {et~al.} 2021, \apj, 914, 117,
  \dodoi{10.3847/1538-4357/abf6ce}

\bibitem[{{Townsley} {et~al.}(2006){Townsley}, {Broos}, {Feigelson}, {Brandl},
  {Chu}, {Garmire}, \& {Pavlov}}]{2006AJ....131.2140T}
{Townsley}, L.~K., {Broos}, P.~S., {Feigelson}, E.~D., {et~al.} 2006, \aj, 131,
  2140, \dodoi{10.1086/500532}

\bibitem[Tram \& Hoang(2022)]{2022arXiv220813195T} Tram, L.~N. \& Hoang, T.\ 2022, arXiv:2208.13195

\bibitem[{{Tram} {et~al.}(2021{\natexlab{c}}){Tram}, {Hoang},
  {Lopez-Rodriguez}, {Coud{\'e}}, {Soam}, {Andersson}, {Lee}, {Bonne}, {Vacca},
  \& {Lee}}]{2021arXiv210509530T}
{Tram}, L.~N., {Hoang}, T., {Lopez-Rodriguez}, E., {et~al.} 2021{\natexlab{c}},
  arXiv e-prints, arXiv:2105.09530.
\newblock \doarXiv{2105.09530}

\bibitem[{{Young} {et~al.}(2012){Young}, {Becklin}, {Marcum}, {Roellig}, {De
  Buizer}, {Herter}, {G{\"u}sten}, {Dunham}, {Temi}, {Andersson}, {Backman},
  {Burgdorf}, {Caroff}, {Casey}, {Davidson}, {Erickson}, {Gehrz}, {Harper},
  {Harvey}, {Helton}, {Horner}, {Howard}, {Klein}, {Krabbe}, {McLean}, {Meyer},
  {Miles}, {Morris}, {Reach}, {Rho}, {Richter}, {Roeser}, {Sandell}, {Sankrit},
  {Savage}, {Smith}, {Shuping}, {Vacca}, {Vaillancourt}, {Wolf}, \&
  {Zinnecker}}]{2012ApJ...749L..17Y}
{Young}, E.~T., {Becklin}, E.~E., {Marcum}, P.~M., {et~al.} 2012, \apjl, 749,
  L17, \dodoi{10.1088/2041-8205/749/2/L17}

\bibitem[{{Yuen} \& {Lazarian}(2017{\natexlab{a}})}]{2017ApJ...837L..24Y}
{Yuen}, K.~H., \& {Lazarian}, A. 2017{\natexlab{a}}, \apjl, 837, L24,
  \dodoi{10.3847/2041-8213/aa6255}

\bibitem[{{Yuen} \& {Lazarian}(2017{\natexlab{b}})}]{2017arXiv170303026Y}
---. 2017{\natexlab{b}}, arXiv e-prints, arXiv:1703.03026.
\newblock \doarXiv{1703.03026}




\end{thebibliography}
\end{document}